# THE SECURE MACHINE: EFFICIENT SECURE EXECUTION ON UNTRUSTED PLATFORMS

Ofir Shwartz, Yitzhak Birk

{ofirshw@tx , birk@ee}.technion.ac.il

Electrical Engineering Department, Technion

Israel

# Table of Contents

# List of Figures

# List of Pseudocodes

# List of Tables

# Abstract

Remote computing services on shared, third party platforms (e.g., virtualization and cloud services) offer advantages to organizations and individuals, putting at their disposal enormous computing resources while permitting them to pay only for the resources actually used. Unfortunately, such environments are prone to attacks by hackers, adversarial users of the systems, or even the owner of the service. Such attacks may address the operating system, hypervisor, virtual machine monitor (VMM), or even the hardware itself. It would therefore be extremely beneficial if users could ensure the security of their programs in such environments, as this would likely lead to a dramatic expansion of their use for applications ranging from research, through finance, and to medical systems. Specifically, the confidentiality of the code and data must be preserved, and tampering with those or with the sequence of execution must be detected.

Although prior works suggested various ideas and architectures, they are missing key features for becoming practical and ubiquitous:

a) Supporting existing application binaries

b) Providing security without significant performance, power or cost penalties.

c) Being scalable to many compute nodes

In this work we present the **Secure Machine, SeM** for short, a CPU architecture extension for secure computing. SeM uses a small amount of in-chip additional hardware that monitors key communication channels inside the CPU chip, and only acts when required. SeM provides confidentiality and integrity for a secure program without trusting the platform software or any off-chip hardware. SeM supports existing binaries of single- and multi-threaded applications running on single- or multi-core, multi-CPU, or multi-node computing environment, and it is also extendable to accelerators (e.g., GPU, smart NICs), which allows the use of secure heterogeneous systems. The performance reduction caused by it is only few percent, most of which is due to the memory encryption layer that is commonly used in many secure architectures.

We also developed **SeM-Prepare**, a software tool that automatically instruments existing applications (binaries) with additional instructions so they can be securely executed on our architecture without requiring any programming efforts or the availability of the desired program's source code.

The development of SeM included the creation of several independent and important building blocks:

To enable secure data sharing in shared memory environments, we developed **Secure Distributed Shared Memory** (**SDSM**), an efficient (time and memory) algorithm for allowing thousands of compute nodes to share data securely while running on an untrusted computing environment. SDSM shows a negligible reduction in performance, and it requires negligible and hardware resources.

We developed **Distributed Memory Integrity Trees**, a method for enhancing single node integrity trees for preserving the integrity of a distributed application running on an untrusted computing environment. We show that our method is applicable to existing single node integrity trees such as Merkle Tree, Bonsai Merkle Tree, and Intel's SGX memory integrity engine.

All these building blocks may be used together to form a practical secure system, and some can be used in conjunction with other secure systems.

SeM is thus an important step towards high performance, cost effective secure computing, which will allow wide usage of those for sensitive application.

# List of Abbreviations

ACK – Acknowledgement response
AES - Advanced Encryption Standard
AI - Artificial Inteligence
API – Application Program Interface
BMT – Bonsai Merkle Tree
CM – Counter Mode (encryption)
CMP – Chip Multi Processor
CPU – Central Processing Unit
DBMT – Distributed Bonsai Merkle Tree
DIT – Distributed Integrity Tree
DMEE – Distributed Memory Encryption Engine
DMT – Distributed Merkle Tree
DSM – Distributed Shared Memory
ELF – Executable and Linkable Format (linux binary)
FPGA – Field-Programmable Gate Array
GCC - GNU Compiler Collection (compiler)
GCM – Galios Counter Mode (encryption)
GHASH – Galios Hash (secure hash)
GPU - Graphics Processing Unit
I/O – Input / Output
IVLCS – Integrity-Verified Local Coherence State
KB – Keystream Block (encryption / decryption pad)
LEP – Legal Engry Point
LSb – Least Significant Bit
MAC – Message Authenticating Code
MEE – Memory Encryption Engine
MESI – Modified / Exclusive / Shared / Invalid (coherence protocol)
MPI – Message Passing Interface
NIC – Network Interface Card
NSS – Non-Secure Stack
NUMA – Non-Uniform Memory Access
OS – Operating System
ParSeM – Parallel Secure Machine
PID – Process ID
RAM – Random Access Memory
RDMA – Remote Direct Memory Access
ROP – Return Oriented Programming
RSA - Rivest–Shamir–Adleman (encryption)
SDK – Software Development Kit
SDSM – Secure Distributed Shared Memory
SeM – Secure Machine

SHEF – Signal Handling Entry Function
SLF – Shared Library Function
SMU – Security Management Unit
SS – Secure Stack
SSS – SMU Saeled Storage
TA – Trusted Area
TCB – Trusted Code base
TCM – Trusted Coherence Manager
TID – Thread ID
TLB – Translation Lookaside Buffer
TSC – Thread Secret Context
VA – Virtual Address
VM – Virtual Machine
VMM – Virtual Machine Monitor

# Chapter 1 Introduction

## 1 Objective

Security is a major concern in any shared (multi-user) computer system, and this concern is growing with the proliferation of shared third-party computing platforms, e.g., public clouds. A traditional key role of an operating system (OS) is to provide isolation among processes, but the trust in those has declined due to complexity, modularity, bugs, cyber-attacks, and in some cases simply not trusting their owner. Ensuring privacy in such environments would therefore be extremely beneficial, enabling a dramatic expansion of their applicability to sensitive applications such as financial and medical systems. Specifically, maintaining the confidentiality of user code and data, and detecting any malicious alteration thereof, are key requirements.

The goal of this research is to provide **security** for **general purpose computing** in a **usable manner**.

**Security** essentially means the ability to send a (secure) program for execution on a remote computer, in which even a privileged attacker cannot inspect its code, data, or temporary data, and cannot alter with those or with the sequence of execution without being noticed. Some alterations are detected immediately, some will be detected as soon as the altered data or state is used, and some (e.g. alterations to results that are stored, not being touched again, and waiting to be shipped) will only be detected by the owner upon completion. (The threat model will be discussed shortly.)

**General purpose computing** essentially means computers and CPUs as we know them today, supporting settings that are currently supported, running operating systems, hypervisors, and virtual machine monitors (VMMs) as we run today (with all their inherent services).

**Usable manner** essentially means supporting existing programs, and running those with small performance and power overheads.

The desired support for existing computing settings yields in a challenge of adding the abovementioned security while introducing only few changes to the system. A direct benefit of this approach is that the lesser the changes, the simpler it is to embed such solutions into existing systems. Additionally, performance and power are often heavily harmed for added security, yet we strive to minimize the overheads in those.

## 2 Threat Model

The user possesses a private computer, which is trusted and is assumed to be completely protected against any attack. In the shared remote computer(s) (“the computer”), an adversary:

- Completely controls the OS/VMM/hypervisor, running in the most privileged ring, including the ability to change or implant code both in advance and during runtime.
- Has physical access to off-CPU signals on the board, memories and I/O, and an ability to monitor, alter, or emulate any of those, including the communication network.
- Before, during and after execution, may try to read or change user code, data and results, or interfere with the OS services that the program receives.
- Cannot physically inspect or alter CPU chip internals (including on-chip cache). This chip (hardware) is assumed to be correct, and its manufacturer is trusted so that silicon-internal secrets stored during manufacturing are not leaked.

**Remark.** Alterations of the untrusted OS services (Iago attacks [55]), if called intentionally by the user program, are unavoidable and can only be detected by the user program. In defense, the user can use shims that verify return values [2,13,14,40], submit an entire verified OS with his code, or use secure boot to validate the authenticity of the OS [15].

User program bugs are not addressed here. Side and covert channel attacks [51,19] are important, but are not addressed here, though solutions suggested in other works are applicable to SeM as well. Denial of service [23] is possible, as the platform owner may simply shut it down.

## 3 Main Contributions

To address the abovementioned goal, we identified several parts of computer systems that need to be considered.

First, we address the case of a single-core CPU that runs single-threaded programs securely on platforms in which everything (hardware and software) is untrusted, except for the CPU and its on-chip cache. For this setting we present the Secure Machine (SeM) (Chapter 2). SeM is a CPU architecture extension that is largely CPU-architecture agnostic, and it requires only small modifications to existing CPUs (mostly between major CPU blocks). By a novel binding of code and data that are under the same cryptographically validated (proven) ownership, SeM assures that only the owning code can access its data in a manner that no other software, privileged as it may be, can break. Using the ownership proof, SeM automatically and quickly switches between security modes. SeM provides secure execution for existing program binaries, and its performance and power are similar to a system with no security.

Next, we addresses parallel execution on single- and multi-core CPUs, and multi-computer settings, where for parallelism we refer to the multi-threading programming model. We therefore add node-to-node secure communication, and hardware assisted secure thread creation, migration, and termination (Chapter 3). Supporting multi-node secure computation, we also address heterogeneous systems containing Secure Accelerators.

We developed SeM-Prepare, a tool that automatically and statically instruments existing binaries, thereby making them SeM-ready (Chapter 5). To our knowledge, SeM is the only secure architecture that accepts *unmodified binaries*, where the ones that come closer (Haven [43] and Graphene [49]) support *unmodified source code*, and thus require the program's source code and recompilation. For evaluation purposes, we developed SeM-Simulator, a tool that runs SeM-ready binaries, simulating SeM's behavior, and collects statistics (Appendix A).

To run multi-threaded programs on a multi-computer environment, a memory sharing model must be set. Since multi-threaded programs share data using a shared memory space, distributed shared memory [80] (DSM) is the native setting to choose. With the assumption of compute nodes that are connected via an untrusted medium, a security layer has to be added to protect shared data in between the compute nodes. For that, we present Secure Distribute Shared Memory (SDSM) (Chapter 6). By hiding the encryption latency behind the inherent communication latencies, SDSM can serve thousands of compute nodes with negligible performance reduction (compared to DSM with no security); by smart allocation of resources, SDSM minimizes the hardware resources and extra work added for encryption.

Finally, although SDSM protects the confidentiality and integrity of data blocks during their transfer between compute nodes, protecting the integrity of a distributed program (for blocks that are not necessarily transferred) requires additional measures. Traditionally, integrity trees are used to protect the integrity of programs running on single compute nodes. We present the Distributed Integrity Trees (DIT), for extending existing integrity trees to support distributed settings, by adding a per-node integrity verified view of the coherence state of memory blocks (Chapter 7). While adding support for distributed settings, DIT adds no overhead on top of existing single-node integrity trees.

Together, these form a complete secure execution system that can be practically adopted into cloud systems, allowing to securely and efficiently run existing (and future) sequential or parallel programs.

# 4 Related Work

In this section we review related work and compare it with SeM's goals.

Proposed solutions for running a secret workload on an untrusted machine rely on trusted hardware, and some also on trusted software. Each type has different inherent properties.

**Trusted software solutions.** Some, e.g. InkTag [14], run a trusted hypervisor; Overshadow [2] and PrivExec [25] use trusted VMMs; Overshadow runs unmodified programs on an untrusted OS, by using trusted shims that are part of the trusted VMM. These shims mediate between the secure program and the untrusted OS, to enable the use of system calls. For thread creation, the shims prepare the new thread's context, and sets the correct register values. Having some similarities in approach, its less strict threat model allows solutions that are inapplicable to SeM. Trustvisor [24] provides application separation on a trusted OS, and Virtual Ghost [40] recompiles the OS to insert hardware abstraction layer code. OP-TEE [85] is an open-source API to develop trusted applications. Although software based solutions provide many insights and tools, they are inherently susceptible to attacks on or by an untrusted service provider; SeM thus differs critically in the assumed threat model.

**Hardware based solutions** require hardware modifications, but offer stronger capabilities, most notably secure compartments. E.g., [4,5,20,41] protect a VM (Bastion [20] is a hardware attested software solution), while SeM protects at process granularity. Finer granularity serves con-current adversarial programs without an OS for each, thus having less code in the trusted code base (TCB), making it more reliable (fewer bugs). XOM [1] protects at process granularity, but suffers from performance and applicability issues, and has vulnerabilities like memory integrity. AEGIS [3] and SecureBlue++ [18] solved its vulnerabilities and improved performance, but unlike SeM, still exhibit large performance degradation. Komodo [84] uses a hardware-software solution, where hardware maintains encryption and cache separation, and software manages virtual enclaves. Sanctum [86] is also a hardware-software solution that aims at protecting against memory access pattern leaks and cache timing attacks, but it does not address physical attacks of any kind. SecureME [28] relies on a verified machine state using secure boot, and SICE [35] relies on trusted BIOS, both are producing a hash for attesting the machine state. However, state verification of the OS and drivers is problematic, as updates lead to new (therefore untrusted) state. Others assume that the machine is mostly maintained against hostile activities [6] (e.g. a rogue employee), or protect only one process at a time [41]. Iso-X [13] isolates at page granularity, requiring explicit programmer guidance for which areas to protect. [13,20,21] do not address hardware attacks. Unlike SeM, none support parallel or distributed computing, or dynamic job migration.

Intel's SGX [22], addressing SeM's threat model, allows an unprotected process to instantiate a small secure memory region, dubbed enclave. Enclaved code and data are protected from software and hardware attacks. SGX2 [54] specifications add dynamic memory allocation, enclave runtime permission management, and lazy loading of code into an enclave. Operations inside an enclave are limited, and costly overhead for entering/exiting the enclave has been spotted as a major performance bottleneck [26,52]. Furthermore, although SGX uses counter mode encryption, its integrity tree protects the data directly [81] (in a Merkle Tree [72] manner) rather than the counters (in a Bonsai Merkle Tree [34] manner). This design choice limits SGX's supported memory range, and results in a reduced caching efficiency for the integrity tree, which thus harms performance. Using a Bonsai Merkle Tree version of MEE is therefore an opportunity for a better future SGX. SGX's SDK directly targets applications developed for it, and these cannot run elsewhere, which breaks the x86/x64 compatibility. Software extensions to SGX, such as HAVEN [43] or Graphene [49], enhance its applicability up to running some unmodified code by adding LibOS to the enclave, and SCONE [27] addresses that using special Linux containers. Eleos [26] and Hotcalls [52] suggest performance enhancements by adding software optimizations, but the resulting performance is still very limited. SGX limits the number of active threads upon the creation of an enclave, so dynamic creation of new threads is limited, though SGX2's dynamic memory allocation will tentatively solve this problem. Furthermore, SGX threads cannot be migrated, even between different cores of the same CMP. In contrast, SeM fully supports threads conventionally, including thread migrations.

# 5 Organization of This Thesis

This rest of the thesis is organized as follows. In Chapter 2 we first present the basic mechanisms of SeM, supporting single-thread single-core, and in Chapter 3 we extend it to support multi-threads and single- and multi-core and multi-node settings. In Chapter 4 we discuss types of supported programs and another alternative of the use of shared library functions. In Chapter 5 we present the SeM-Prepare tool. Chapter 6 presents the Secure Distributed Shared Memory, and Chapter 7 presents Distributed Memory Integrity Trees. Finally, Chapter 8 offers concluding remarks.

In Appendix A we present our SeM simulator, and in Appendix B we provide a detailed list of SeM instructions.

Throughout the thesis we use the term "cache blocks" when referring to "cache lines".

# Chapter 2

# The Basic Secure Machine (SeM)

## 1 Introduction

In this chapter we describe the basic mechanisms of the Secure Machine (SeM), addressing single-threaded programs running on a single-core CPU.

We consider a user with a trusted private computer, wishing to run his program on a remote computer such as a public cloud, which concurrently serves multiple (possibly mutually adversarial) users. The user sends a program, comprising code and data, for execution. It can be provided as files on disk or via a network, and program output is collected similarly.

The remote computer's system software may be adversarial, and the only trusted hardware is its CPU chip (with its in-chip caches). Support for identity authentication is assumed (a certificate), merely enabling the establishment of a secure virtual communication channel between the user's trusted private computer and the remote trusted CPU chip. This is done using a secret signature key stored in the trusted CPU chip by some trusted agent, possibly the chip's manufacturer.

In this setting, and without requiring OS modification, we wish to enable a user program to run conventionally on the remote machine: switch context in and out while maintaining its state, allocate memory dynamically, use I/O, and invoke system calls. All this while ensuring its confidentiality and integrity: its secrets (code, data, temporary values, and data communicated via I/O) cannot be discovered by any adversary (confidentiality); and its results, including any information that it sends to the outside world, are unaltered, or an alteration is detected (integrity). (Nonetheless, misbehaving untrusted OS services that are called by the user program can usually only be detected by a user program.)

Overshadow [2], some of whose ideas we adopt, is a comprehensive solution that provides some of the above (e.g., securely running an unchanged application); however, as it is based on a trusted virtual machine monitor (VMM), it does not address an untrusted owner (which can manipulate the VMM, or track and modify memory). Another is Intel's SGX [22]; although it addresses an untrusted owner, its key challenges are applicability to existing programs, resource efficiency, and performance reduction that is partly caused by expensive security mode switch operations (EENTER, EEXIT). Software extensions to SGX include PANOPLY [56] (helps develop SGX applications), Haven [43] (for Windows) and Graphene [49] (Linux); these two accept unmodified applications but impose various restrictions on the programs, require a large TCB, and substantially degrade performance. SCONE [27], a small-TCB container-based solution for running unchanged applications, comes closer to achieving the goal. However: the user needs to know security-related aspects of the service to create an image from application binaries; customized support for libraries that use system calls must be developed (currently only libc); performance drops by tens of percents; and it is unclear whether it supports signals and exceptions. Chapter 1 reviews additional related work.

We present the Secure Machine (SeM), an extended CPU architecture enabling secure computing on a computer that is managed by an untrusted entity, jointly addressing the aforementioned needs without any restrictions on the flow of the system or on the untrusted OS/VMM/hypervisor. (By abuse of terminology, we use the term SeM to refer both to the architecture extension and to the resulting extended architecture, with the meaning being obvious from the context in each case.) SeM can easily be integrated into any CPU architecture, and incurs only tiny performance, power, and area penalties. We use prior-art memory encryption and integrity [34], and add a novel cache and register management layer, as well as setup and termination capabilities.

In a nutshell (and ignoring setup), a secure program invokes its trusted instructions, and accesses its trusted data. When the need arises (e.g., OS kernel code for performing a context switch), it invokes an untrusted instruction, at which time the secure program's registers are immediately hidden by SeM at a tiny cost (Sec. 2.4). The trusted cached blocks are only accessible by same-process trusted code, so the trusted memory space is protected (Sec. 2.3). When a trusted instruction is subsequently invoked (e.g., returning to user code), the registers are restored immediately. As done in previous works, memory blocks that are fetched or evicted are automatically protected by a memory encryption and validation layer.

SeM's main novel protections are its ability to automatically hide register values on the first invocation of an untrusted instruction, and its ability to block untrusted memory-access instructions from accessing trusted cached blocks. SeM's main novel performance benefits are its ability to hide and restore the registers' data in a single clock cycle, and fast context switches without flushing the cache. SeM's novel applicability benefits are that the CPU is largely unchanged, the security mechanisms are hidden from the program and the unmodified OS, and the programmer need not modify the application code.

We designed and implemented *SeM-Prepare*, a tool running on the user's trusted computer for preparing existing binaries for SeM (running offline or in the course of program submission to the cloud, sometimes referred to as *application deployment*). This tool is described in detail in Chapter 5.

Then, we run SeM-Prepare on the SPEC CPU2006 benchmark suite [57] binaries to make them SeM-Ready. We designed and implemented *SeM-Simulator* to execute the resulting binaries, thereby demonstrating completeness and correct results, and showing overheads to be negligible. The simulator is described in detail in Appendix A.

This chapter focuses on the core architecture and on the protection of the user's process from (software and off-chip hardware) adversaries. Extensions to support multi-thread, -core and –node settings, as well as task and thread migration are discussed in the next chapter.

The key building blocks contributed by this work are:

- A novel hardware-maintained secure process context management, allowing efficient and secure switching between programs, and between the program and the OS.
- Secure Access, a novel method for cache access control, coupling authenticated instructions and authenticated data; this allows unencrypted code and data of adversarial programs to securely co-reside in cache.
- An automatic tool for preparing existing binaries for SeM; no programming efforts are required.

# 2 The SeM Architecture

## 2.1 Overview

The basic Secure Machine (Figure 1) comprises a single-core multi-user computer. SeM's main hardware is the Security Management Unit (SMU). It exclusively manages and controls access to the CPU registers (Sec. 2.4) and caches (Sec. 2.3) in an on-chip physical domain dubbed the Trusted Area (TA), and serves as a gatekeeper between the TA and the untrusted world.

To ensure confidentiality, the user's code and data are encrypted whenever outside the TA and, for integrity, are signed using a Message Authentication Code (MAC). We use a counter mode (CM) technique for memory encryption, and a lightweight secure hash for authentication (MAC); GCM [33] is an authenticated encryption technique that provides both. We use Bonsai Merkle Tree (BMT) with a TA-resident root hash [34] to keep the integrity of the CM seeds. All these are widely used in previous works, and have been proven safe and efficient (in performance and memory footprint) [33,34]. SeM is agnostic to the memory encryption and authentication techniques, as long as these provide memory confidentiality (when desired) and integrity breech indication (mandatory).

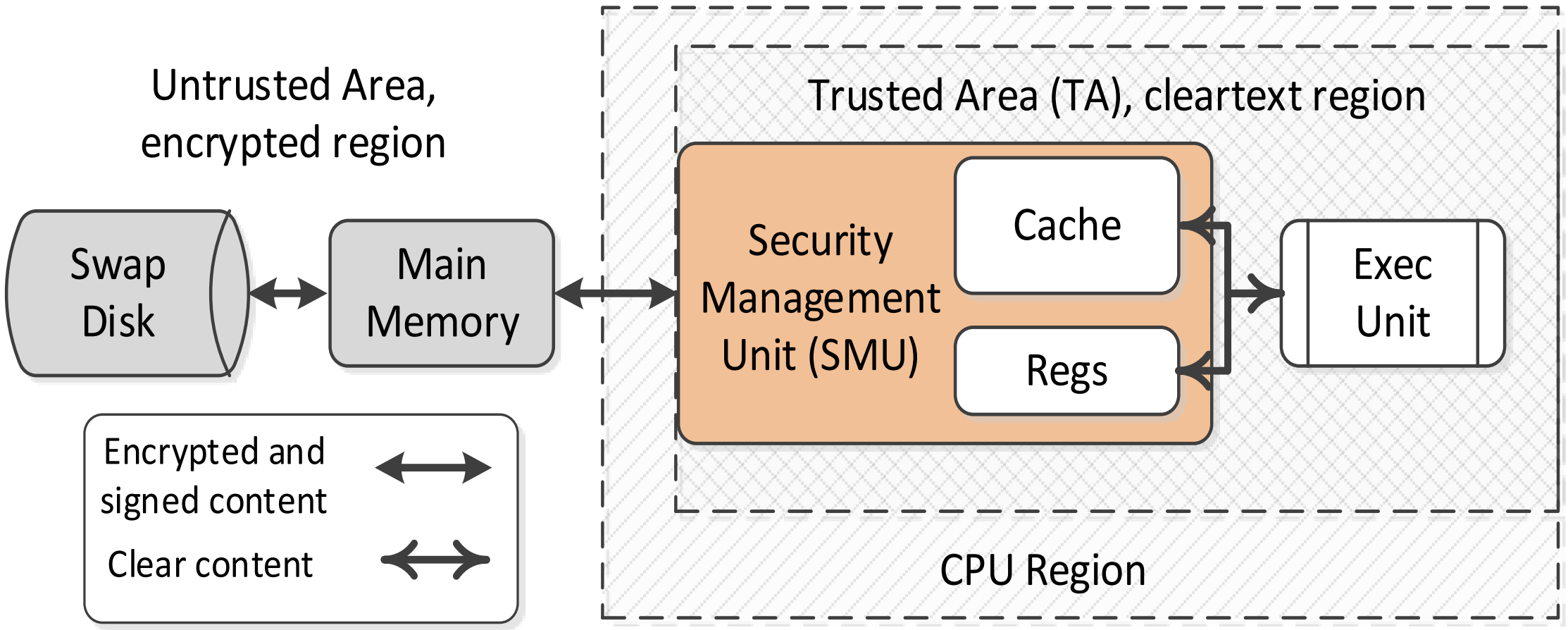

*Figure 1: SeM hardware block diagram*

CM encryption protects the memory at cache block granularity by assigning a seed value (commonly 64 bits) for each block's virtual address; these seeds are cached. BMT is a hash tree used for maintaining the integrity of the seeds, so that an old data block with its corresponding old seed cannot be injected into memory. The BMT blocks are also cached, so only missing BMT nodes (rather than the entire hash tree) need to validate when fetched. Each CM counter protects an entire data block, and each BMT hash (last layer) protects an entire CM counters block, so both natively experience far better locality than normal data blocks. The performance implication of those is small [33,34], and is not unique to SeM.

Both CM encryption and BMT require memory for metadata (encryption seeds and a hash tree). This memory need not be protected, because an attacker is unlikely to inject correct values without holding the secret keys [33,34]. These small regions are allocated and zeroed at the secure program's request, and discarded when the program finishes or crashes; if the OS fails to cooperate, an error is detected upon access. The SMU performs these operations using metadata (e.g., secret encryption and authentication keys) stored securely for each secure program during its setup. As in many other secure architectures [3,7], CPU debugging (which exposes detailed state), is disabled for a secure program.

SeM can be tailored to either physically or virtually addressed caches. (By "virtually addressed caches" we refer to virtually indexed caches with either virtual or physical tags.) When fetching a missing block, its virtual address is known in both cases. When evicting, a virtually addressed cache can store the updated seed immediately, but a physically addressed cache needs to perform a reverse TLB lookup to find the correct seed to update; this can be done in the background, without delaying the actual eviction. From here on we assume virtually addressed caches.

The rest of the computer is largely unmodified. The OS can start and stop processes, switch among them, and perform any conventional OS task, but it only accesses the cache under SMU supervision. Likewise for the hypervisor or any layer between a user program and the actual CPU hardware.

**The flow (e.g., submitting to the cloud, part of the 'secure cloud deployment')**: A user program's binary (in his own trusted computer) is statically linked with shared library functions that it requires (similarly to [55]), and is then automatically instrumented with some additional instructions (explained later). Next, it is encrypted and signed, and is then sent to SeM through an untrusted medium. To execute the program, a secure connection is established between the user's computer and the SMU to securely store the program's settings (e.g., keys) in the SMU. This enables the SMU to provide each secure program with encryption, decryption and authentication services for code and data using the program's unique keys. The program is then executed, using these keys. Upon completion, the user may collect the encrypted output and validation information from the SMU.

Many previous works required attestation of the machine's cumulative state [28,35]; this state is very hard to verify, as it varies among systems and changes with system updates. (E.g., each OS update modifies OS executables, resulting in a different state hash.) In SeM, we use simple attestation to authenticate the existence of a genuine SMU, regardless of the state of the machine. This is easily doable using a publicly provable signature [11] (Sec. 1); e.g., by requesting the SMU to sign a requestor-generated random number.

SeM runs an untrusted management program for direct communication with the (possibly remote) user. Data passing through the management program is safe, as it is encrypted and the decryption keys are only known to the SMU.

## 2.2 Security Management Unit (SMU)

The SMU is SeM's core hardware. It resides in the CPU chip, situated between the last level on-chip trusted cache and the rest of the memory system, and between the L1 cache and the execution unit. It creates a boundary between the TA (comprising the execution unit, registers, trusted caches, etc.) and the untrusted domain (comprising optional untrusted cache levels and everything that resides off-chip). The SMU's main roles are:

Securely store and manage cryptographic keys;

Hide and restore register values upon switching between different modes of operation (secure / non-secure);

Enforce the memory access control;

Decrypt (encrypt) cache blocks upon entry into (eviction from) the TA, and maintain their integrity;

Figure 2 depicts the SMU: on the left, it is connected to the untrusted levels of the memory hierarchy, and on the right ---to the TA (execution unit and on-chip caches). The SMU comprises encryption and decryption units for both symmetric (e.g. GCM) and asymmetric (e.g. RSA [11]) ciphers, signing and signature validation units (e.g. GHASH and RSA), a key table, and a small storage, dubbed the SMU Sealed Storage (SSS), for temporary data. The asymmetric cipher is used to establish a secure connection between the user's computer and the SMU for sending the program's encryption and authentication keys to store in an SMU table.

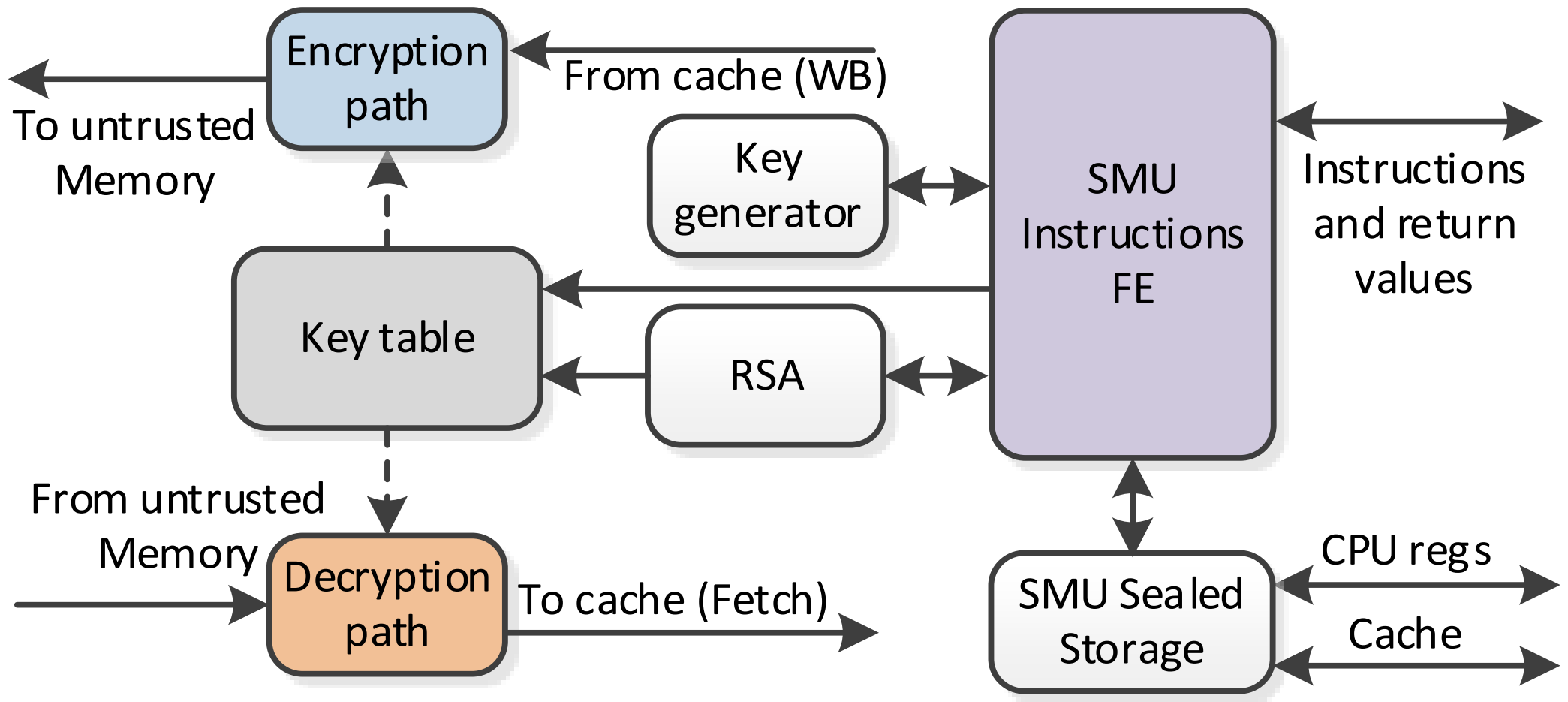

*Figure 2: SMU block diagram*

When a program first launches, it attaches its process ID to this SMU table entry (Sec. 5). The program's code and data are encrypted using symmetric CM encryption (GCM), and are signed using a secure MAC (GHASH). As will be explained in detail in Chapter 6 [36], whenever the seed is zero, the memory block's virtual address serves as the seed for encryption pad creation. Since the seed memory is initialized to zeros, this obviates the need to supply initial seeds with the program. Also, non-zero seeds are concatenated with '1' during encryption pad creation, so initial and runtime encryption pads are never the same. Upon a last-level cache miss in the TA, the SMU uses the symmetric decryption and authentication units to decrypt and authenticate incoming blocks. Modified cache blocks are encrypted and signed upon eviction from the TA, preserving their secrecy and integrity.

Enforcing memory access control ensures that only the (same-process) secure code can access its secure data. This is the core of SeM's protection against software attacks, which nonetheless allows blocks belonging to mutually adversarial applications to co-reside in the cache, unencrypted (Sec. 2.3). This requires that upon initiation of a new secure process, the SMU clear existing secret cache blocks of the same PID (possibly of an old secure process). Sec. 3 discusses some attacks on SeM, including an attack by an adversarial OS.

The SMU operates in two modes: *Trusted* and *Untrusted* (Sec. 2.4). In *Trusted mode*, it expects to run only the secure program, namely run secret instructions. In *Untrusted mode*, it expects to run untrusted code, such as a non-secure application or the OS, where the latter may run from within the context of the secure application (e.g., during an interrupt or a system call). The SMU uses these operating modes to provide the register access control. Switching between modes and register maintenance are discussed in Sec. 2.4.

The SMU table holds the keys and configurations for the secure programs. Each table entry contains (Table 1):

| | |
|---|---|
| **PID** | The process ID of the secure program using this entry.– |
| **Skey** | A symmetric key for memory encryption |
| **Mkey** | A symmetric key for memory authentication, if needed (e.g., GHASH uses Skey for both) |
| **Root Hash** | The root value of the BMT for integrity |
| **Process Hash** | Secure process hash, used to connect the secure process with its table entry |
| **First LEP** | The address of the first secure instruction |
| **Sig LEP** | Signal handling entry point |
| **Error Status** | Holds the error code |

***Table 1: SMU table entry content***

Upon launching of a process containing the Process Hash (more in Sec. 5), its PID is stored. If a table entry with this PID already exists, it is erased, and any secure remnants of the same-PID program are removed from the TA. The table entry must remain in the SMU throughout the execution of the secure application, even when it is not active (for cache evictions, if required), so the size of this table limits the number of concurrent secure applications. However, a typical SMU table entry is 256 bits, allowing many secure programs to run concurrently using a small in-chip memory.

The SMU executes special instructions, required for SeM's operation, these are described in detail in Appendix B. Throughput this thesis, these instructions cannot be used to gain OS privileges nor can they bypass the OS's process separation, so they cannot harm other processes. They serve to perform security operations for the secure program, and only in its address space and security domain. For security reasons, these are treated as fences in out-of-order CPUs. Some are automatically added to the program's code as it is submitted for execution (sparsely), and some are used by the untrusted setup application.

## 2.3 Secure Access

We now present a novel cache access management approach that allows adversarial applications' blocks to concurrently reside in cache unencrypted, while maintaining complete isolation. SeM runs multiple unrelated processes. Our encryption and authentication scheme is based on per-secure-process secret keys stored inside the SMU, which decides whether to grant a given program encryption and decryption services for any given cache block and whether to grant it access to cache (cleartext) blocks. We discuss unified caches for instructions and data, but separate ones behave similarly.

Instructions and data required for execution must be fetched into the CPU's clear cache, residing in the TA. If there is an SMU table entry containing the current process ID, the encrypted block is decrypted using the corresponding decryption key, and its MAC is checked. If correct, the **clear** block is considered *authentic* and is stored in the TA with an *Auth=True* mark; else, the **originally fetched** block is stored in the TA with *Auth=False*, and is considered *non-authentic*. This is done at cache block granularity, and only upon cache miss. Upon eviction from the TA, authentic blocks are encrypted and signed, while non-authentic ones are simply evicted (Figure 3). Wishing to support integrity only, encryption and decryption can be bypassed. A block's *Auth* bit is reset upon cache block eviction and purging, and is propagated between the clear cache levels with the block itself. Any program, privileged as it may be, only gets decryption services by the SMU using its own private keys, if exist. Consequently, although the operating system can access any of the program's private memory outside the TA, secrecy is ensured (ciphertext).

Decrypting a memory block with GCM requires its seed, which does not exist for untrusted code that runs under the context of the secure application. Therefore, the SMU shall only perform decryption attempts for memory blocks that have the required data.

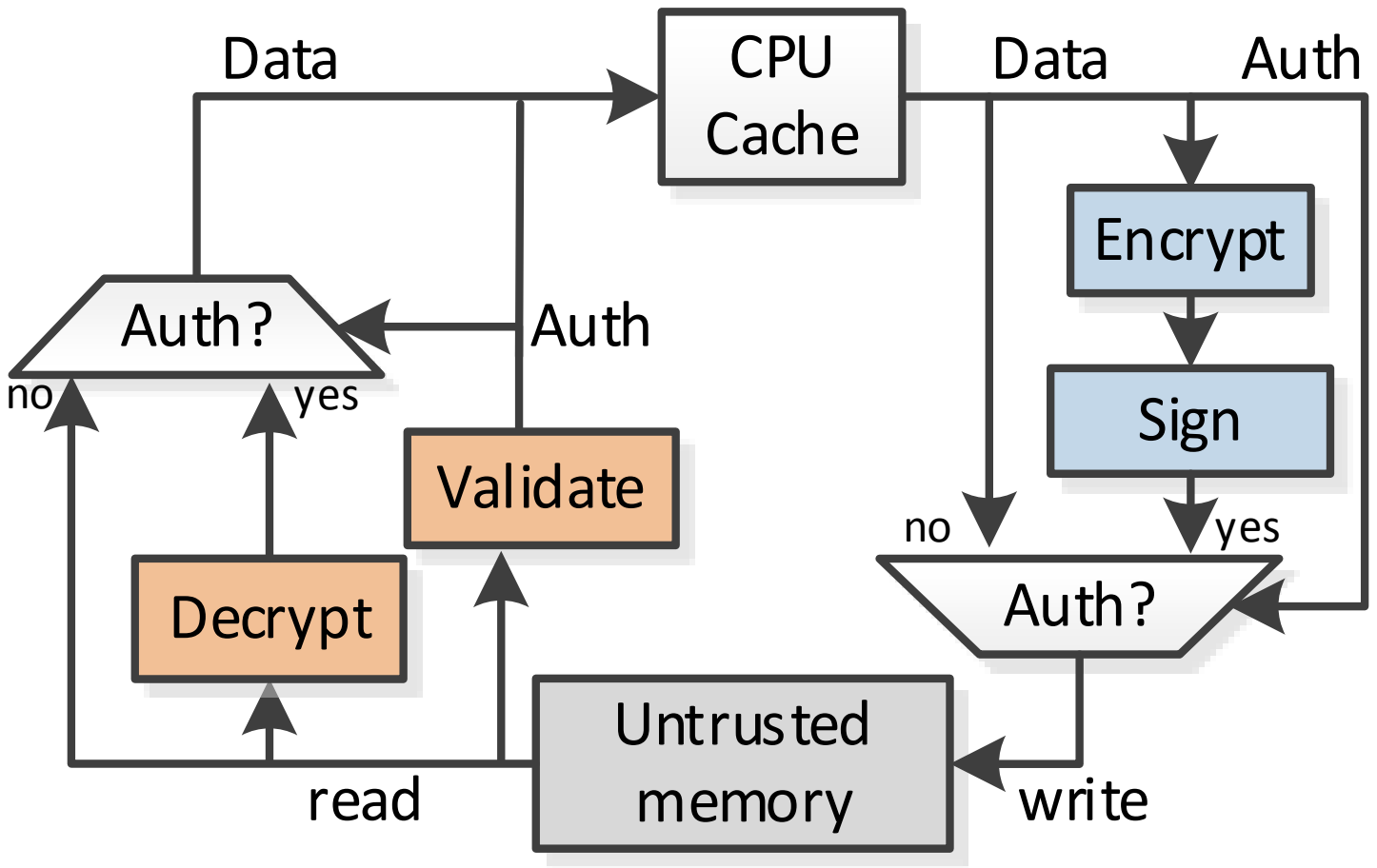

*Figure 3: SMU encryption and decryption paths*

The cache contains clear-text instructions and data, which may belong to unrelated processes and to different users. Each cache block's tag includes its PID, providing inter-process isolation. Attacks by a malicious OS are discussed in Sec. 3.

For data confidentiality in the clear cache, we employ *Secure Access*: authentic data blocks can only be accessed by (same-process) authenticated load/store instructions, and non-authentic blocks can only be accessed by non-authentic instructions. Upon violation of *Secure Access*, the process halts and an error is declared.

## 2.4 Mode Changing and Stack Management

We now present a novel mechanism for automatic register hiding and maintenance, using the SMU modes. Any program starts running in *Untrusted mode*. If and when its secret code starts to execute, the SMU switches to *Trusted mode*. Interrupts may occur, suspending the secure program. Their handlers must run in *Untrusted mode* (their code is untrusted), so secret information is not leaked. Later, to resume execution, the SMU reverts to *Trusted mode*.

Every program starts with a conventional non-secure stack, allocated by the OS. Secure programs also require a secure stack (protected for secrecy and integrity) for managing function calls in Trusted mode, so it is allocated (by the non-secure code) and initialized (by the secure code, as soon as it starts); these instructions are automatically added into the binary. Switching modes also switches between stacks.

A fetched instruction inherits the authenticity mark of its L1 cache block, and the SMU switches modes automatically to match the mark of the invoked instruction. When changing to Untrusted mode, the SMU first stores the contents of the registers (the secret context) in the SMU Sealed Storage (SSS), clears them, and changes the stack pointer to the non-secure stack). It also stores the address of the next authentic instruction to execute, dubbed the Legal Entry Point (LEP), the PID of the running process, and sets a validity mark for the content of the SSS (Pseudocode 1.) Then, the non-authentic (untrusted) code may execute safely. (In out-of-order CPUs, the untrusted instruction is delayed until the last trusted instruction fetched is committed and the registers are hidden, and the LEP is the address of the next instruction.)

***SMU_ChageToUntrustedMode*** *(NextLEP)*
    *LEP=NextLEP*
    *Store secret context into SSS*
    *Set SMU.SSS.valid = True*
    *Clear registers*
    *Mode=Untrusted*

***SMU_ChangeToTrustedMode*** *(InstAddr)*
   *If (InstAddr==LEP) and (SMU.SSS.valid) and*
     *(SMU.SSS.PID ==PID)*
        *Restore registers from SSS*
        *SMU.SSS.valid = False*
        *Mode=Trusted*
   *else*
        *Report error and halt*

***Pseudocode 1: Operations performed by the SMU (by hardware) during an automatic mode switch***

Attempting to execute an authentic instruction in Untrusted mode only succeeds if its address matches the process' LEP and the data in the SSS is valid and matches the PID. If so, the SMU restores the register values and the secure stack pointer (the secret context) from the SSS, and changes the process to Trusted mode; else the program halts and an error is declared. In both cases, the SSS is invalidated. (Pseudocode 1.) By so doing, the SMU verifies that the secure program has resumed from its expected point of execution with the correct register values. (In Sec. 3 we add an LEP for handling signals.) Upon initiating a secure program, the SMU creates an empty secret context in the SSS (with the first LEP from the SMU entry); only during the first switch to *Trusted mode*, the register values are preserved (not restored), but the entry point is enforced.

Switching to Untrusted mode is fast: the registers may be cleared by simply switching a register window to a pending set of erased registers. The switched-out set of registers acts as the SSS, along with the LEP (which must be known for fetching the next instruction) and the PID. Switching to Trusted mode is also done instantly. Verifying the validity mark and comparing the PID are simple operations. Also, register values are restored by switching a register window.

### 2.5 Sharing Data with Untrusted Code

To receive services by untrusted code, such as some OS system calls, a secure application may need to reveal some of its data. The following SMU instructions allow **only trusted code** to bypass the *Secure Access* mechanism, so these instructions must be authentic to run.

- SMU_StoreNA(address, data) – stores data into a memory block regardless of its Auth status, and sets its Auth bit to False, making it accessible to untrusted code.
- SMU_LoadNA(address) – loads data from a memory block regardless of its Auth status, for importing untrusted data by trusted code.
- SMU_InitA(addr, size) – stores zeros into an entire memory region of size size that starts at address addr; sets Auth bit to True for in-cache blocks; and signs and encrypts blocks that are not, used in conjunction with a write-no-allocate cache policy. Used for initializing allocated memory, so it is accessible by trusted code.

## 3 Interactions with the Operating System

### 3.1 Operating System Services

SMU modes and *Secure Access* ensure that confidentiality is preserved even with unexpected invocation of untrusted code. However, the secure program runs concurrently with other (possibly adversarial) programs, so its context must be securely evicted (and later restored) on context switches. Also, although shared library functions are statically linked into the binary when prepared for SeM, system calls invoked by these functions must still be allowed to execute. Furthermore, dynamically allocated memory must get initialized to be used under *Secure Access*. Lastly, signals may be invoked and must be handled. All these are OS services, and are discussed in this section, including the required novel hardware.

### 3.1.1 Context Switch

At context switches, the OS (untrusted) modifies the page table register (e.g., CR3 in Intel Architecture). A hardware watchdog normally exists, which invokes microcode upon page table register modifications [58]. We use this mechanism to also call *SMU_EvictContext* to evict the switched-out content of the SSS from the SMU into the process' memory (cache), and *SMU_RestoreContext* to restore the secure context of the switched-in secure program from memory back into the SSS. When evicted from the TA cache, the process' memory protection is applied.

In 64-bit systems, the size of the SSS is roughly 350 bytes, similar to a thread control block, requiring ~40 cycles for these instructions, which is negligible relative to the thousands required for a context switch [10].

Resuming execution of a secure program entails attempting to execute its next authentic instruction, which causes the SMU to verify its address and the SSS content (Sec. 2.4). If the OS refrains from updating the page table register on context switch, then the SMU evict and restore calls will not be invoked. Having multiple secure programs running on the machine, the PID check upon changing to Trusted mode will fail, causing the SMU to halt the secure program and to report an error. In any case, information is never leaked. *SMU_EvictContext* and *SMU_RestoreContext* are only required for secure programs; if called for a non-secure program, they finish immediately.

### 3.1.2 System Calls

External library functions are usable in SeM by statically embedding them in the SeM-ready binary on the user's trusted computer. This assumes that the library functions are trusted to execute on the user's trusted computer, and we want to ensure that they are essentially unmodified when being used on SeM. Embedding these statically must be done recursively, so that embedded function A calling external function B will result in also embedding B.

Many (normally external) services become trusted by static linking, but some external services require OS assistance at the far end by performing a system call. The code of the system call is provided by the OS, so it is untrusted (like the OS itself). Allowing system calls to run is a crucial requirement for the correctness of the program.

Every OS and CPU architecture has its convention for invoking system calls. We focus on Linux [10] with System V AMD64 ABI [37], though the ideas can be adapted to any OS / architecture. Linux system calls are invoked via the *syscall* instruction, with the system call number passed in register *rax*. System call arguments are passed via the registers *rdi, rsi, rdx, r10, r8, r9* (in order). Invoking syscall puts the return address in *rcx*, and the system call's return value is passed in *rax*. Unlike with function calls, the remaining registers are guaranteed to preserve their values.

Because it is embedded into the SeM-ready binary, the *syscall* instruction is authentic (trusted), but the code of the system call is untrusted. Yet, to allow argument passing to the system call, the mode change for the system call code must not clear all registers; specifically, it must keep the required arguments and *rcx* (return address) in place. When done, the OS jumps back to the address in *rcx* (which is an LEP), and the secure programs resumes. We also require that switching to Trusted mode (if the last switch to Untrusted mode was for a system call) not restore the value of *rax*, since it holds the system call's return value.

Different system calls require different numbers of arguments. We must supply exactly the required number of arguments, since supplying fewer will harm its operation, and supplying more may reveal secret information. The number of arguments required for a system call can be discovered in the binaries (during the preparation phase) by noting the value of *rax* prior to *syscall* instruction (which always receives an immediate value prior to the *syscall* instruction) and comparing it with a list of known system calls. (In most cases, *eax* is set instead of *rax*, but because of the small range of system call numbers the outcome is the same, just with a smaller instruction.) Then, we replace *syscall* with *SMU_syscall(argnum)* instruction, *argnum* being the number of arguments for the system call. The SMU will keep the required argument registers untouched on the next switch to Untrusted mode, and then invoke *syscall*. The next switch to Untrusted mode is likely to occur on the first instruction of the actual system call code (since it is untrusted).

A context switch occurring right after *SMU_syscall()* only delays the actual execution of the system call, but the OS should restore its registers (syscall arguments) before switching back (though SeM does not enforce that since it is executed in Untrusted mode, similarly to the system call).

To avoid hard-coding the syscall convention in hardware, we use an SMU instruction that sets the syscall convention, using a bitmap of registers. This must be set per secure program (to prevent one secure application from attacking another), and must run as trusted code at the beginning of the secure program, so it is automatically embedded during the instrumentation process.

Some system calls require pointer arguments for memory buffers or data structures (e.g. *sys_write* for file access*)*. This requires a buffer copy to/from the untrusted memory before invoking the actual system call. We use dedicated wrappers for these, using *SMU_StoreNA() and SMU_LoadNA().* The same holds for system calls that require complicated structs to be passed (e.g., serialization / deserialization for *recvmsg()*), in which case the wrapper must implement the required logic to perform such copy, sometimes replicating some of the original system call logic. In any case, it is always the secure program that reads/writes from/to its trusted memory region and writes/reads to/from the untrusted memory, and the system call always accesses the non-secure memory.

Chapter 5 contains an in-depth discussion of the instrumentation process.

The above approaches for passing the syscall arguments are similar to Overshadow [2] and SCONE [27], yet instead of *shim* functions we simply avoid clearing the desired registers on a mode change. In complicated system calls, our approach is similar to Overshadow's.

### 3.1.3 Dynamic memory allocation

Dynamic memory allocation is commonly required. Newly allocated memory must be accessible to the secure program, and its integrity must be kept.

When allocating a new memory block, its content is usually zeroed by the OS (for security reasons); yet, when read by a secure program as if it is signed and encrypted, it will most likely fail the signature test and will be considered non-authentic, which would forbid storing new data into it by authentic store instructions (see *Secure Access*). Therefore, a newly allocated memory block is initialized (zeros) by the *SMU_InitA* instruction. When completed, memory blocks of the new region that are outside the TA contain zeros, signed and encrypted correctly, and blocks that happen to be in the cache contain zeros with *Auth=True*.

```
void* sem_malloc(size_t size)
    void* retval;
    retval = malloc(size);
    if (retval)
        InitA(retval, size);
    return retval;
```

***Pseudocode 2: sem_malloc() implementation***

We do that by implementing a *sem_malloc(size)* wrapper for *malloc()*, and replacing every call to *malloc()* with a call to *sem_malloc()* in the SeM-ready binary. *Sem_malloc()* essentially invokes *malloc()* with its original parameters; if successful, the newly allocated memory block is initialized by *SMU_InitA*. Pseudocode 2 shows a suggested C implementation for *sem_malloc().*

When allocating a new memory block, the untrusted operating system must make sure that the memory required for this block's metadata (CM encryption seeds and BMT hashes) has also been allocated. The seed memory is considered part of the OS memory (like the page table). Avoiding seed memory allocation will result in an error when accessed (if not allocated), as the SMU assumes its existence for allocated memory, so malicious or erroneous OS behavior will be caught.

Previously allocated memory that was freed and then reallocated has seeds that are not necessarily zeros. Nonetheless, when initializing the reallocated memory, its seeds will be updated correctly upon cache eviction.

Memory allocation may map a new virtual address to a previously used physical block. Without initialization, fetching a new memory block into the TA may require the seed of its previously mapped virtual address to validate correctly, so a MAC error is likely to occur. Therefore, every new memory allocation requires initialization.

The *free()* process is unchanged, and a freed block will simply become available for future allocation.

### 3.1.4 I/O Access

I/O access is crucial in many applications. Two main considerations must be addressed when discussing I/O in SeM: 1) Passing memory buffers from/to the I/O device; 2) Data secrecy. We address files as the I/O mechanism, but the ideas can be adapted to other entities.

File reads and writes are implemented over the *sys_read* and *sys_write* system calls (regardless of the actual API in use), both untrusted. They receive a file descriptor, a buffer pointer, and size. *Sys_read* natively receives a pointer to the secure memory, but it cannot write there. Similarly, *sys_write* is natively required to read from the secure memory but may not. These are examples of more complicated system calls, for which we use trusted wrappers, implanted during the automatic instrumentation of the application. The read wrapper allocates a non-secure memory region (by simple *malloc()*, if wasn't done before) and then reads data into it. It should then perform decryption and validation into the secure memory. Being a trusted function, it may approach both secure and non-secure regions. The write wrapper works the other way around.

Encryption and decryption are done by special functions, using a key that was supplied with the program when preparing it to run on SeM. These also require in-memory state to maintain I/O integrity. For performance and power efficiency, we suggest using the dedicated cryptographic instructions available in modern CPUs (such as Intel's AES-NI [59]). Pseudocode 3 shows a suggested C implementation for the read wrapper.

```
ssize_t sem_read(int fd, void *buf, size_t count)
   if ((!non_secure_buf) || (non_secure_buf_size < count))
         free(non_secure_buff);
         non_secure_buf = (void*) malloc(count);
         non_secure_buf_size = count;
     int size_read = sys_read(fd, non_secure_buf, count);
     int retval = decrypt(key,non_secure_buf, buf, size_read);
     if (!retval)
         error;
     else
         return size_read;
```

***Pseudocode 3: sem_read() implementation***

### 3.1.5 Signal Handling

Signals may be sent to any program during execution. In some cases, signal handling by the program is required (for specific signals), so the program registers a dedicated signal handler for these signals. When this signal is raised, the OS invokes the dedicated signal handler. When finished, it returns to the OS (by returning to *rt_sigreturn* system call that was pushed by the OS into the stack as the return address), and the program resumes execution from the point at which it was interrupted. The approach we use to handle signals is similar to the one of SGX [22].

To allow dedicated signal handlers for secure programs, SeM must support an additional LEP. In this LEP we place a trusted signal handling entry function (SHEF) that identifies the signal, runs the desired signal handler, and returns. If signal handling is required, a SHEF is automatically added to the program during its preparation for SeM, and the SHEF's address is set as a *sig LEP* in the SMU entry.

Signal handler registration is done using the *sys_rt_sigaction* system call, so we use trusted syscall wrapper (automatically instrumented). The wrapper copies the *sigaction struct* to an untrusted memory buffer, and changes the handler's address to the SHEF's.

It also registers an entry in a dedicated signal mapper that holds the address of the dedicated signal hander for this specific signal.

Prior to setting a dedicated signal handler, signals are treated by the OS. After registering a signal handler, when the registered signal is raised the OS calls the SHEF, supplying information of the signal. If a dedicated handler exists in the signal mapper, then it is called. Calling the SHEF by the OS (untrusted code), the mode changes to Trusted but the register values are not restored, allowing the OS to pass signal information. To keep the signals handler's confidentiality, the SHEF uses a dedicated secure stack for handling signals; before returning to the OS, the registers are cleared, and the stack is restored to the one of the OS.

## 3.2 Hostile OS Attacks and SeM's Resilience

An attacker may try to read or change code or data, rerun the secure program for various purposes, or even manipulate the secure context data or the flow of execution. These are all directly addressed by SeM's mechanisms. In this section we wish to discuss two privileged code attack examples that require further attention.

**Forged identity attack**: Consider a privileged attacker (e.g., OS) that tries using the PID of a secure program to access its clear cache. The cache natively allows same-PID cache access, but *Secure Access* allows secret (authentic) data to be accessed only by same-process authentic instructions; the attacker must therefore also properly encrypt its attacking code in order to gain access to the secret data. Not knowing the secure program's secret keys, this is impractical.

**Iago attacks**: These attacks use carefully chosen system call return values to manipulate library functions that are statically linked into the secure (and therefore trusted) program. [55] shows that manipulating *brk()* system call return value can cause *malloc()* to allocate an intended secure memory in a non-secure region, thus exposing data stored in it. In SeM, however, dynamically allocated memory always becomes secure memory, so data stored in it is always protected against leakage. Moreover, Iago uses return oriented programming (ROP) [60] to redirect the secure program to the attacker's code. However, if the ROP target is untrusted then automatic mode change will prevent any data leakage. Finally, Iago attacks may be defeated by checking system calls' return values [13,14,40], which can also be done in SeM.

# 4 Discussion

**Integrity with no Confidentiality.** Program confidentiality is maintained by the mechanism of memory encryption, and its integrity is maintained by the mechanism of signature validation. Without the integrity mechanism, one cannot ensure that its own program was indeed executed, because an attacker may forge data and instructions of its own, even ones that may intentionally uncover secret data; furthermore, *Secure Access* tightly depends on the integrity mechanism, so the in-chip per-process secure compartment cannot be maintained without it. Alternatively, SeM can operate in a mode during which **only** the integrity of a program is preserved, protecting against attempts to interfere with its correct execution. This mode is very similar to the one containing encryption, where the encryption and decryption blocks are simply bypassed in the SMU. Requiring no encryption also means that no encryption seeds are used (therefore not allocated), thus unlike the use of BMT, the integrity tree in use must protect the data blocks themselves (Merkle Tree).

**The Size of the Integrity-Protecting Metadata.** The memory integrity protection mechanism uses a MAC for protecting each memory block, and a hash tree (with a root in-chip hash) for protecting the encryption counters. Assuming that all the protecting metadata currently resides in the off-chip (untrusted) memory, a cache miss will cause the fetching of the encryption counter and its validating nodes of the hash tree, and once the cache block was decrypted, it is validated using its MAC.

The sizes of the MACs and hash values in the hash tree are important for quantifying the probability of a successful attack. Furthermore, only a single failing attack is allowed, since an integrity error will immediately halt the program and no further attempts will be allowed.

The MAC directly protects the memory block against an attacker that modifies its encrypted version while in the untrusted memory. Because we use counter mode encryption, flipping a bit of a memory block that resides in the untrusted memory will result in flipping the same bit in the decrypted version. Since the MAC is calculated using a secure hash (and its key is assumed to be unknown to the attacker), the probability for successfully modifying the encrypted block and going unnoticed is $1/size_of(MAC)$. Therefore, the size of the MAC should be determined to achieve a desired probability. For evaluation purposes, we chose a 16 bit MAC.

The hash tree protecting the encryption counters, where the leaf hash directly protects a block of counters, and the ancestor hashes protect their children hash blocks. Reverting a memory block with its MAC and its counter to an old value (ensuring that the MAC will validate correctly) will commonly result in a new leaf hash, which will most likely result in new hashes until the root of the tree (which resides in-chip, therefore impossible). Wishing to keep the hash tree unchanged, the success rate of reverting the memory block with its MAC and counter to old values is $1/size_of(hash)$. Similarly to the MAC, the size of the hashes should be determined to achieve a desired probability. For evaluation purposes, we chose 16 bits hashes.

# 5 Quantitative Evaluation

In this section, we evaluate SeM's performance reduction, increase in power consumption and additional hardware (area).

## 5.1 Performance Evaluation

We developed SeM-Prepare, a tool that automatically and statically instruments existing binaries, thereby making them SeM-ready. We also developed SeM-Simulator, a tool that is able to run SeM-ready binaries, simulating the SMU’s behavior, and collecting statistics. These are described in detail in Chapter 5 and Appendix A, respectively.

SeM's overhead on program execution is in memory access (encryption/decryption), memory allocations (secure initialization), and system call wrappers (only when wrappers are needed). The system (OS) oriented overhead is for the context switch (merely ~1.02X the non-secure context switch, Sec. 3). All these have a negligible effect on unmodified performance-critical elements such as the cache (flushes are not required) and branch prediction, and none at all on the execution units. Also, mode changes impose no overhead (Sec. 2.4).

We instrumented the integer programs of the SPEC CPU2006 benchmark suite [57] using SeM-Prepare, and then ran it on SeM-Simulator. SPEC CPU2006 was chosen because it targets the causes for overheads in SeM. Although SPEC CPU2006 also has floating point benchmarks, we only use SPEC’s integer benchmarks, because SeM is agnostic to the type of numeric operations that it performs.

Performance is measured relative to the corresponding non-secure application. Overhead caused by memory encryption and authentication, including memory accesses for fetching missing GCM seeds, their BMT hash authentication, and cache contamination, comes inherently from the chosen memory encryption and authentication technique (and its implementation), as in any secure architecture; we therefore rely on previous works' simulations [33,34] for these.

Figure 4 shows the performance penalty with and without memory encryption relative to no security at all (running the benchmarks unchanged). The mean penalty is under 1.9%, of which 1.8% is for memory encryption, so SeM adds merely 0.1%. (I/O traffic and allocated memory are also in Figure 4.) Thus, no other solution (past or future) can do much better.

All programs successfully changed modes, invoked system calls, used files, and allocated secure memory. Finally, we verified that the SeM-ready benchmark results matched the original ones. Besides evaluating SeM's performance, this also serves as a strong indication for the applicability of SeM to existing binaries, without requiring programming effort.

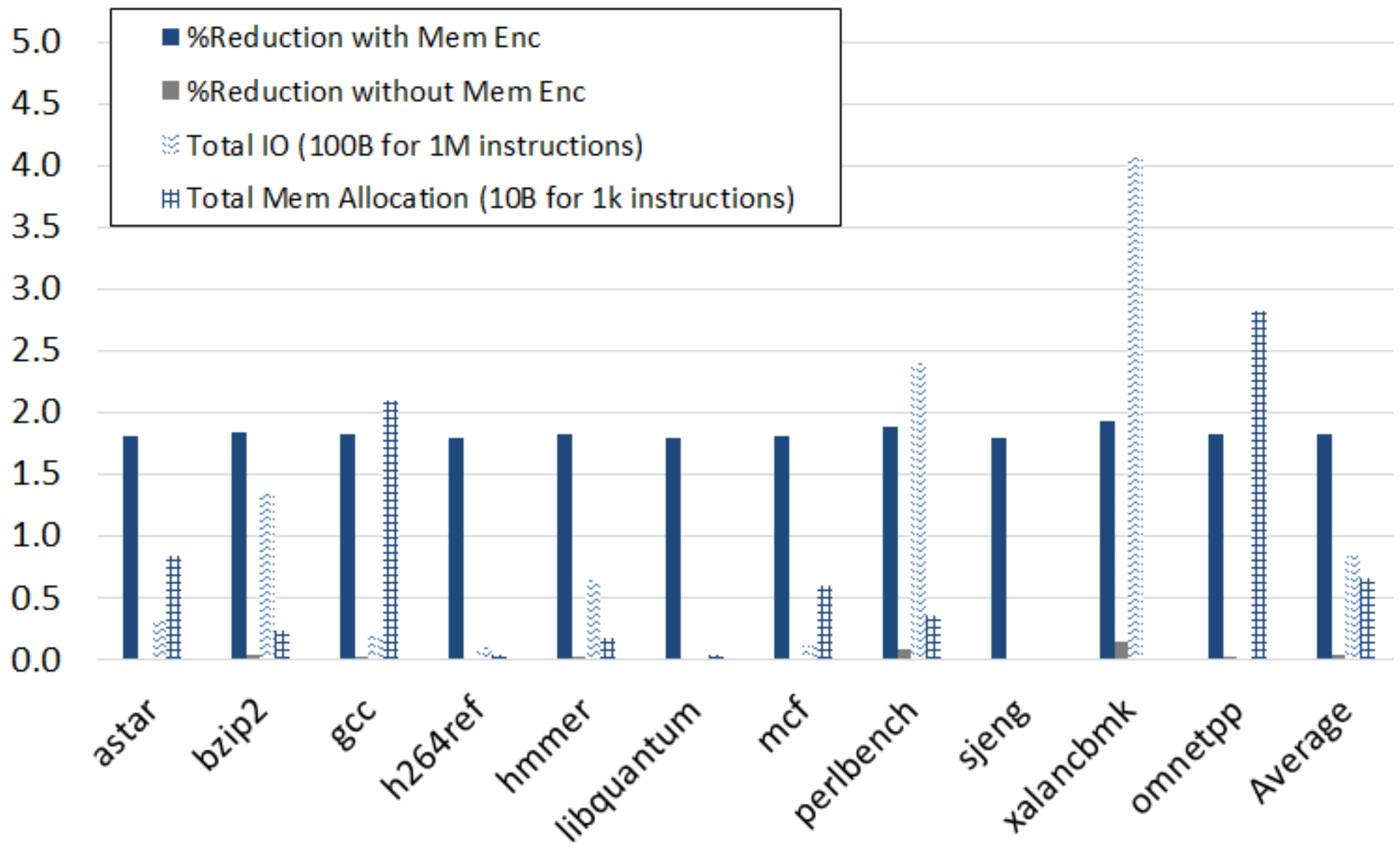

*Figure 4: Results for running SPEC CPU 2006 on SeM - percents of performance reduction, memory allocations, and I/O operations, per instruction.*

## 5.2 Memory, Area, and Power Overheads

**Additional memory footprint** depends on the chosen size of the seeds, MACs, and hashes, and also on the size of cache blocks. To estimate the memory footprint overhead, we use commonly used settings. We assume cache blocks of 64 Bytes, 64-bit per-block counter, 16-bit MACs and 16-bit hashes, and a 64-bit system. Each counter block protects 32 memory blocks, and each hash block protects 128 counter blocks, so a hash block protects a memory region with the same 44 LSbs. Therefore, the hash tree is composed of 7 levels, including the in-chip root hash; each level is 128 times smaller than the previous (excluding the in-chip root hash which is 16 bit). In view of the extreme size drop of each added hash level relative to the previous (a factor of 128), we only count the first (larger) level of hashes. Therefore, using M bytes of memory will result in:

$$\frac{M}{block_{size}} \cdot (Counter_{size} + MAC_{size}) + \frac{M}{Block_{size} \cdot \frac{Block_{size}}{Counter_{size}}} \cdot (hash_{size}) \text{ [Bytes]}$$

Where $\frac{M}{block_{size}}$ is the number of data and instruction blocks in use, and $\frac{M}{Block_{size} \cdot \frac{Block_{size}}{Counter_{size}}}$ is the number of counter blocks in use. Putting the numbers in the above term we get ~16% of memory overhead for the encryption counters and BMT.

**Hardware additions to the CPU, including extension for parallel execution (Chapter 3)**: an extra set of registers used for fast mode switching, ~0.5kB (24k gates). Each cache block requires extra metadata bits for Secure Access; assuming 64 Bytes per cache block and 10 bits for metadata, this yields in a <2% of cache size overhead. An SMU table entry is 32 Bytes, so a 3.2kB RAM suffices for 100 concurrent secure applications (equivalent to 25.6k gates [12]).

For SDSM: 10 outstanding inactive TSCs, each 350 Bytes (similar to a thread control block), result in a 3.5kB RAM (28k gates). SDSM uses 10 cache entries for outstanding encryption pads (each 240 Bytes), which requires 2.4kB RAM (19.2k gates).

The only non-negligible cryptographic cores are two AES-256 cores (used for both memory encryption and SDSM) and one RSA core. Implementations using 51.2k gates for AES-256 [8] and 2.6k for RSA [9] have been reported. AES area can be reduced by using AES-128, which was shown to be completely safe [17]. AES-128 core's latency is half that of AES-256, so only 1 core is needed (time-shared). An AES-128 core's area is also about half, yielding a 4X area reduction in total AES core area.

The overall estimated area overhead is under 150k gates. With 1.3B transistors for a modern PC CPU [67], and an average of 4 transistors per gate, SeM's area overhead is less than 0.02% of a PC's CPU, and much smaller of a server's.

**Power consumption.** Most of SeM's additional power consumption is for the encryption/decryption. [8] reports 76.8nJ for 256 Byte AES-128 encryption (130nm), and [9] reports 454µJ for RSA decryption (130nm). RSA decryption is only needed once per secure application execution and is thus negligible; AES encryption occurs only upon a secure application's cache miss, so additional power consumption is small. Also, future crypto cores will require even less energy.

# 6 Conclusions

The Secure Machine (SeM) is an extended CPU architecture that uses a novel hardware based security management unit (SMU) and a software tool, enables running a program securely even on a platform with unchanged and untrusted OS, Hypervisor, VMM, and hardware other than the CPU chip. Existing binaries are automatically instrumented to run on SeM as part of the submission to the secure cloud, requiring no programming efforts. SeM-Prepare does this by analyzing the binaries, statically linking external libraries, and adding wrapper functions for memory allocation and required system calls, and finally encrypting and signing. This essentially allows running any application (new or existing) on SeM.

SeM reduces performance by at most 2% relative to no security at all, and 95% of the reduction stems from memory encryption and authentication, which are not unique to SeM.

In the next chapter, we take the next stop, extending the basic SeM architecture to permit parallel workloads (multi-thread, -core and -computer).

# Chapter 3 Parallel and Distributed Secure Execution

## 1 Introduction

The Secure Machine (SeM) provides basic secure processor services with untrusted and unmodified OS, hypervisor, and VMM: securely maintaining secret keys, memory encryption and integrity preservation, fast switching between Trusted and Untrusted execution modes (protecting data in the process), enforcing the correct flow of a secure program, and access control to cached (cleartext) data. However, SeM's program flow enforcement mechanism prevents the OS from creating new or alternative flows of execution, which does not allow creating new threads.

In this chapter we extend SeM's capabilities to multi-threaded programs that run on single core CPUs, CMPs, multi-CPU, and multi-computer systems, and even secure accelerators. For convenience, throughout this chapter we refer to these extensions and to the resulting CPU architecture as ParSeM. ParSeM supports both single- and multi-threaded secure programs, and features secure process and thread migration. All this while supporting existing binaries, without restricting the OS, and only adding one instruction to it. ParSeM supports most existing application binaries (languages, programming methodologies, etc.); details in Chapter 4.

We designed and implemented SeM-Prepare, a tool running on the user's trusted computer for preparing existing binaries for ParSeM (running offline or in the course of program submission to the cloud, sometimes referred to as *application deployment*). Then, we run SeM-Prepare on the PARSEC benchmark suite [48] binaries to make them SeM-Ready. We designed and implemented *SeM-Simulator* to execute the resulting binaries, thereby demonstrating completeness and correct results, and showing overheads to be negligible. These tools are described in detail in Chapters 5 and Appendix A, respectively.

A multi-threaded program that uses use-level threads creates its own threads and manages them, thus appearing single-threaded to both the underlying OS and the hardware, so it can run securely even on SeM. However, only kernel-level threads may benefit from hardware parallelism (multi-core, multi-CPU or multi-machine), so ParSeM must address them.

Support of multi-threaded applications raises several challenges: 1) thread creation,

migration, and termination normally require the services of the underlying OS, yet it is untrusted; 2) ensuring a shared address space for all threads, even though the untrusted OS controls the virtual memory system, and the memory space may span across many compute nodes connected via an untrusted medium.

Overshadow [2] addressed thread creation and termination assuming that the VMM is trusted. A dedicated (trusted) VMM was suggested, running shim functions to mediate between the trusted program and the untrusted OS. Thread migration was not supported. However, when all platform software is untrusted, new challenges arise; e.g., similar functions cannot be used to guard the creation of a new thread on the secure program's behalf (even if embedded into the secure program), since these must be privileged to access certain registers (e.g. the page table pointer). Another important work, Intel's SGX [22], which does not trust any platform software, does not support conventional threading. (See Chapter 1 for a detailed discussion of related work.)

Supporting multi-node (and even multi-chip) computation moreover motivates the ability to run a program securely on heterogeneous platforms, which contain accelerators (GPUs, Smart NICs, FPGA, etc.).

To enforce correct invocation of parallel programs, ParSeM provides hardware-assisted secure thread creation, migration, and termination, by extending the ISA with instructions that are executed by ParSeM's hardware security management unit (see Appendix B for a detailed list of instructions). These ensure that privileged operations cause no harm, and can be trustfully executed by the secure program. Furthermore, we support heterogeneous systems whose accelerators meet a small set of requirements that we define.

For securely fetching missing memory blocks from other trusted nodes, ParSeM uses a secure distributed shared memory layer (SDSM) connected through an untrusted medium, that uses address independent encryption for the transport only (unlike ParSeM) (Chapter 6). In SDSM, the directory must be trusted. The directory may be implemented in hardware or software; it may attest its authenticity or be provided as a trusted user program (protected by SeM). We also assume a distributed integrity-preserving layer for maintaining the memory integrity of a distributed program (Chapter 7).

## 2 The ParSeM Architecture

ParSeM uses the secure single-thread execution and setup capabilities of SeM, i.e. each CPU of ParSeM has an embedded SMU that protects its private Trusted Area (TA) via *Secure Access* and mode changes. A CMP has multiple cores on the same trusted chip, so

CMP internal core-to-core communication is done in the clear; therefore, the SMU's encryption and decryption functions are carried out between the CMP's last level cache (LLC) and the memory interface. The same SMU table is accessible by all the CMP's cores (Figure 5).

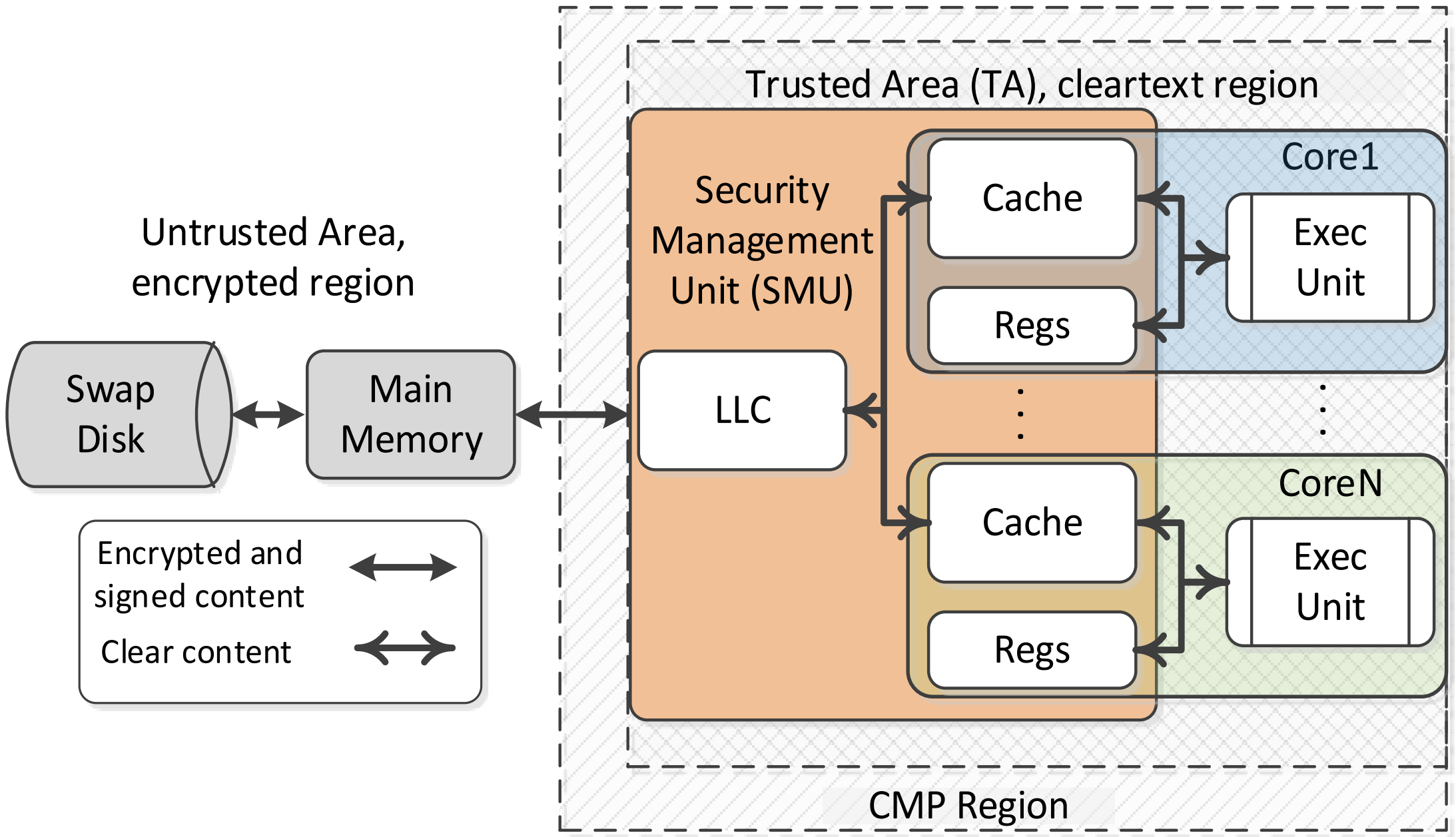

***Figure 5: SeM hardware block diagram***

To establish a distributed Trusted Area (TA), i.e., a set of TAs in which each participating SMU stores the secret configurations of the secure (distributed) programs and their runtime state, we add secure SMU-to-SMU communication.

To support multithreaded secure programs, we add thread creation and migration SMU instructions, and extend the SMU to support per-thread (instead of per-program) secret context eviction and restoration. All the SMU instructions (CPU ISA extensions) are performed by the trusted CPU using dedicated hardware and/or assisted by trusted and atomic CPU microcode.

Preparing a program for ParSeM is similar to preparing a program for SeM: shared library functions are embedded (statically linked) into the binary, some additional instructions are instrumented into it, and then finally it is encrypted and signed. Chapter 5 discusses that in further detail.

A parallel program begins as a single-threaded process running on one of the cores.

Then, additional threads are created, either by the main thread or by sibling threads. These threads may run on the same core/CPU or may be migrated to a different one. Eventually only the first thread is expected to remain, and it terminates the program. This raises several challenges: ensuring that new threads are created correctly; ensuring that all the threads share the same memory address space; creating a distributed TA securely; safe thread migration; and maintaining the integrity of a distributed program. Also, enabling the use of a distributed TA on execution platforms such as accelerators (GPUs, Smart NICs, etc.). We next address these challenges: in Sec. 2.1 we discuss multi-threading in a single core, and then extend that to multi-core and multi-machine settings.

## 2.1 Secure Thread Creation and Execution

A new thread shares the process ID and address space with its parent thread, as do sibling threads, so the SMU will serve its memory access requests without further changes. If the OS initiates a new thread with the wrong process ID, then upon the first memory access of the new thread (its first attempt to read or modify the program's memory) an owner-ID mismatch will occur (for Secure Access), which will cause a detected error. Each thread has a unique register set and stack; a new thread is launched with the register values of its creator (mostly), and with empty secure and non-secure stacks.

Thread creation is initiated by secure code (of the *parent thread*) that belongs to the secure program, which normally calls a shared library function (e.g., *pthread_create()*). Being a shared library function, it is embedded in the binary during its preparation for ParSeM. Regardless of the thread creation method, the *clone* system call is eventually used (on Linux, or an alternative on other OS); it receives an argument of flags and a pointer to the new stack; the stack is assumed to be allocated by the code prior to the system call invocation, and it will be used as the non-secure stack for the new thread. Since the system call's code is untrusted (untrusted OS), this results in an automatic mode change to *untrusted*, as part of which the SMU stores the register values (and clears the registers) and the LEP into the SMU sealed storage. However, unlike a single-threaded program that leaves *Trusted mode*, here both the original and the new thread will return from the *clone* system call to the same LEP, differing from one another only in the stack pointer and the return value. From here on, the parent thread will continue; the new thread will allocate its secure stack, and will jump to its starting point.

Creating a new thread requires an SMU instruction, *SMU_NewThread(),* for creating a new inactive thread secret context (TSC) containing the parent's register values (except for the return value register, which is not restored), and the LEP is set as the address of the instruction following the syscall instruction (*SMU_syscall (argnum)*). (The actual offset may be configured beforehand per compiler.) *SMU_NewThread* is instrumented into the binary just before the *clone* system call, and it returns the ID of the created TSC (SCID). The TSC does not leave the SMU until it is attached to the new thread.

When *clone* returns, its return value is checked. If this is the new thread (TID>0), a secure stack is allocated and initialized, and the thread's function address is pushed into it; *clone* normally performs this push, but being untrusted it cannot manipulate the secure stack. If *clone* fails (either for legitimate reasons or due to an attack), the new inactive secret context will be deleted using *SMU_NewThreadDelete(SCID)*, which is instrumented into the binary as conditional code just after the call to *clone*. Pseudocode 4 shows the pseudocode and (automatic) additions for the original invocation of *clone*.

```
…
new_stack=malloc(…) // non secure stack
…
SCID=SMU_NewThread()
clone(flags, new_stack )
if (rax>0) // clone return value – new thread only
    rsp = malloc(..)  // secure stack, including secure init
    push(thread_func_addr) // normally done by clone
else if (rax<0) // clone return value - failure
  SMU_NewThreadDelete(SCID)
…
```

**Pseudocode 4: Additional code inserted before and after the clone system call. New code is in grey background**

*SMU_NewThread()* and *SMU_NewThreadDelete()* must be *authentic* to run, and a program may only initiate or delete a new TSC for its own PID, so a malicious program (even if privileged) may not initiate new secure threads.

Finally, when the new thread is set to run, the OS (untrusted code) calls *SMU_NewThreadAttach(TID, addr)*, which attaches the new thread ID with the inactive TSC (at address=*addr*), so it will serve to keep this thread's consistency for the rest of its

```
SMU_NewThread()
  SC=New(SMU.SealedStorage.SecretContext)
  Store registers into SC
  SC.LEP = SMU_NewThread.rip+offset  // offset is the size of
                      //  SMU_NewThread + syscall instructions
  SC.ParentTID = TID
  Return SC.ID

SMU_NewThreadDelete(SCID)
  SC=SMU.SealedStorage.FindInactivePTID(SCID)
   If (SC) // if found inactive context for this LEP
         Delete(SC)
   else
     Report and error and halt

SMU_NewThreadAttach(TID, addr)
  SC=SMU.SealedStorage.FindInactiveTSC(addr)
  If (SC) // if found inactive context for this LEP
      SC.TID=TID
      SC.SetActive();
  else
      Report error and halt

SMU_ThreadDelete(TID)
   DeleteTSC(TID)
```

***Pseudocode 5: SMU thread creation and termination instructions, performed by the SMU***

execution. Pseudocode 5 shows the functionality of the new SMU instructions.

If the OS fails to call *SMU_NewThreadAttach()* for the new thread, the thread will use a NULL TSC, which will fail verification due to an illegal entry point. If the OS calls *SMU_NewThreadAttach()* for an already existing thread, it will fail for an illegal entry point. If the OS also sets the existing thread's entry point to match the expected LEP, then the new thread will run **instead** of the already running thread, which is merely a denial of service to the already running thread, therefore outside of our scope. Further OS attacks on ParSeM's threading mechanism are discussed in Section 2.3.

**Remark.** Since ParSeM does not protect against denial-of-service, we do not enforce fairness in scheduling; if fairness is important, it may be checked for using other measures, e.g. using barriers to make sure that all the threads have reached specific points of execution.

Note that there is no temporal order guarantee for attaching new threads with inactive contexts. Multiple threads may be created by different parents, and may be attached to their context and set to run in any order; however, a new thread will be attached to its correct context and will run from an LEP.

When a thread finishes (using any threading method), it eventually invokes the *exit* system call. To discard the thread's TSC, we add *SMU_ThreadDelete(TID)* just before its invocation, and then *exit* discards the thread by the OS (as in untrusted threads). If the OS wrongfully attempts to resume the thread, it will attempt to execute without a TSC (when scheduled), and the SMU will declare failure upon scheduling.

Threads can also die for other reasons, such as by receiving a KILL signal, or for dividing by zero. In that case, the entire process dies and the OS merely needs to clean after the secure program, both for its active and inactive threads. The SMU thread deletion instructions do not require to be authenticated to run, so the untrusted OS may perform these operations.

When a node dies, no cleanup is needed. Furthermore, maintaining threads in an environment where nodes may crash is managed directly by the secure program, and it is orthogonal to the added security of ParSeM.

A thread's secure execution is preserved similarly to the single-threaded SeM: memory secrecy and integrity are kept as before; suspension and resumption of a thread's execution are done via a context switch, but with the protection of per-thread (instead of per-process) context. (A secure memory region is allocated at the beginning of the program for TSC storage.)

Shared address space is ensured by the same mechanism that ensures integrity of a single thread's memory access: blocks being evicted from the TA are signed and encrypted, and the encryption counter is assigned per virtual address, so fetching the wrong block into the TA (or serving two threads with different page mappings) causes a detected MAC integrity error.

## 2.2 Secure Process and Thread Migration

Migration is the core of distributed multi-threading. Although process and thread migrations are initiated by untrusted entities (such as the OS), they should not expose secret information; the only damage potential is denial of service, which is unavoidable with an untrusted OS controlling the scheduler.

Secure thread migration between cores of the same CMP does not require any special treatment, as the same SMU (with its SMU table entries) serves them all. Counter and DBMT data are fetched between the CMP cores via the in-chip cache coherence mechanism, which is trusted. Migration of non-secure processes and threads between CPUs does not require special treatment. The challenge is thus inter-CPU migration of secure threads and processes.

To allow secure thread migration between CPUs, an SMU table entry must be securely moved between SMUs, and sometimes securely duplicated, along with the thread's TSC.

We require secure direct communication between SMUs. Prior knowledge is assumed to exist inside the SMU for providing identity authentication of other SMUs, so each SMU must store the trusted ParSeM authentication keys in a protected non-volatile memory. These keys may be updated using a rare update procedure signed by the SMU's vendor. The update mechanism allows secure SMU-to-SMU communication between different generations of SMUs, or even across CPU vendors. (Cross vendor secure communication is also used in secure accelerators, Sec. 2.4.)

**The Flow of Migration**

Thread migration is initiated by an untrusted entity such as a load balancer, which instructs an untrusted local privileged agent to perform the actual migration. Then, the source SMU establishes a secure channel to the target SMU to verify its authenticity and exchange symmetric keys for encryption and authentication, like the creation of an SMU table entry in SeM.

We distinguish between migrating an SMU table entry of a non-running secure process (before being invoked or after finishing) and migrating a thread of an already running secure process; within the latter, we act differently for a non-active thread (has not yet started its execution) and an active one.

**SMU table entry migration (for a non-running process)** is done by *SMU_MigrateEntry(Phash, Target)* instruction (invoked by untrusted code), with the process hash and target CPU (machine) as parameters. After the source SMU establishes a secure channel with the target SMU, it sends the SMU table entry (encrypted and signed) to the target, which creates a new local SMU table entry (Pseudocode 6). For the target SMU, this is similar to creating a table entry for a new secure program. When done, the source SMU discards its table entry. Finally, the local untrusted agent moves the secure program to the target machine using simple (already encrypted) file copy.

**Migrating a thread of a running process** requires a different approach: its data should be moved using its shared address space (without harming the existing threads). Migration is initiated by the *SMU_MigrateThread(PID, TID, addr, Target)* instruction (invoked by untrusted code, and only performed when the thread is suspended), with parameters: process ID, thread ID (for an active thread), thread entry point (for a non-active thread), and the target CPU (machine). This is an SMU instruction, executed by the source SMU.

First, it opens a secure channel to the target SMU. It then sends the (encrypted and signed) SMU table entry to the target. If the thread has yet to start running, the source sends the new inactive TSC encrypted and signed (using keys exchanged between the SMUs), and discards it locally. If it has started, its secret context is available in the process' shared address space when suspended, so it will be migrated by SDSM. See Pseudocode 6 for migration instructions.

The target SMU creates a new SMU table entry unless one with the same PID and keys already exists (because another same-process thread is already running there). The DBMT root hash is not sent as part of the SMU table entry, as no counter-protected data is being moved. Finally, the thread executes as usual on the target machine, causing the memory system to fetch all the required data and instructions using the shared address space over SDSM.

The SMU table entry is sent through an untrusted medium. An attacker that tries to modify it while being sent will fail the signature validation at the receiver, causing the SMU to halt the secure program and report an error.

```
SMU_MigrateEntry(Phash, Target)
   {EncKey,SigKey}=SMU_SecureChannel(Target)
   Send EncAndSign(SMU_Entry(Phash),
                   EncKey,SigKey) to Target
   Discard SMU_Entry(Phash)

SMU_MigrateThread(PID, TID, Faddr, Target)
   {EncKey,SigKey}=SMU_SecureChannel(Target)
   // without the root hash
   Send EncAndSign(SMU_Entry(PID),
                   EncKey,SigKey) to Target
   If (!TID)
      SC=SMU.SealedStorage.FindInactiveLEP(Faddr)
      If (SC) // if found inactive context for this LEP
         Send EncAndSign(SC,EncKey,SigKey) to Target
```

***Pseudocode 6: SMU thread migration instructions, performed by the SMU***

Even when no active threads remain in the source SMU, this table entry must be kept for future evictions of data that still resides in its cache and for data blocks that are currently owned by this CPU. It may only be discarded when no thread or data of the secure program remain in this CPU.

**Cloud Applicability.** Cloud services normally have gateway computers that accept new job submissions and direct them to target machine(s). Migrating an SMU table entry is useful for initiating a new secure program's SMU table entry on the gateway computer without running it, and then the gateway's SMU may trustfully migrate the secure program's SMU table entry to execute elsewhere. The gateway's SMU acts as a *trusted key vault* that may pass its knowledge to a target SMU, as long as the target SMU's vendor is trusted by the source SMU's vendor. When finished, the SMU entry may be migrated back to a gateway computer, so the status result is collected by the user.

## 2.3 Security Evaluation of ParSeM

We now discuss several attacks on ParSeM that are related to its beyond-SeM features.

**Malicious thread creation.** Here, OS privileged code creates extra threads in order to cause incorrect execution of the program. When a thread is set to run, the SMU automatically loads its TSC from memory. Each thread has its own TSC, protected by the memory encryption subsystem, so duplicating memory blocks will cause a detected error when fetched.

Another flavor of this attack entails duplicating the TSC to the same virtual address on a different machine, so different page tables are used. When the duplicated TSC is fetched into the cache for running the duplicated thread, there is no private counter associated with this TSC on this machine, so decryption will fail. When trying to also duplicate the associated counter, the DBMT will detect a counter integrity error because this counter is not supported by the current DBMT on the duplicate machine.

**Modify the start point or argument of a new thread.** A new thread is maliciously started from an arbitrary address, instead of the instruction following the *clone* call; however, running the thread's secure code requires its TSC that enforces the next instruction to be an LEP, or else switching to Trusted mode will fail, the program will be halted and an error will be reported. Also, modifying the thread's arguments requires either modifying data in the secure memory (protected by memory encryption), or invoking the thread through a different trace than intended (and manipulate the registers meanwhile), which is protected by the LEP upon switching to Trusted mode.

**Erroneous return value to the clone() system call.** Since our *clone()* wrapper code uses the *clone*'s return value to determine its operations, we must consider wrong return value by the *clone* system call.

Case 1: clone failed, but the return value is success, so a pending thread context stays available. This is essentially a denial of service attack, since the program expects a new thread to run, but it will never do so. If a future *clone* will use the long-waiting pending thread context, then the denial of service has finally ended and we simply 'shift' the denial of service to the newer thread. Because this attack must cause at least one thread to suffer from an infinite denial of service, it is outside the scope of this work. (It should nonetheless be noted that security is not compromised.)

Case 2: *clone* succeeded, but the return value is fail. This causes a race condition with two possible cases:

a) The new thread is attached to its pending context after the secure program has deleted the pending context. In that case an error is most likely to occur, since the pending context has been deleted. It will not occur if a new thread was set to run meanwhile (using the same address), so the error is simply shifted to the last thread in such a sequence of events.

b) The new thread is attached to its pending context before deletion. In that case the thread is ready to run (and may already start running). However, the creating thread calls *SMU_NewThreadDelete()* to delete the inactive context. In the discussed case this context is no longer inactive, so *SMU_NewThreadDelete()* will declare an error and the program will be halted.

**Unshared address space.** Here, different address spaces are presented to two same-process threads, so when accessing the same address they will read different values. If the threads run on the same CPU, cache coherence protects the data internally, so this must be done by manipulating evicted memory blocks. The counter and DBMT are updated on every eviction, and counters are validated when being fetched back into the TA, regardless of the last writer, so the most recent value must be fetched on the next read, or else an error will be detected.

If the threads are running on different CPUs, then although memory blocks may reside in more than one CPU, they must be unmodified. Modifying a block causes SDSM to invalidate all its other copies. Also, fetching a missing block from another machine is validated by SDSM. Undetected violation of the same-process address spaces is thus impossible.

## 2.4 Secure Accelerators

Heterogeneous systems often comprise accelerators alongside CPUs [46]. Being architecture agnostic, ParSeM’s approach is also applicable to accelerators. Unlike CPU cores, accelerators normally do not have their own OS; they run parts of the program, and oftentimes of multiple programs concurrently. They normally comprise a management entity, execution elements, and on-board memories (sometimes shared with the CPU), and normally use isolated memory regions rather than the process’ address space.

We define the trusted area as the accelerator chip. We require an SMU in the secure accelerator chip for establishing a secure communication channel with other SMUs (or users), and to store secret keys. The SMU uses these keys for encrypting and decrypting the memory content of the related job when entering and exiting the accelerator chip, such that no data enters or leaves it chip without the SMU's supervision.

Since accelerators do not run an OS, there is no privileged code that may interfere with the secure job. We assume that the trusted accelerator chip is correct and performs its operation trustfully. A key requirement for an accelerator that runs one job at a time is to clear the previous job's traces in between jobs. If running multiple jobs simultaneously, another key requirement is that the accelerator must enforce complete separation among different jobs. The SMU is used for identifying the jobs and job replacement time and it produces a wipe signal when required, but the accelerator's internals must be modified to accept such signal and work accordingly.

To our knowledge, a critical obstacle in the way of such a solution is that the management of many accelerators still heavily depends on the CPU. If this CPU is untrusted, then it may damage our security assumption on the accelerator. However, having a small CPU as part of the accelerator (e.g., in the same chip) just to perform the controlling operations may solve this issue. This small embedded CPU should have SMU capabilities, and it merely acts as another trusted node. In view of the strength and area of accelerators, and the fact that only little is required from this CPU (controller), adding it is actually a simple and effective solution.

Consider, for example, GPUs and Smart NICs. They run programs that access local untrusted memory, portions of which may be cached internally. The program must be protected using conventional memory encryption, performed by the SMU. The data may belong to the program itself, so the internal unencrypted memory content is protected by the SMU using *Secure Access*. Alternatively, the data may consist of a stream that comes from the untrusted world (e.g. the network, or a sensor), so the actual program must be prepared for the ParSeM accelerator using unprotected memory access (see Appendix B for instructions). In any case, the registers used for computation are cleared between different jobs.

Another example is FPGAs, which are commonly used for pipelined calculations. The FPGA configuration program is signed (and optionally encrypted), so it must be validated correctly by the SMU to be applied to the FPGA. This also triggers a wiping process for all the internal registers and memories. For calculations over protected data only, the stream of data must be validated by the SMU when entering the chip; this can be done by assigning continuous addresses to the data stream. Calculation over unprotected data is possible if stated so by the image, which causes the SMU to bypass its security checks during runtime.

A new job is sent to the accelerator with a randomly chosen encryption key (using the secure link between SMUs), and the memory image is encrypted with this key using counter mode with zeroed initial seeds, alongside an integrity signature (if required). The accelerator's SMU then creates a local tree for integrity tracking, verifies the signature, and zero-initializes a local seed structure. When done, it saves the results signed and encrypted in memory. Besides the aforementioned cleaning, the accelerator internals and normal operation remain unchanged, so we do not impose any requirement on the accelerator's programs. Furthermore, existing accelerator programs (binaries) are also supported, where even instrumentation is not required (unlike programs for ParSeM CPU). The only addition required is security attributes (discussed above), which reside in the program metadata.

Preparing a program to run on ParSeM with compatible accelerators requires calling a shim for the accelerator management functions for code insertion. This code opens a secure channel with the accelerator’s SMU, signs and encrypts a package for the accelerator, and finally accepts the protected results from it. This shim must be tailored per accelerator, presumably by the accelerator’s manufacturer. Since the accelerator’s internal operation is unaltered, and the added protection only acts as a gateway for encrypted data and internal separation of unrelated jobs, ParSeM's secure accelerator method is applicable to many types of accelerators. Also, because most of the overheads occur between jobs, we do not expect a significant implication on the actual job's performance, so a quantitative evaluation is left as future work.

# 3 Performance Evaluation

We extended SeM-Prepare and SeM-Simulator to support the instructions and added for ParSeM, and these are described in detail in Chapter 5 and Appendix A, respectively. Since running parallel workloads on distributed settings assumes the existence of SDSM and DBMT, the performance results take into account the overheads caused by those as well.

ParSeM's overhead is in the program’s memory allocations (for secure initializations)

and access (encryption), its use of file access functions (for cryptography), and its mode changes. All these have a negligible effect on performance-critical elements such as the cache and branch prediction, and none on execution units.

SeM-Simulator allows running long programs (>$10^{12}$ instructions), yielding statistically significant results. We instrumented the PARSEC benchmark suite [48] using the SeM-Prepare tool, and then ran it with its biggest dataset (native) on SeM-Simulator, with many (32, 64, 128, 256) threads, each runs on a single-threaded node. (*facesim* and *blackscholes* do not scale to 256 threads.) All programs successfully spawned threads, changed modes, invoked system calls, used files, and allocated secure memory.

Figure 6 shows the performance loss relative to no security at all (running the benchmarks unchanged) with and without memory encryption, including SDSM's overhead. The mean performance penalty is 2.2% - 3.1% (for 32 - 256 threads), of which 1.5% - 2.3% is caused by memory encryption and merely 0.8% by ParSeM itself. Performance reduction for the PARSEC *dedup* program is much higher than the rest, and without it the mean reduction drops to 2.0-2.7% with memory encryption and 0.17-0.18% without. Given such a low penalty, no other solution (past or future) can do much better.

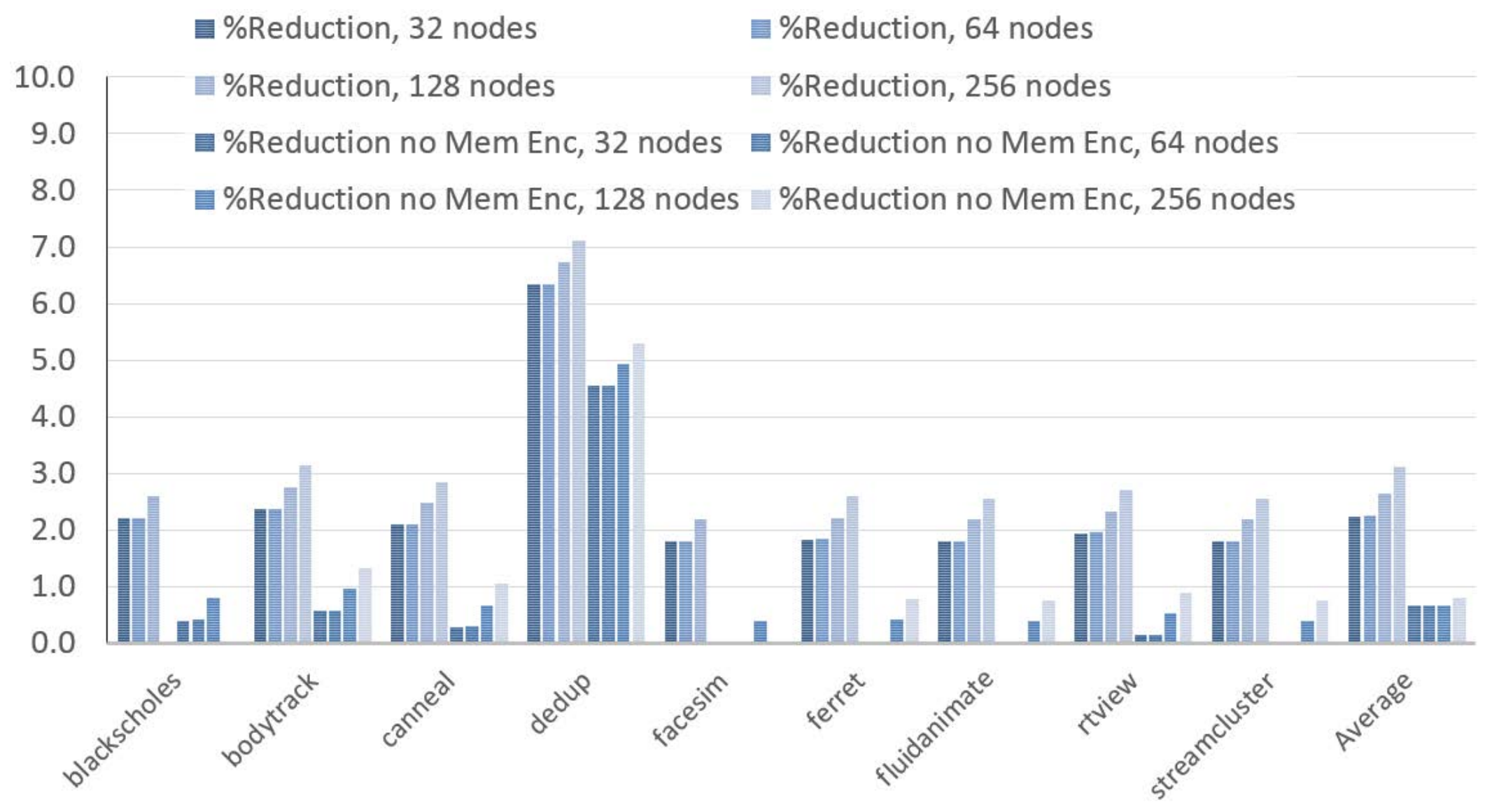

***Figure 6: ParSeM performance overheads vs. node count, running PARSEC benchmark suite***

Table 2 shows the number of thread creations by various PARSEC benchmarks during their entire run time for various numbers of nodes, as well as the normalized mode-switch counts (per 1k instructions). From Figure 6 that depicts performance in conjunction with Table 2, it is clearly evident that mode changes have only a negligible implication on the actual performance; the overhead caused by thread creations is negligible, since these are infrequent relative to total amount of work. Figure 7 shows memory allocations and I/O accesses normalized to the number of instructions, for the various benchmarks. Both are clearly correlated with ParSeM's performance reduction, and our tests show that software encryption for I/O is at the core of this reduction. E.g., *dedup* requires an order of magnitude more I/O accesses and memory allocations, causing a 2-3X bigger performance reduction than the rest. This may be sharply mitigated by hardware support for secure I/O.

| | 32 Nodes | | 64 Nodes | | 128 Nodes | | 256 Nodes | |
|---|---|---|---|---|---|---|---|---|
| Benchmark | Thread Creations | Mode Switches (per 1k instr.) | Thread Creations | Mode Switches (per 1k instr.) | Thread Creations | Mode Switches (per 1k instr.) | Thread Creations | Mode Switches (per 1k instr.) |
| blackscholes | 32 | 9.581 | 64 | 10.907 | 128 | 7.543 | --- | --- |
| bodytrack | 33 | 1.3 | 65 | 1.385 | 128 | 0.953 | 257 | 1.013 |
| canneal | 32 | 8.512 | 64 | 9.572 | 128 | 7.094 | 256 | 7.934 |
| dedup | 98 | 1.091 | 194 | 1.357 | 386 | 0.697 | 770 | 0.75 |
| facesim | 31 | 3.009 | 63 | 2.56 | 127 | 2.014 | --- | --- |
| ferret | 130 | 2.168 | 258 | 1.947 | 514 | 2.157 | 1026 | 1.808 |
| fluidanimate | 32 | 8.712 | 64 | 9.558 | 128 | 10.788 | 256 | 10.306 |
| rtview | 32 | 0.337 | 64 | 0.339 | 128 | 0.333 | 256 | 0.33 |
| streamcluster | 192 | 0.323 | 384 | 0.691 | 768 | 1.973 | 1536 | 4.674 |

***Table 2: Statistics of events that occurred during PARSEC execution on ParSeM for 32, 64, 128, and 256 compute nodes settings.***

ParSeM was not measured against prior art because no prior art supports distributed parallel execution; moreover, with ParSeM's minor performance reduction and negligible overheads, none can do noticeably better on common features.

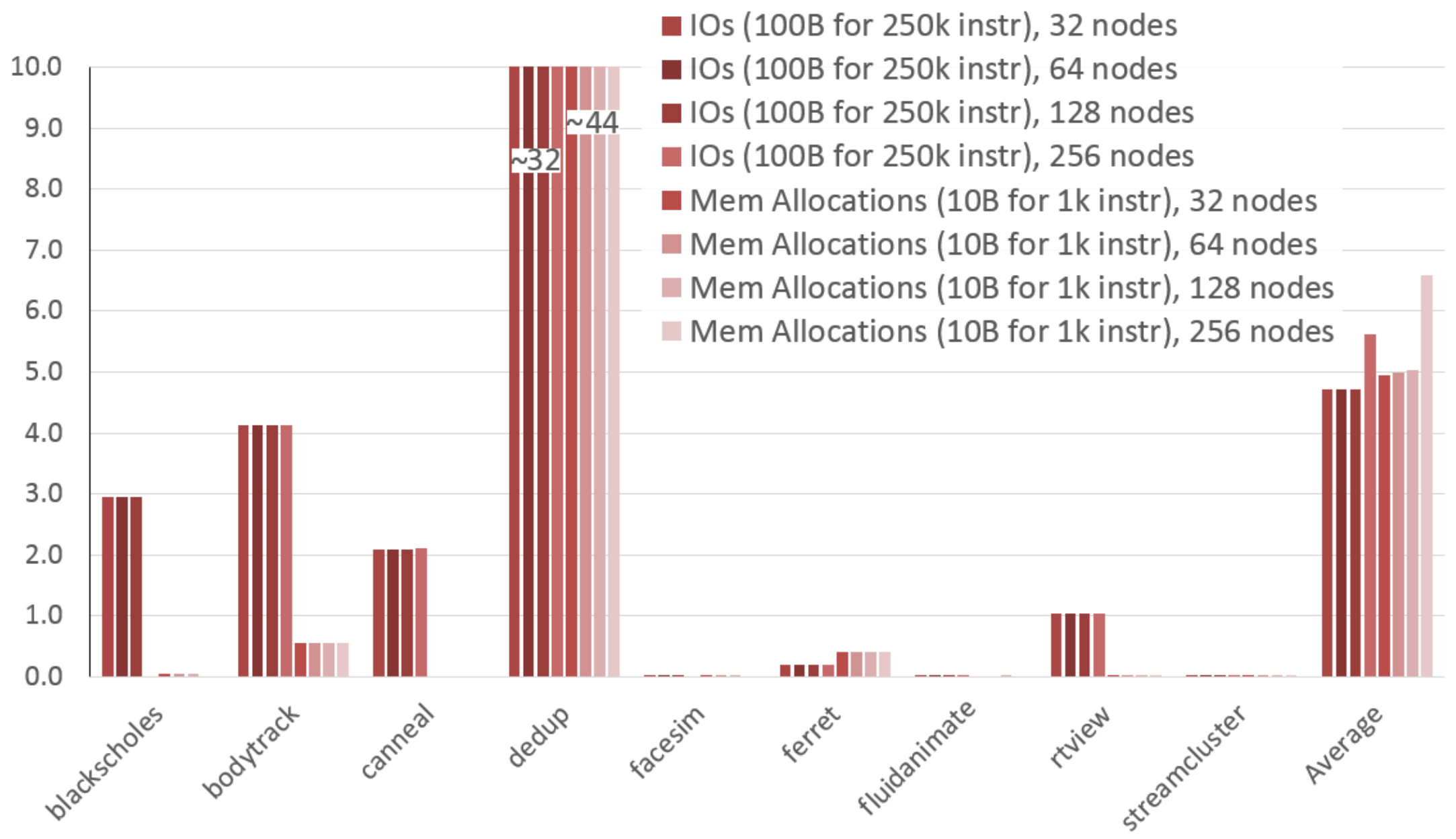

*Figure 7: PARSEC benchmark suite I/O and memory allocations*

# 4 Conclusions

ParSeM is a hardware based solution that protects against software and off-chip hardware attacks, including a hostile OS, hypervisor and provider. It can protect at process granularity, and allows secure programs to use unlimited, dynamically allocated memory. By adding hardware assisted thread creation, termination, and migration, it enhances SeM to fully support parallel programs on single core CPUs, CMPs, and distributed platforms (including dynamic thread migration). Additionally, ParSeM supports heterogeneous systems with secure accelerators that support a small set of requirements.

Existing program binaries are automatically instrumented on a trusted computer by SeM-Prepare, simply replacing syscall instructions and embedding trusted shims, shared libraries are embedded, and then encrypting and signing the ParSeM-ready binary. This allows running existing program binaries securely. Lastly, only one instruction is added to the untrusted OS.

ParSeM degrades performance by < 3% relative to no security. Moreover, in most cases, 95% of this is due to memory encryption and authentication; in I/O intensive programs, much of the degradation is due to software assisted encryption, and may be sharply mitigated using hardware.

Jointly providing security, performance, backward compatibility, and furthermore supporting parallel and distributed programs and even accelerators, ParSeM thus constitutes a major step towards convenient and efficient secure computing on untrusted platforms, including public cloud systems. Topics for further study include hardware support for secure I/O, further study of the use of secure accelerators, direct support for scripting and bytecode supported languages, and using RDMA with ParSeM.

# Chapter 4 Supporting Programming Models and Shared Libraries

## 1 Scope of Support

SeM support existing binaries with some limitations. Embedding all shared library functions into the binary, we assume that calling untrusted code intentionally is only done by system calls. Indirect calls must only aim at locations within the same parts of the code (trusted indirect calls to trusted code, and untrusted indirect calls to untrusted code), or else arguments will not be passed. This may limit dynamically loaded libraries.

Programs that use the *mmap()* system call are currently not supported, because *mmap()* uses a memory region that is accessed both by the untrusted OS and by the application. This prevents a secure initialization as done for secure memory regions, so SeM cannot use that securely.

Nonetheless, running the PARSEC benchmark suite correctly (including C and C++ parallel workloads from various fields), and also other workloads such as AI, compiling and interpretation, and video and data compression, suggests that this limitation may not be severe.

SeM assumes a multi-threaded system with distributed shared memory, though programming for message passing interface (MPI) [53] is possible by running many single-threaded instances of the secure program, which are also supported in the basic single-thread single-core solution.

Since SeM target accepting program binaries, using languages such as Java or server side scripting languages requires preparation of the runtime environment for SeM in order to ensure its confidential and unharmed execution, and the actual program is then run as its input. Direct support for such languages remains a topic for future work.

Finally, the SeM benchmark results matched the results with unchanged binaries. This strongly indicates correctness, as well as the applicability of SeM to existing binaries.

## 2 Shared Libraries

Using shared libraries like libc [32] for execution is common. Thus far, preparing a binary for SeM required the embedding of shared library functions (SLFs) into it. Besides the obvious drawbacks of larger binary and memory footprint for the secure process, this also grants the SLF's code access to the entire memory space of the secure program.

Though accidental data leakage is prevented by SeM's security mechanisms, intentional code manipulation by a malicious shared library provider cannot be ruled out. Embedding SLFs into the binary ensures that the **same SLF code executes on both the user's trusted computer and on SeM**; however, the same code may behave differently in different environments (e.g., private computer vs. VM). On the other hand, SLFs that are not part of the secure binary may have been manipulated beforehand, and can return undesired results, which therefore must be actively checked by the secure program. A tradeoff between approaches. Therefore, we want to also allow the use SLFs that are not statically embedded into the secure binary, and it is critically important to ensure that these can only access their required data.

We now present a new method that enables the use of untrusted SLFs by a secure program. We start by introducing new stack operations, and then dive into the details.

## 2.1 Stack Operations

A secure program uses a non-secure stack (NSS) for non-secure function calls that run from its context, and a secure stack (SS) that serves function calls within the secure code. Each is natively accessed (by *push*, *pop*, *memory[rsp+offset]*, *call*, *ret*, etc.) as if it were the only stack.

To use the NSS for communication between the secure and non-secure code, we add two SMU instructions that allow secure code to bypass the *Secure Access* mechanism:

- SMU_PushNA – pushes data into the NSS, regardless of its Auth status, and sets its memory block’s Auth bit to False.
- SMU_PopNA - pops data from the NSS, regardless of its memory block’s Auth status.

  These instructions use the non-secure stack pointer, and update it accordingly. They must be authentic to execute, and must therefore run in Trusted mode.

## 2.2 Using Untrusted Shared Library Functions

We wish to enable a user program provided as binaries to use shared library functions (SLFs) while running securely. These SLFs are not part of the secure program, so we do not enforce their correct execution. Also, we do not modify them.

Each OS has a convention for passing arguments to and getting return values from functions. These conventions (e.g. System V AMD64 ABI [37], Microsoft x64 [38], IA-32 fastcall [44]) address the number of arguments passed via **registers**, and which registers to use. Passing of more arguments than the number of dedicated registers always uses the stack, and the function’s return address is pushed into the stack by the *call* instruction, so returning is simply done by *ret*.

**First Change to Trusted mode.** With no embedded SLFs, the program initially runs untrusted code (e.g. libc functions) before calling *main()*; then, a trusted function is called by untrusted code and the secure phase of the program begins. When done, it returns to its caller, so the return address (originally stored in the NSS) must be available in the SS. Therefore, the first mode change to *trusted* for a *call* instruction (untrusted code calling a trusted function) should pop the return address from the NSS and push it into the SS instead. This allows a secure program to finish correctly in *Untrusted mode*.

**Inter-Mode Calls.** Although the registers are automatically cleared when switching to Untrusted mode, some of them must sometimes remain untouched. For that, we add the *SMU_CallNoSec(i)* instruction, where *i* ranges from 1 to the maximum number of arguments passed via the registers in the chosen architecture (E.g., 6 in AMD64 ABI). This instruction is similar to *call*, with three minor differences: 1) (only) the first *i* register arguments will not be cleared in the next switch to *Untrusted mode*; 2) the return address is pushed into the NSS instead of the SS; and 3) the next switch to *Trusted mode* will not clear the register used as return value (*rax* in AMD64 ABI).

**Argument Passing.** Consider code generated by a GCC compiler [45]. (Our methods are easily adaptable to binaries produced by other compilers.) When calling SLFs, GCC uses a dedicated per-SLF .plt function to locate the address of the requested SLF. When called for the first time, the .plt uses the dynamic loader to find the address of the SLF. Upon subsequent calls, it redirects the program directly to the SLF.

Whenever an SLF is not part of the secure program, its code is *non-authentic*. Therefore, when called by the secure program, the SMU will change to *Untrusted mode* and will clear the value of the registers, making them unavailable to the SLF. Furthermore, arguments passed via the stack (using *push* instructions) are natively pushed into the SS, so these are also unavailable to the SLF. Therefore, transferring arguments to untrusted code requires special attention.

Whenever an SLF's arguments can be passed via the registers, we simply replace the original call instruction to the .plt function with *SMU_CallNoSec(i)* to the same target (where *i* is chosen per SLF). This allows argument- and return-value passing via the registers, without revealing any additional data.

For SLFs that require more arguments than the available argument registers, or if simple argument passing does not suffice (more later), we use automatically-created trusted shims: we create a per-SLF trusted shim, and embed it in the secure program (must be encrypted). It uses *SMU_PushNA* to store the additional arguments into the NSS, and then calls the SLF's .plt function using *SMU_CallNoSec(i)* (where *i* is the maximum number of register arguments for the target architecture). Finally, a call to the SLF's shim replaces the original call. (Figure 8.a)

When the original binary code invokes the SLF, it prepares its arguments according to the target OS's convention (in registers and in the stack), so our approach simply bridges the security gap for the SLF. When the SLF finishes, it invokes *ret*, which returns to the trusted shim (if used) or to the original calling point. Using *SMU_CallNoSec(i)* ensures that the return address is available to the SLF, and that the return value is available to its caller. (Figure 8.b)

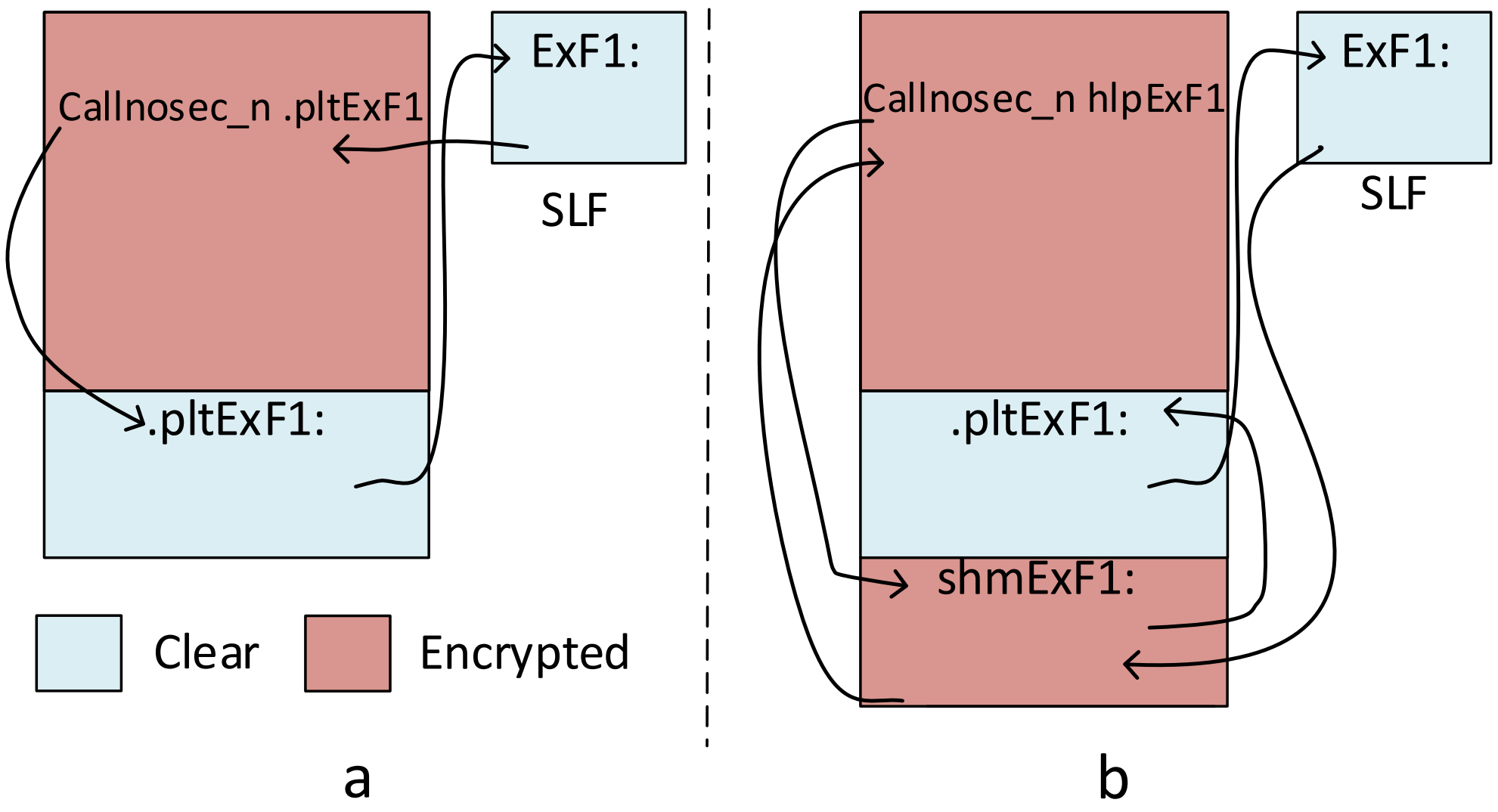

***Figure 8: Program flow for a shared library function (SLF) that accepts arguments: a. by the registers only; b. by the registers and the stack.***

Pseudocode 7 depicts the modifications to the original binary as well as implementation of the trusted shims, with the secure parts (encrypted) on a grey background. It targets the System V AMD64 ABI function calling convention, used in modern 64-bit Linux systems [10], and assumes integer arguments; implementing for floating point arguments is similar.

The trusted shim is added automatically to the original binary on the user's trusted computer. This is done by scanning the .plt section for per-SLF entries, analyzing the invoked SLFs (their binary, source code if available, or public documentation), and locating the SLF calls. The shims must match the number of arguments required for the SLF.

**Pointer arguments.** The above approach is suitable for values passed directly between

| | |
|---|---|
| **Original binary:** #secure<br># store argument #1-#6 into the registers<br>mov rdi, arg1<br>…<br>mov r9, arg6<br># store argument #7-#N into the SS<br>mov [rbp+..] arg7<br>…<br>mov [rbp+..] argN<br>call ~~*.pltExF1*~~ *shmExF1* | ***shmExF1:*** #secure<br># put arguments in NSS<br>…<br>SMU_PushNA [rsp+(N-6)*4] #argN<br>…<br>SMU_PushNA [rsp+4] #arg7<br>SMU_CallNoSec(i) *.pltExF1*<br>ret |
| ***.pltExF1*:** #non secure<br>… | |

***Pseudocode 7: Using a trusted shim for argument passing (new code is underlined). Encrypted parts are with grey background***

secure and non-secure code. However, when pointers are passed, the actual data exists in the secure memory. Possible approaches to address this case:

1. Embed SLFs that get pointers as parameters into the original binary, thereby making them part of the secure code. This should also be done recursively for SLFs that they call (with pointers).
2. Copy the data to non-secure memory using a trusted shim.

Both approaches assume prior knowledge of the type of the SLF's arguments, so that appropriate decisions can be made. The complexity of the second approach is similar to the one used in complex system calls, discussed in Chapter 2.

**Additional code insertion.** We support automatic insertion of additional code into the trusted shims for SLFs that require more than just argument passing. E.g., buffer copying from and to the secure memory, software protected I/O functions (including encryption and decryption), etc. It may also be used for secure thread creation (Chapter 3) instead of embedding the thread management SLFs, and for initialization of secure memory allocated by the secure code such that it is accessible by the secure program (due to *Secure Access*).

This is also usable against Iago attacks. E.g., to protect against a malicious *malloc()* SLF that reassigns an already allocated memory region (causing the secure program to overwrite its own data), simple code is added to the memory allocation (and freeing) trusted shim for checking the uniqueness of the allocated regions.

We designed and implemented a SeM-Prepare tool, which accepts the original binary and also code in C for additional code insertion, and compiles and embeds it into the desired location in the trusted shim.

# Chapter 5 SeM-Prepare

We implemented SeM Prepare, to prepare an existing binary for SeM (hence making it SeM-ready). To our knowledge, this tool makes SeM the only secure architecture that supports existing binaries, where the ones that come closer [43,49] only support unmodified source code, and thus require the program's source code and recompilation. In this chapter we focus on the preparation (instrumentation). The SMU instructions for the setup and finish processes, to be used by an untrusted management program running on the untrusted machine, are listed in Appendix B.

SeM Prepare is an automatic tool for converting previously-compiled Linux binaries (ELF) into SeM ready Linux binaries, and a similar tool can be used for any OS. We used GNU BinUtils [N] to encode and decode instructions, and ELFIO [N] to manipulate binaries, i.e. reading, generating, and instrumenting ELF files. New instructions were encoded using multi-byte NOP instructions.

SeM Prepare adds two new sections to the binary: '*nosec_init*' and '*sec_init*'. *nosec_init* is set as the new entry point instead of *main()*. First, *nosec_init* connects the secure program with its already existing SMU entry by invoking *SMU_SetPID*(*phash),* where *phash* is either chosen by SeM Prepare or supplied by the user. (*phash* is the same value as in the SMU entry, and is not secret.) It then allocates memory for a secure stack and for the encryption counters and integrity tree metadata (Chapter 2), and calls *sec_init* that initializes the secure stack by *SMU_InitA*, updates the stack pointer, and calls *main()* (so the secure code begins). The address of *sec_init* is the *First LEP* in the SMU table entry. See Pseudocode 8 for the sequence of events.

SeM-Prepare may either embed shared library functions into the secure binary, or it may use them as shared resources on the target machine. Each requires a different approach.

```
nosec_init:
   SMU_SetPID(phash)
   SecStackPointer=Malloc(stack_size)
   EncCounters=Malloc(counters_size)
   IntegrityTree=Malloc(integrity_size)
   call sec_init

sec_init:
   SMU_InitA(SecStackPointer, stack_size)
   SP=SecStackPointer
   call main()

……. // the original program
```

***Pseudocode 8: Initialization code added into the binary***

### Embedding Shared Library Functions

SeM-Prepare embeds shared functions into the binary, and then updates the target of *call* instructions with the new embedded location. Originally, a binary that uses shared functions invokes a helper function to locate the address of the shared function, using the dynamic loader. Considering binaries that were compiled using GCC, the .plt section contains the shared function locators (e.g. for "printf" shared function, the printf@plt is called by the program). These plt functions are not required anymore, and calls for those are simply replaced with a direct call to the embedded function. This must also be done recursively for the embedded functions themselves, since one shared function may call another. Additionally, shared functions may require shared functions from other libraries, so these should be embedded and redirected to as well.

Then, SeM-Prepare looks for *syscall* instructions in order to identify the system call in use; this is done by locating an immediate assignment to eax just before the *syscall* instruction. SeM-Prepare then queries a directory of system calls (consisting of 314 system calls in 64-bit Linux systems) to identify the number and type of arguments. It then replaces the *syscall* instruction with the *SMU_syscall*(argnum) instead. For system calls with pointer arguments, the syscall instruction is replaced with a call to a shim that copies the required data to the unprotected memory by using *SMU_StoreNA* instructions, and it then invokes the desired system call by *SMU_syscall*, with the pointer argument redirected to the unprotected copy. Additionally, some shims may perform encryption and decryption instead of simply copying the data to/from the unprotected memory, e.g. for file accessing system calls (*sys_write* or *sys_read)*. New shims are put in a new section in the binary that is called *secure_shim*.

Being closely related with the thread creation and termination, the *clone* and *exit* system calls require special instrumentation to support secure hardware-assisted thread management. This is done by adding *SMU_NewThread()* just before the *clone* system call to create a pending context for the new thread (See Pseudocode 9), and *SMU_ThreadDelete(TID)* just before the exit system call to discard the thread's secret context.

```
…
new_stack=malloc(…) // non secure stack
…
SCID=SMU_NewThread()
clone(flags, new_stack )
if (rax>0) // return value
    rsp = malloc(..)  // secure stack, including secure init
   push(thread_func_addr) // normally done by clone
else if (rax<0) // clone return value
  SMU_NewThreadDelete(SCID)
…
```

***Pseudocode 9: Additional code inserted before and after the clone system call. New code is in grey background.***

**A technical remark:** adding and changing instructions in an already compiled binary is not an easy task. Data may be placed in the code section (and accessed by other instructions), and function pointers may be used. Because of these, we strive to edit the binary **in place** where instructions are not moved from their original location.

A syscall instruction is 2 bytes long, where converting it to SMU_Syscall(argnum) requires additional bits for the argument. However, we observe that unused bits in the original binary may be used: an immediate value assignment to eax is 5 bytes, where many of these bytes are zero (since only 314 syscall types exist). We therefore use the MSB of the eax assignment to encode the argnum for the SMU_Syscall instruction, and SeM CPU should reduce its value back to zero for invoking the actual system call. There are cases in which eax is not assigned with an immediate value but it is assigned with a value from another register (e.g., r9d), a 3-byte instruction. In every case that we checked, the other register is assigned by an immediate value that is easily backtracked in the code, and it may be modified accordingly.

For cases in which we are required to wrap the system call, we replace the syscall + eax assignment (5-7 bytes together) by a call to a new function that we embed in the ELF file (a call instruction is 5 bytes), that may perform additional code. When done, we simply *ret* to the original location.

The suggested approach is of course a workaround for the commonly used x64 ISA, which allows implementation using current ISA and binaries. Future ISA may enable simpler modifications.

Next, SeM-Prepare implants a shim for memory allocation functions (malloc / calloc), that initializes the allocated memory as part of the secure program's code so it is accessible for the secure program (See Pseudocode 10).

```
void* sem_malloc(size_t size)
    void* retval;
    retval = malloc(size);
    if (retval)
        InitA(retval, size);
    return retval;
```

***Pseudocode 10: sem_malloc() implementation***

**Non Embedded Shared Library Functions** SeM-Prepare scans the binary for shared function calls; for GCC compiled binaries, the plt functions are found in the .plt section, and each plt function is responsible for a specific shared function. Since shared functions are publically known, we query a database of shared functions in order to identify the argument needs of each, similarly to the system call database. For shared functions that only require register arguments, each call directed to a plt function is replaced by a *SMU_CallNoSec(i)* instruction, where *i* is the number of register arguments used by this shared function. (See Pseudocode 11 for an example.)

```
Original binary: #secure
…
# prepare arguments in registers
mov rdi, arg1
…
mov r9, arg6
call .pltExF1
SMU_CallNoSec(6) .pltExF1
```

```
.pltExF1: #non secure …
…
```

***Pseudocode 11: Calling a shared function that only accepts register arguments. Encrypted parts are with grey background, and new code is underlined.***

**A technical remark:** similarly to the issue of converting syscall instructions into SMU_syscall (argnum), we wish to convert *call* instructions into SMU_CallNoSec(i) without moving the original code. Since the .plt section always exists at the beginning of the address space of the file, these *call* instructions always use a negative offset to address the plt function. Additionally, since the binaries are of limited size (we checked binaries that are up to ~100MB), this offset commonly requires far fewer bits than it is addressed with. We therefore use the top 3 bits of the offset to encode the number of registers to be kept (*i*) on SMU_CallNoSec(i) instructions.

The suggested approach is of course a workaround that allows implementation using current ISA and binaries. Future ISA may enable simpler modifications.

For shared functions that require more than the number of arguments that can be passed using the registers (>6 in System V AMD64 ABI), and for shared functions that take pointer arguments, we use a similar approach to the one used in embedded shared functions. A shim is automatically prepared and added into the binary, and the original call to the plt function is replaced by a call to the shim. It copies non-register arguments into the non-secure stack using SMU_PushNA instructions (see Pseudocode 12 for an example); for pointer arguments, it copies the required memory content to the untrusted memory using SMU_StoreNA, and it then calls the original plt of the shared function.

Remark: when not embedding shared functions, the secure hardware-assisted thread

```
Original binary: #secure
…
# prepare arguments in registers
mov rdi, arg1   # pointer
…
mov r9, arg6
push arg7
call .pltExF1
call .shimExF1
…

.shimExF1
pop rcx
SMU_PushNA rcx   # push the 7th argument into the non-secure stack
SMU_CallNoSec(6) .pltExF1
ret

…
```

```
.pltExF1: #non secure …
…
```

*Pseudocode 12: Calling a shared function that accepts seven arguments, the first six are passed via the registers, and the seventh via the stack. Encrypted parts are with grey background, and new code is underlined.*

management instructions cannot be instrumented just before or after the threading system calls. To keep this approach simple, we always embed thread-managing shared functions.

**Additional Code Insertion:** As discussed in Chapter 4, some non-embedded shared functions require more than just argument passing; e.g., memory copying to or from the unprotected memory space, encryption and decryption functions, etc.  We use the trusted shims for this purpose: SeM-Prepare accepts **additional** code in C, complies it, and implants calls to it from the shim functions before or after the actual plt function call. Such code is likely to be supplied with SeM-Prepare for commonly used shared libraries or by the shared library vendor, still keeping the flexibility of a power user to set an existing binary to match its security needs.

Finally, the resulting binary is encrypted and signed, leaving the *nosec_init* and .plt unencrypted. The keys are either provided by the user or chosen randomly; these keys are crucial for the secrecy and integrity of the secure program.

Remark: for the simulation, we used a version of SeM-Prepare that does not encrypt the binary; the simulator differentiate between the secure and non-secure sections using their names, so the security boundaries are preserved.

A secure program is only prepared once, and the secret key's owner may send it to SeM for execution multiple times.

# Chapter 6 Secure Distributed Shared Memory

## 1 Introduction

Adding support for multi node secure execution on SeM raised the need for securely sharing data between trusted nodes that are connected via an untrusted medium (Chapter 3). In this chapter we address this need.

### 1.1 The Problem

Security and privacy in computer systems are major concerns, especially when running programs on third-party systems such as public clouds, in which the user may be unwilling to trust the system provider and thus its hypervisor, operating system and even most of its hardware.

Various studies [1,2,3,13,28] and industry efforts (E.g., SecureBlue++ [18], Intel SGX [22]) have attempted to maintain the secrecy of a program running on a platform of an untrusted owner. They all share a basic element: using encryption to preserve program secrecy.

To execute an encrypted program, every secure system has a system-internal *trusted area* (TA) holding the data (and similarly, code) as cleartext, because operations that can be performed directly on encrypted data are currently very limited [61]. The TA commonly includes the processing core as well as data and instruction caches, and related control mechanisms. Information is encrypted by the user before being provided to the machine in the first place; it is decrypted upon entering the TA (into the cache), is encrypted automatically upon eviction, and is decrypted automatically whenever fetched back into the cache. Unfortunately, all this comes at the cost of added latency.

For encryption of data being placed in (untrusted) memory, *counter mode encryption* [62] has been proposed. Data is encrypted using a keystream block (KB) [63]. The KB is calculated as a complex cryptographic function of the block address, a secret key, and a counter seed. Upon cache miss, the requested block's address is known; assuming the presence of the key and the seed in the TA, the decryption KB is calculated while the missing block is being fetched, requiring only a bitwise XOR with the arriving encrypted block. The memory access latency is thus used to hide that of the KB preparation. However, data encryption upon cache eviction has remained slow, as the KB calculation traditionally [62,64] requires the address of the block being evicted, which is not known in advance. Alternatively, one can pre-calculate and store in the TA dedicated KBs for some of the cache lines (100% memory overhead for those).

Most past studies assigned little importance to encryption latency of evicted blocks, claiming (rightfully) that a write buffer within the TA can be used to hide this latency. However, in a many-core system, eviction may take place in response to a request for the cache block by another compute core. If the cores are connected through an untrusted medium (E.g., PCB), the data must be encrypted for eviction, adding noticeable latency to block fetching. We therefore focus on this problem.

**General setting.** We consider a multi-node distributed shared memory (DSM) system with a (per process) shared address space. Each node consists of a core with its private cache and memory, and the memory coherence of the system is managed by a central or distributed directory. The directory may be implemented in hardware or in software, but it must be trusted (discussed later). At any given time, a given block may reside in multiple private memories and/or caches with read-only permission. Once write permission is granted, a block may only reside in one private memory and its local cache. When a block is needed by a core that does not have it in its own memory, the core's hardware turns to the directory for assistance.

Due to unacceptable communication latency, many DSM systems scale to thousands of computing nodes by sacrificing memory coherence. However, recent technologies (such as Compass-EOS' icPhotonics [65]) allow low-latency communication that may enable coherent DSM systems with thousands of cores at the box or rack level, simplifying the programming model for many massively parallel applications. Fast security support for such systems is thus of interest.

Throughout this chapter, we use *core* to denote a single threaded execution unit along with its private caches, security access control, and cryptography primitives. (Single-threaded merely for facility of exposition.) We refer to the core requiring a missing data block as the *requestor*, and to the core holding this block as the *sender*. Although each core serves as both *sender* and *requestor*, we discuss these roles separately.

## 1.2 Threat Model

We consider a DSM system for many cores connected through an untrusted medium. A parallel program (that shares data among its threads) runs on the system, requiring a common key to be stored in the executing cores. A setup process is assumed to exist for securely distributing and storing such keys in the cores. The cores are trusted, and we assume that they are correct and their internals are physically inaccessible for snooping. An attacker with physical access to the system can inspect and record any off-chip signal and message. It can also change messages, replay old ones and initiate new ones. This applies to both data, commands, control, and coherence management messages (inter-core or between cores and main memory).

All other software, including other concurrently running applications, operating system and hypervisor are assumed to be hostile. We rely on a secure CPU architecture (such as Intel SGX) enforcing by hardware correct process separation, permissions and data integrity using compartmentalized state within the TA; data alteration by hardware, as well as software security issues are treated by other layers, as part of a secure architecture (such as SGX). Furthermore, we do not create new problems in that respect. Denial of service of any kind is outside the scope, as an attacker with physical access may simply power the system down. Side channel attacks are also outside the scope, but we do not introduce new vulnerabilities. These settings are common in real world scenarios, and similar settings were addressed in [16,66,68].

In this setting, we strive to provide fast, scalable security support for directory-based coherent distributed shared memory. Security includes preserving the secrecy of the user's program and data, and detecting any alterations thereof.

## 1.3 Our Contributions

We present a new approach for supporting secure coherent distributed shared memory (SDSM), which provides support for secrecy and integrity of inter-core communication. SDSM scales to thousands of cores while maintaining good performance. It can be added to secure CPUs using any variant of counter mode encryption, running either a trusted or an untrusted OS, such as SGX [22]. By using a TCM (a trusted coherence manager, comprising a trusted directory with added functionality), exploiting native latencies of the DSM system, and using a simple adaptation technique, we are able to dramatically reduce wasted work relative to prior art; also, SDSM scales with essentially constant per-core hardware resources. We assume a write-back cache and inclusive main memory, updating only modified blocks. We do not focus on any particular coherence mechanism, but consider MESI [69] as a common yet simple example.

The specific contributions of this chapter are:

- A new approach for using seeds in counter mode encryption with block-address independent KBs, obviating the need to supply initial seeds while preventing initial KBs from being reused during runtime.
- A new seed management and distribution protocol for avoiding wasted KB pre-calculation work, and exploiting DSM systems' communication latency for hiding that of the KB calculations.
- Smart allocation of hardware resources to obtain a secure and scalable DSM with essentially constant per-core resources.
- Establishing the need for a trusted coherence manager (TCM) to ensure correct coherence status and messages.

The remainder of the chapter is organized as follows. Section 2 provides an overview of memory encryption and related work; Section 3 presents our contributions; Section 4 evaluates them, and section 5 offers concluding remarks.

# 2 Background and Related Work

This section reviews related work on counter mode memory encryption, mechanisms for memory encryption in multi-core settings, encryption seed management and encryption latency reduction schemes.

## 2.1 Memory Encryption

Many systems [16,66,68] use Galois Counter Mode (GCM) [24], which is an authenticated variant of counter mode encryption. GCM relies on a running counter, with KB generation requiring a long computation, similar to counter mode encryption. For simplicity, we will consider the original counter mode as our encryption algorithm, but the ideas and results are easily adaptable to any of its variants.

The use of keystream blocks for memory encryption simply entails encrypting k-bit data *D* using a k-bit pseudo-random secret *R* by performing a bitwise XOR: $E = D \oplus R$. XOR is reversible: $D = E \oplus R$, and is fast to execute.

Recent implementations of counter mode encryption [70] use AES [71] block cipher to generate a pseudo-random number $R = Enc(P, k)$, where *P* is a block-related seed, and *k* is a symmetric secret key. *P* is commonly defined as *P* = *VA*||*S* [62], which is the concatenation of the block's virtual address (VA) with *S*, a counter based seed. Having a unique VA ensures that each block has a unique set of *P* values, so AES guarantees that using the same key *k*, *R* is unique per block and does not repeat as long as *S* doesn't. The seeds may be stored in the clear, as no attacker can reproduce *R* without knowing the secret key *k*.

Only *S* must be stored per evicted block, along with negligible-sized metadata for locating it based on the block address. Together, their size is only a small fraction of *R*'s, resulting in reasonable storage overhead. (See [1,34,72,73] for seed storage and caching details.)

*Enc* may be any block cipher algorithm, and *P* is padded with zeros up to the required size of *Enc*'s input. If *Enc*'s output is shorter than the data block, we use multiple *Enc* blocks $R_i = Enc(P_i, k)$, where $P_i = P \,||\, block_index$, and concatenate all $R_i$'s to form a KB of the required size [62]. The encryption's strength is the same as the block cipher's. The seeds are commonly initialized to 0, obviating the need for supplying initial values while maintaining a unique KB for each block.

Using a proper design, the VA and seed of a missing block are known in the TA at the time of a fetch request. (For simplicity, all the caches are assumed to be in the TA.) The calculation of the *Enc* function, typically shorter than the main memory latency, is initiated concurrently with the data fetch request, so the latency calculating the KB required for decryption of the fetched block is hidden by the memory latency.

## 2.2 Multi-Core: KB Pre-Generation and Encryption Latency Reduction

In multi-core settings, fast memory encryption is essential. Without it, remotely requested blocks will suffer from high eviction latency, which adds to their fetch latency. This problem was first addressed in [74] and subsequently in [75], but only for bus-based shared memory multiprocessor architectures, which are very restrictive and non-scalable.

In order to address distributed directory-based shared memory settings, [68] suggested that each sender core pre-generate a seed and use it to pre-calculate a block-address independent KB. Upon eviction, this KB may be used to encrypt the block that is being evicted on the fly.

A later work [66] added a central trusted global counting controller for all the processing cores, providing a trusted running counter for generating the seeds. It also added three buffers in each core: outstanding pre-calculated KBs (sender), KBs of recently fetched blocks (for re-encryption upon eviction if unchanged), and outstanding KBs for incoming blocks (requestor); this provides greater flexibility with stressful workloads. Its threat model assumes tamper-free memory management messages, allowing only data messages to be attacked, and that no management or coherence message is ever lost.

In order to hide the decryption latency, a requestor must receive the seed that served to produce its encryption KB so as to permit calculation of the KB before the encrypted data arrives. [66,68] suggested sending the outstanding seeds to all cores in the DSM, so they can prepare the KBs in advance. As any KB may only be used to encrypt a single data block, in an N-core system each useful block transfer is accompanied by N-2 wasted seed transfers and KB calculations; this unacceptable waste of energy, memory and bandwidth moreover grows with the number of cores. Limiting the cache size for seeds and KBs at the requestors is also problematic, as it would reduce hit rate as cores are added.

In [16], a DSM scheme with no KB pre-generation is discussed. Each block modification triggers the creation of an outstanding KB, kept temporarily with its new seed. Due to the large area needed per KB entry (roughly the size of a cache block), only a limited amount of buffer space is used for these KBs; if not found in the buffer, a lengthy process is required for fetching the current seed from memory and calculating a new KB. (KBs must not be kept outside the TA!)

Lastly, [16] protects the integrity of seed and control messages by using a delayed timestamped message authenticating code (MAC), calculated using a cryptographic keyed hash function [67] with the program's private key. It is sent back as ACK to these messages (piggybacked onto the next message) without adding latency to the critical path, so these messages either arrive correctly or a tamper event is declared. Data integrity is protected using GHASH [77], which is a lightweight MAC of GCM. We adopt the same integrity mechanisms in our work.

All previous works failed to provide schemes that scale to many-core settings. Their total hardware resource requirement grows quadratically with the number of cores (because each core holds a set of every other core's KBs), or else the efficiency of the existing resources drops dramatically as the core count increases. Previous works did not discuss the trust model for the coherence manager (such as a directory) for producing correct query replies. Finally, they did not address multiple concurrently executing secure applications; there, the limited resources cannot be replicated per application, so intelligent management is a must. We address all these issues.

### 2.3 Seed Management

In [66,68], block-address-independent KBs are used; this allows them to prepare one KB in advance and use it to encrypt the next evicted block. As the secret key is the same for all memory blocks, preventing KB reuse requires the use of different seeds for different blocks. To this end, a global counter was suggested, such that each seed value is only used once at runtime [66].

For initial delivery to the secure machine, the data and instructions of a secret program are encrypted using initial KBs. Concatenating the block-address with the seed to form the KB (see III.A.) allows the use of *zero* as the initial seed value, obviating the need for providing the seed with the program. However, using block-address independent KBs raises a new issue: supplying initial unique-per-block seeds with the program (while forcing the GCC to refrain from using these values) will cost additional storage. [66,68] did not address the seed repetition problem presented here, nor did they discuss how initial seeds are supplied. We will present a new method that uses *no* initial seeds while still distinguishing among KBs of different blocks and avoiding KB reuse at any time.

## 3 Secure Distributed Shared Memory (SDSM)

In this section we present SDSM, our scheme for providing fast and scalable security support for directory-based distributed shared memory. We first present its methods and building blocks, and then show how these are put together.

### 3.1 Seeds and Keystream Blocks

Our goal is to use block-address independent KBs, while using unique KBs both at the beginning and during runtime. We present a simple yet novel approach for choosing seed values, deriving KBs form these seeds, and a heuristic for when to do so. Our scheme hinges on the observation that block-address independence is only required for modified blocks during runtime (for KB pre-generation), not for the initial encryption. This may be used in any secure system or CPU in conjunction with any variant of counter-mode encryption.

We use $R = ENC(P, k)$ as the KB for encrypting the blocks, with $P = f(S, VA)$ padded by zeros to match the required input size of ENC. The function $f$ is defined as follows:

$$f(S, VA) = \begin{cases} 0 \dots 00 \,||VA \quad if\ S = 0 \ \backslash\backslash\ initial\ seed\ value \\ 0 \dots 01 \,||S \quad otherwise \ \backslash\backslash\ VA - independent, \end{cases}$$

with the seeds $S$ at least the size of VA. Each block is initially encrypted using its VA as the $P$ parameter of $ENC(P, k)$; in subsequent encryptions, $P$ is VA-independent and consists of the value of the seed concatenated by '1', so the resulting KBs differ from the initial KBs. Upon decryption, checking if the seed is equal to zero (and acting accordingly) takes negligible time. The seeds are initialized to zero. During runtime, the seeds are simply generated by a running counter, which increments upon each eviction of a modified block. (In Section 3.4 we discuss assigning and using these seeds with many secure CPUs, where we also increment the write counter for non-modified blocks.)

## 3.2 Process-aware Keystream Block Pre-generation

DSM systems usually serve multiple applications concurrently; yet, past work didn't consider concurrent secure processes. In SDSM, each sender core receives seeds from the write counter or directory (discussed in Section 3.4), but pre-generates its own outstanding KBs using these seeds. Requestor cores generate KBs based on senders' seeds (received from the directory) and on the keys shared with them.

Each sender uses a dedicated cache for holding outstanding pre-calculated KBs (as in [66]), which are subsequently used for encrypting remotely-requested blocks. Each secure process uses its own secret key (shared by its threads), so the sender should prepare outstanding KBs **per process**. Each sender core monitors past block requests, and learns from which of its secure processes they were requested recently. Then, the sender prepares outstanding KBs for these processes, favoring those that were asked for more blocks recently.

Assuming constant per-core resources (regardless of the number of cores), and that each core may run many secure programs concurrently, this approach helps in better utilizing the core's resources (compared both with generating KBs for every process running in the system and with so doing for every process that currently has active blocks in its cache).

## 3.3 Trusted Coherence Manager (TCM)

Any distributed coherent memory system has a managing entity (such as a directory) that keeps the status of memory blocks and responds to queries about it. Some previous works ([66,68]) assumed that the related management messages are delivered correctly; [16] suggested a method for ensuring message integrity of both counter and coherence messages. However, to our knowledge, an adversarial coherence manager (directory) was not considered before. We next do so.

***Proposition:*** a trusted coherence manager is mandatory.

*Proof:* Consider cores A, B, and C. Cores A and B share a block for read. Core A wishes to modify this block, but the adversarial coherence manager refrains from sending an invalidation message to B. When C requests this block, it may get an old version from B. [16]'s inter-core message integrity is thus insufficient with an adversarial coherence manager. □

Any coherence manager may be used; it simply needs to be placed within a trusted area. A distributed network of TCMs may be used, wherein each manages an address-based fraction of the memory, as long as all these are trusted and we ensure that messages are verified for integrity and their reception is acknowledged. We also use the TCM for managing a per-process universal (write) counter for supplying unique seeds, and in the next subsection we will discuss using the TCM as part of the seed management system. With a distributed network of TCMs, for each secure process the seed-generation range must be partitioned among the TCMs to avoid duplication. Each TCM is thus assigned a unique portion of the seed-value space and, for every process, a unique portion of the virtual address space.

## 3.4 Putting it all together

We now present our hardware requirements and scheme for fast and scalable secure data sharing for a coherent DSM system, incorporating the aforementioned building blocks (some for correctness and others for performance and efficiency). Specifically, we present a scheme whereby 1) KB calculation latency is hidden from the requestor; 2) little work is done for KB calculation; and 3) a small amount of hardware resources suffices even for large systems.

We consider a DSM system built of many secure CPU cores (referred to in the threat model). Each core includes a trusted area used for the following tasks: it securely stores the application's secret keys (Section 1.1.); it implements GCM encryption, and treats seeds as described in Section 3.1. KB calculation time is assumed to be less than the core's **local** memory access time (though not required for correctness). It has a small cache for outstanding KBs (for sending), allocated per process as described in Section 3.2, and a **single** KB entry for receiving blocks (unlike [66]'s KB cache). It implements [16]'s integrity mechanisms (described in Section 2.2.), taking its latency out of the critical path. Figure 9 depicts our core architecture. An additional memory encryption path may exist for local memory access (possibly address dependent, unlike SDSM), and it is architecture specific. One or more TCMs may serve the cores, and each TCM is responsible for a subset of the memory address space. (Figure 10).

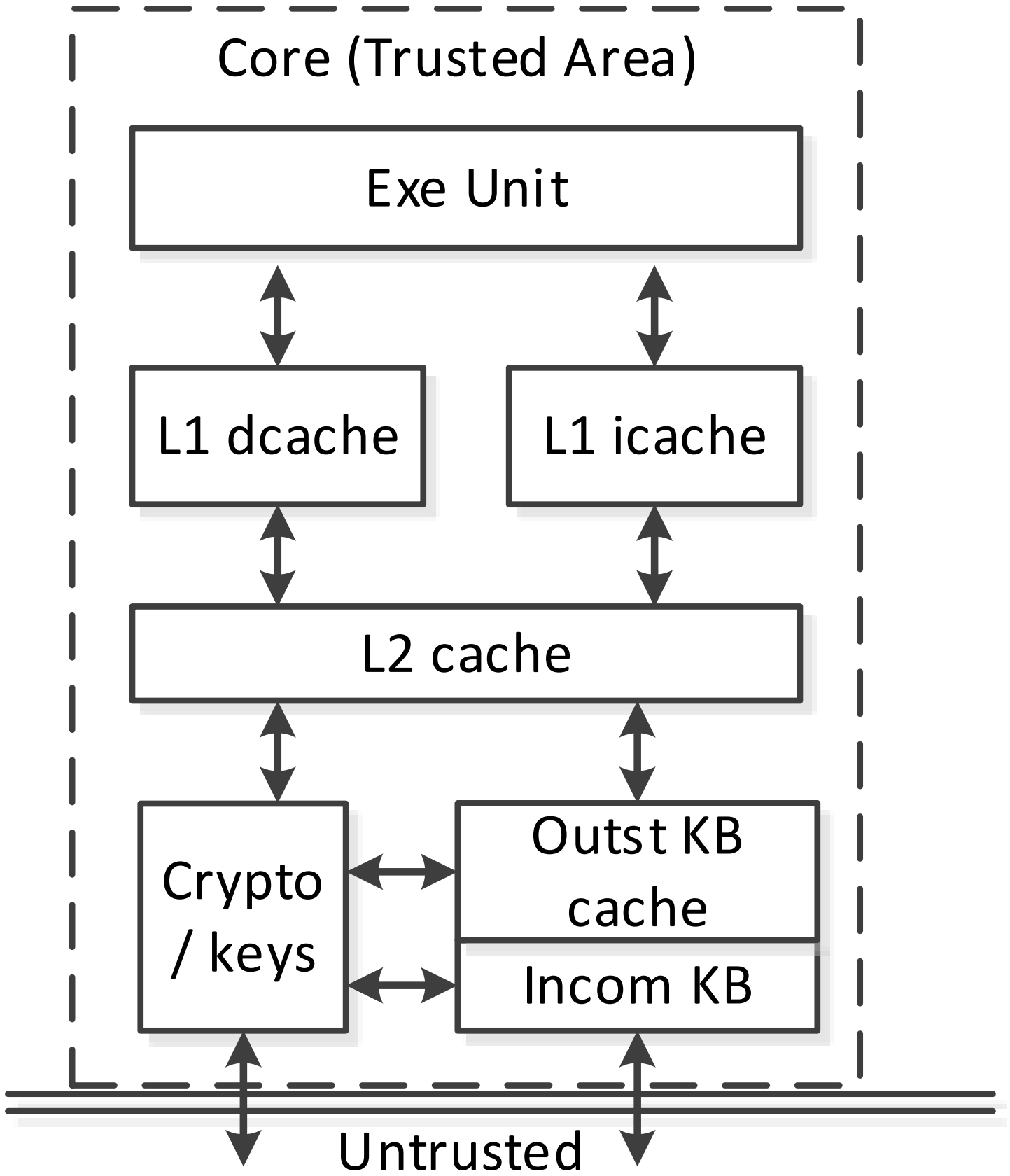

*Figure 9: SDSM core architecture*

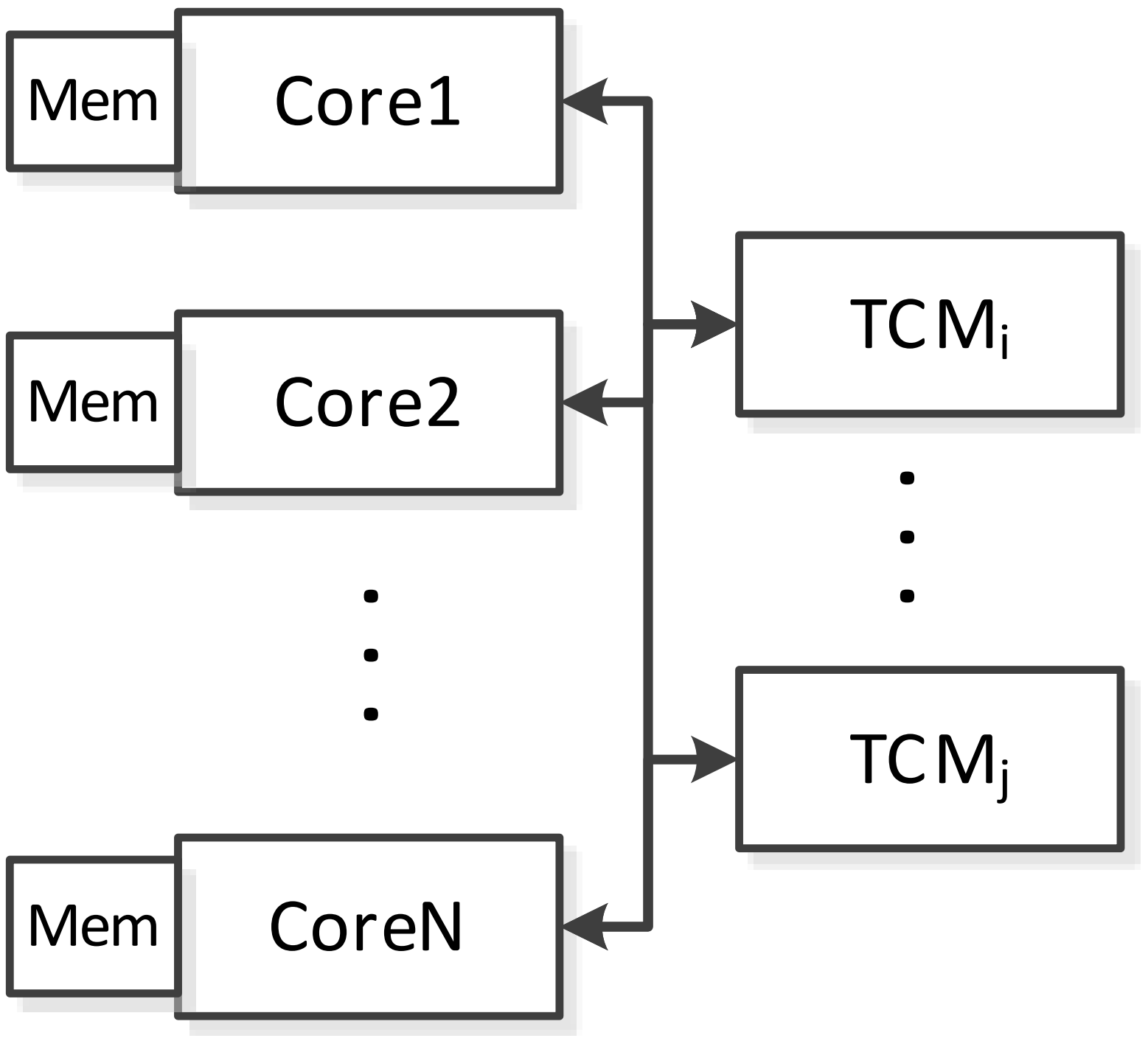

*Figure 10: SDSM system architecture*

**Seed Management in SDSM**

Each cache miss results in a request sent to the requested block's home TCM for checking the block's status. This request also states its purpose – read or write. The TCM forwards the request to one of the cores presently possessing a current copy of the block. (When in shared read mode, the TCM may have a choice.) Considering the communication latency of requests between different cores and the time to calculate the KB, the KB calculation latency can be hidden from the requestor if 1) it gets the seed before it gets the encrypted data such that its KB calculation won't delay its block decryption, and 2) the block owner uses pre-calculated KBs to instantly encrypt blocks that are being evicted. We do this as follows:

**(1)** The TCM generates unique seed(s) (using a per-process counter) at a sender's request (for seeds), stores them in its (TCM's) cache, and sends a copy to the sender. Each sender maintains a short list of (block-address independent) seeds, and prepares outstanding KBs for future evictions.

**(2)** A modified block that is evicted from its owner's cache is encrypted using an outstanding KB (calculated using a seed previously received from the TCM).

**(3)** Once the TCM receives a block request, it sends the first in line (oldest) outstanding seed of the block's owner directly to the requestor (latency reduction), and forwards the request to the block owner, suggesting the sender's seed to be used. (This addresses race conditions among simultaneous requests). If the block is in the sender's cache, it promptly encrypts it using the pre-generated KB, and sends it to the requestor (Figure 11). If it isn't, the sender loads it from its memory, decrypts (using its cached seed with no added latency), re-encrypts with the pre-generated KB, and sends to the requestor (Figure 12). The sender may avoid re-encryption and simply send the currently used seed to the requestor (with the encrypted block); however, the requestor will not be able to use a pre-generated KB, resulting in additional undesirable latency.

Because we use the inherent communication latency to hide the seed transfer latency and KB generation latency at the requestor, the requestor only needs to prepare a useful KB upon need at no latency cost.

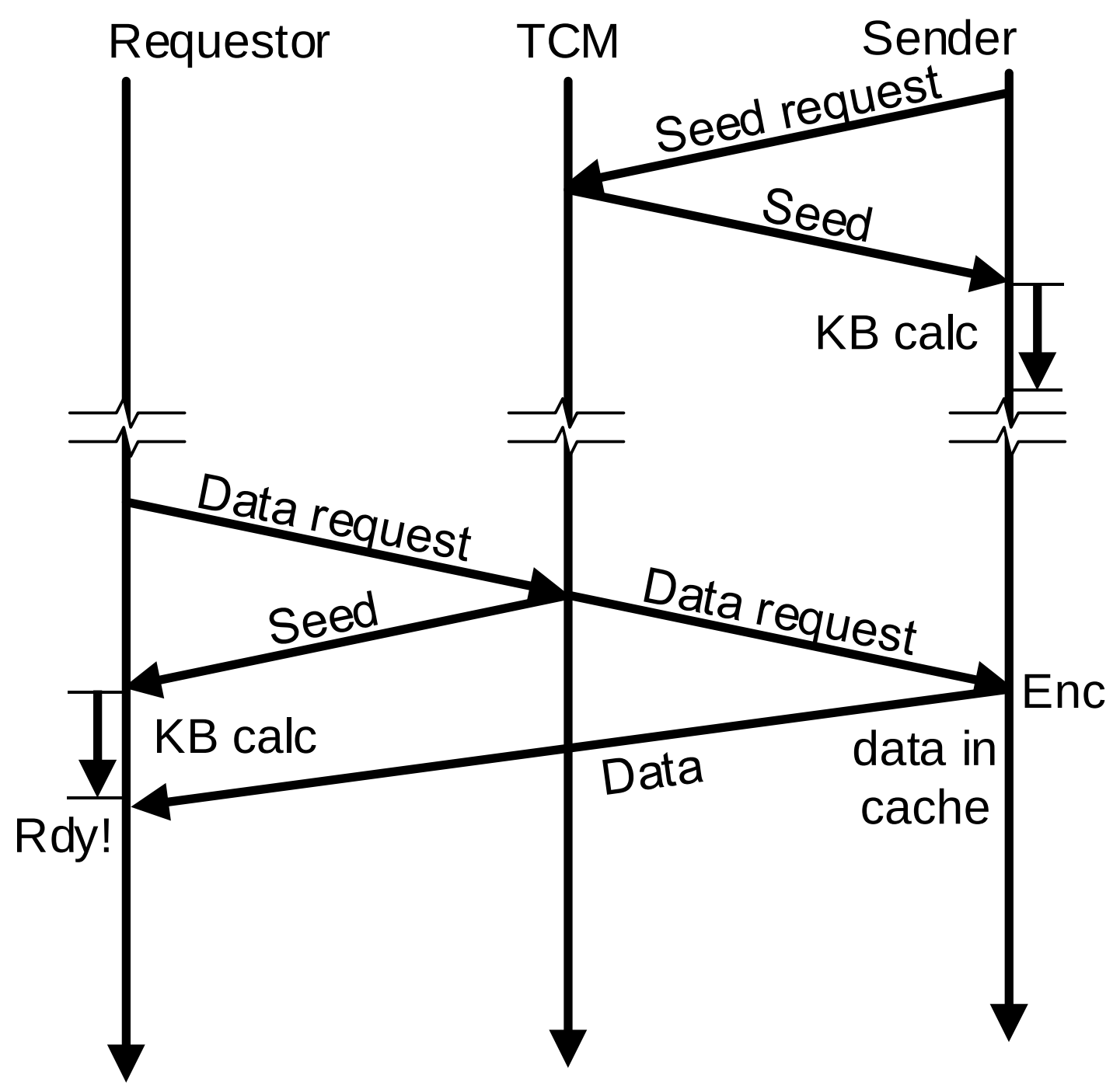

*Figure 12: Order of events on a read request that encountered a local core miss: sender cache hit*

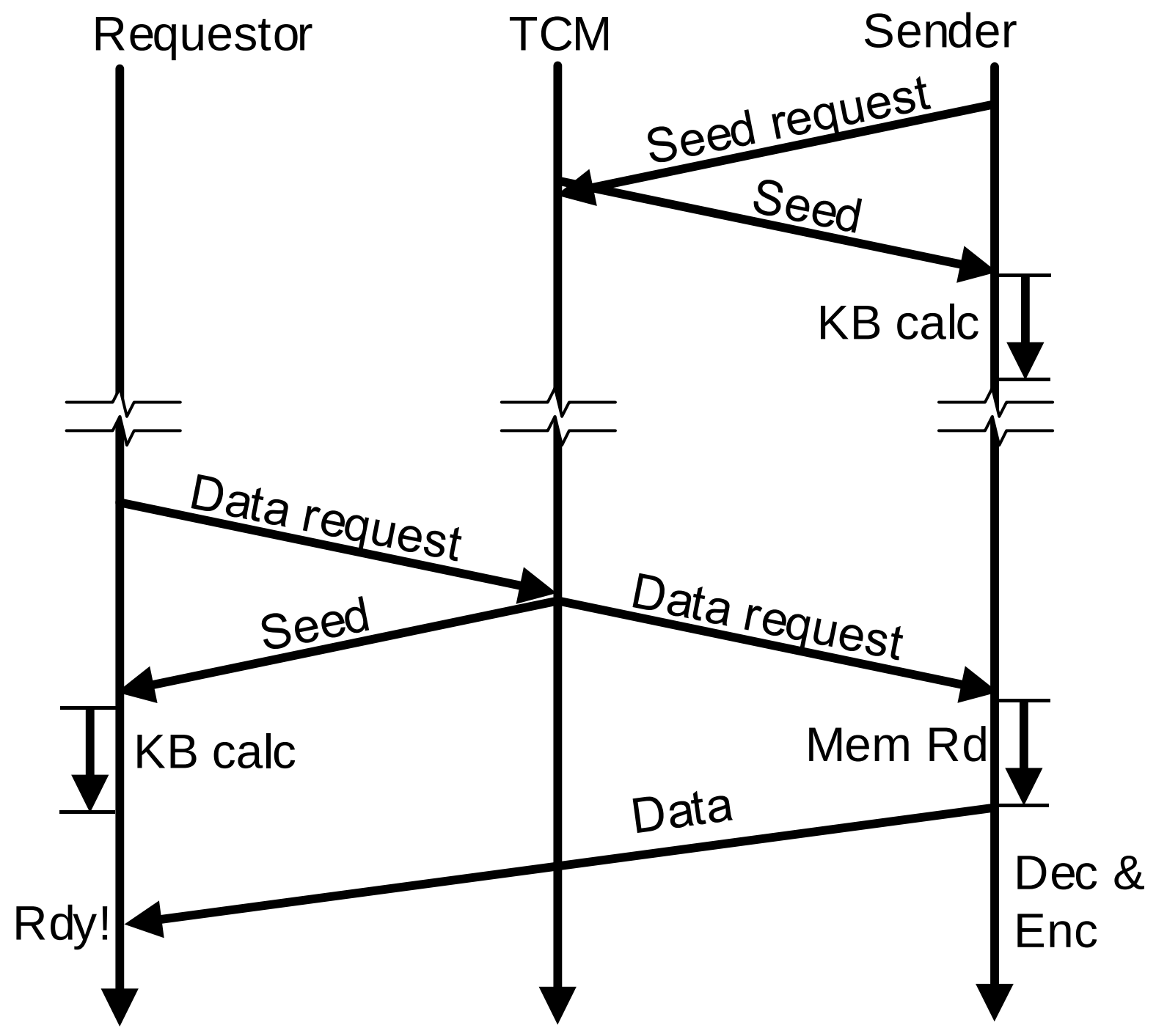

*Figure 11: Order of events on a read request that encountered a local core miss: sender cache miss*

**Remark:** Whenever the requested block is not in the sender's cache, there is an alternative approach to decrypting and re-encrypting. Here, the sender may send the block unopened (and notifies the requestor for its origin), either similarly to a cached block (via its cache or some buffer, without altering it), or else remote direct memory access (RDMA) can be used to copy the block from the sender's memory to the requestor's. The TCM does not know whether the block is in the sender's cache or main memory, so unlike with the original approach, the requestor must prepare two potential KBs for any request (for the outstanding and the block-specific seeds). Therefore, the TCM should maintain a current list of all the block-specific seeds. Although this approach may save some latency by saving decryption and encryption and optionally using RDMA, it results in more space and communication to be used, and requires other extra work to be done. Note also that, with KBs available, decryption and re-encryption by the sender are carried out on the fly and can be pipelined, even down to single-bit granularity. We therefore prefer our original approach. (Moreover, RDMA usually requires operating system support, so the full benefit of this approach requires a hardware-only managed RDMA service.)

Unlike previous work, we do not use a per-core dedicated KB cache for decrypting incoming blocks; instead, the TCM holds the seeds in its cache. Consumer CPUs presently have up to 8MB caches, and server CPUs may have ten times more [78]. Using 8-byte seeds and only half of the TCM cache for seeds (the rest is used for coherence management), we have roughly 0.5M seeds available in the TCM cache. Assuming 10 outstanding seeds per sender, as far as cache goes, one TCM can serve up to 50k cores while providing seeds from cache. The compute requirement of a TCM is similar to that of a conventional directory.

The number of cores per TCM may vary to match the expected workload and required TCM hit rate. Considering an extremely high hit rate for seeds in the TCM cache, our approach shows similar performance to that of a directory based system with no security at all.

# 4 Evaluation

In this section we evaluate SDSM for performance and scalability, and compare it with state-of-the-art work. We ran the PARSEC benchmark suite [48], focusing on applications that can scale to hundreds of threads. We used Pin [39] to capture the benchmarks' activity (instructions and addresses), and we developed a simulator that runs these activities on a multi-node environment, including our seed management and communication layer. The simulator is described in detail in the next section. Each benchmark was executed with no security layer as the baseline for performance measurements, with [16]'s scheme (which to our knowledge has the most recent results published for similar settings), and with SDSM. The performance results are normalized to the baseline. In our tests we used 100 clock cycles for core-to-core latency, and 80 clock cycles for calculating the keystream block [16]. We used 10 outstanding KBs per sender. Unlike [16], which assumed higher communication latency as the core count grew, our evaluations of all the schemes assumed the same latency for every setting, so the security related overheads are fully exposed.

We clearly see (Figure 13) that SDSM scales easily to a thousand cores with less than 2.5% performance reduction, and less than 0.8% performance reduction with 256 cores. [16]'s performance drops by 22% with 1,000 cores, and by 16% with 256 cores. We saw no performance improvement when increasing the number of outstanding KBs per sender beyond 10. In [16], a sender only caches block-specific KBs for a few recently modified blocks, assuming that these are going to be requested soon. This approach suffers from a KB miss rate that increases as the number of potential requestors increases. We only assign a KB for a requested block upon request, so we practically never miss any KB.

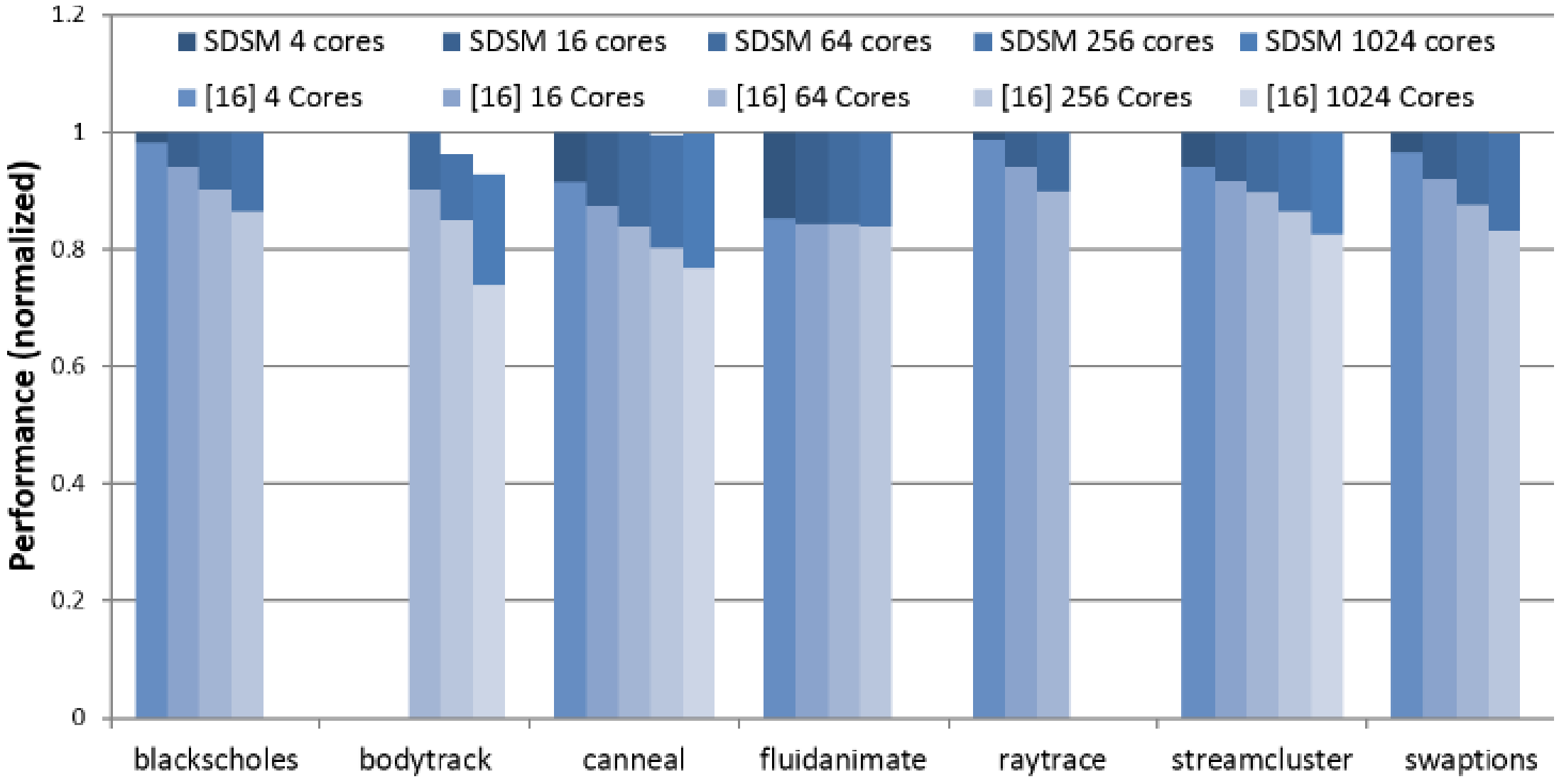

***Figure 13: Running PARSEC benchmark suite: SDSM performance relative to [16], normalized to no security***

Next, we used synthetic benchmarks to assess the system's behavior with various miss rates and core counts. Figure 14 shows that SDSM exhibits less than 0.3% performance degradation for any core count and miss rate, whereas [16] is sensitive to both (up to 25% performance reduction).

We repeated with three different communication latencies: 50, 100, and 200 clock cycles. As expected, the performance penalty for calculating the keystream blocks (by the prior art schemes) drops as the communication latency rises (Amdahl's law). We also evaluated the extra traffic caused by our scheme and found it to be similar to [16].

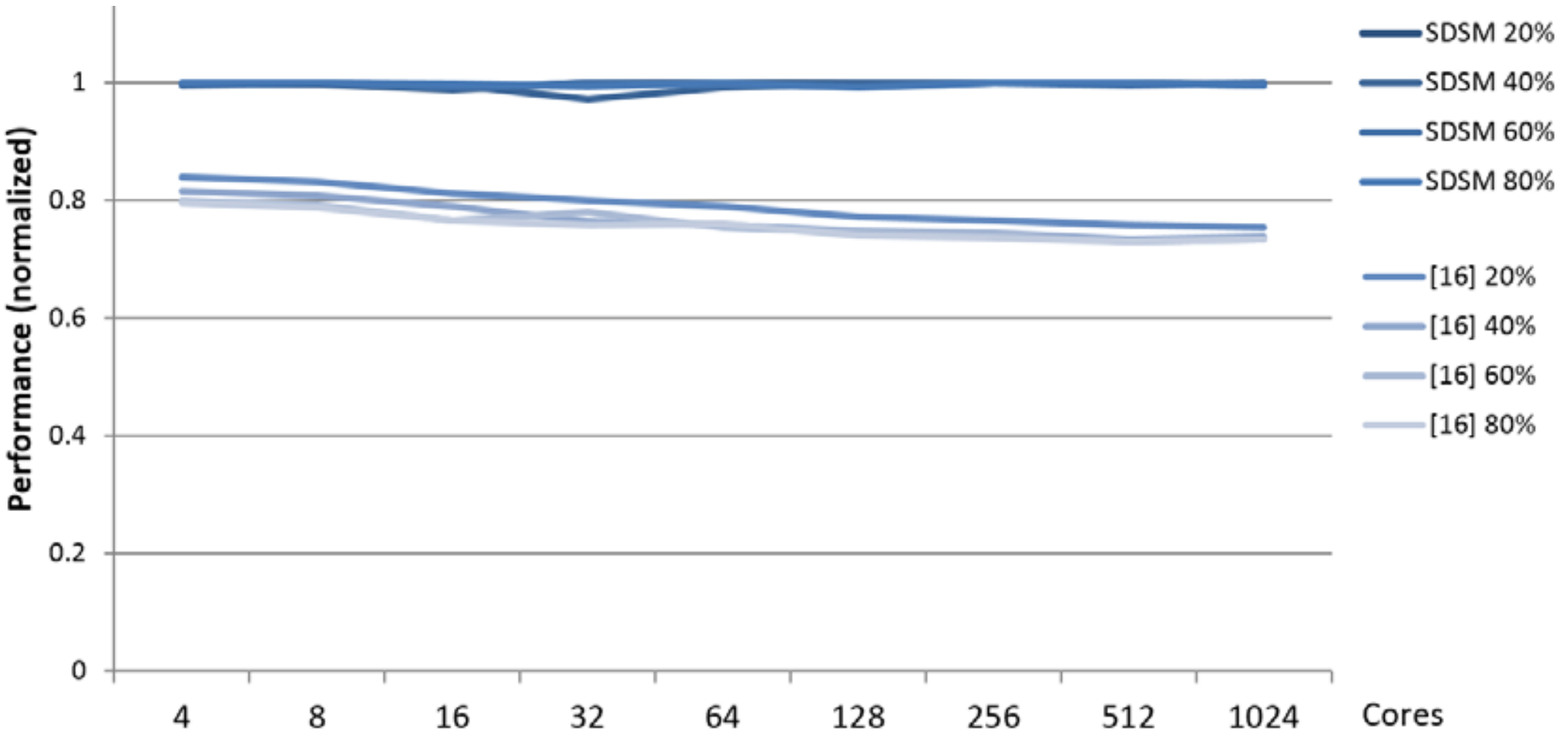

*Figure 14: SDSM vs. [16] normalized performance comparison for synthetic benchmarks with various node miss rates*

**Overheads.** Storage overhead is smaller than previously suggested architectures for the same settings [16,66,68]. The main reason is that we only keep a small cache for outstanding KBs (holding 10 entries) that will be used effectively, and only a single incoming KB per core. The number of cores assigned for TCMs is the same as would be used for a conventional directory, so we do not create new overheads there.

# 5 Implementation

In this Section we describe the simulator implemented for evaluating SDSM. This simulator provides a multi-node execution environment, where only the memory accesses are at stake. However, the level of memory intensiveness of the chosen applications affects the final performance results, so our simulator simulates the memory system accurately, and uses other operations (e.g., ALU operations) as spacers between those.

We used Intel Pin [39] to implement a tool that converts a multi-threaded binary into a sequence of instructions of the form: **ALU**, **Load [address]**, **Store [address]**, where every thread has its own sequence. Then, we used this tool to simplify the PARSEC benchmark suite [48] into these sequences of instructions.

We then built a simulator that takes these multi-threaded sequences as input, and runs them in a multi-node environment. In this environment, every node has its local memory, and requests for missing data (node-miss) are sent to a managing directory, alongside the type of request (read / write). The directory locates the current owner of the data and redirects the request to this owner. Besides running its local thread, every node accepts remote requests for missing blocks, serves these requests with the requested data, and acts on invalidation requests. For coherence, we chose the MESI protocol.

We assume that operations are blocking, but an out-of-order version of the simulator may be developed as well. Each operation takes a configurable number of clock cycles (ALU, memory access, node-to-node latency), and the end result is the total run time in clock cycles (until completion), alongside statistics that were collected.

On top of this basic simulator, we added a security preserving layer that implements SDSM algorithm, and [16] algorithm. This layer modified both the compute nodes and the managing directory.

For SDSM, we manage a list of per-node outstanding KBs (and seeds that were used for their creation), where seeds are fetched from the directory. A node that serves a missing data request uses the seed suggested by the directory for this request, and sends the data encrypted with the seed to the requestor. To save time, every missing data request that was forwarded by the directory also contains a new seed to be used on future evictions. Such arriving seeds result in KBs that will only be ready after a configurable number of clock cycles (to simulate real life operation).

The directory manages the list of per-node seeds, and requests for missing blocks are replied to the requestor with the seed-to-be-used. Each requesting node that gets a seed-to-be-used starts calculating a new KB for it that will only be ready after a configurable amount of time. If the requested data arrives before, it has to wait for the KB for the data to be decrypted.

For [16], KBs are prepared per virtual address (not address independent), outstanding KBs are only prepared for blocks that are modified, and a storage space is used to keep those. If a block is evicted and no matching KB was found, then a seed is chosen and a virtual address dependent KB has to be prepared; to present [16] in the best light, we chose to send the seed right away to parallelize things with the requestor as much as possible. Unmodified blocks need to be fetched from the memory with their encryption seed, so these are sent together to the requestor.

The statistics that are collected: total clock cycles, average number of per-core instructions that were executed, per-core per-type (ALU, Load, Store) average number of instructions that were executed, node-misses, hits in wrong state (data is already there, but needs write permissions), number of requests served, number of requests that were redirected (directory forwarded but the data was already evicted for race conditions), number of requests that were served right away, number of requests that were delayed for eviction awaiting a KB, average delay (clock cycles) for those requests, number of requests that were delayed for fetch completion awaiting a KB, average number of delay cycles for those requests, statistics on the directory activity, statistics of the traffic.

The simulator may be configured to run the baseline (memory is shared on the clear), to support [16] security method, or to support SDSM security method.

# 6 Conclusions

We presented SDSM, a novel approach for creating a scalable, secure distributed directory-based shared memory system. Exploiting native latencies of the DSM system, we are able to scale to thousands of cores, with a tiny performance degradation relative to non-secure DSMs. We are also able to avoid redundant work (relative to previous work), and thus save energy and make better use of memory space. SDSM will enable the construction of massively parallel secure and efficient directory-based coherent memory systems.

Future work includes more detailed study of per-core resource requirements (e.g., energy and traffic) for an N-core parallel task, optimizations for non-uniform memory access (NUMA) systems, and supporting dynamically changing systems.

# Chapter 7 Distributed Memory Integrity Trees

Adding support for multi node secure execution on SeM raised the need for preserving the integrity of the distributed memory, where the local compute nodes and their memories are connected via an untrusted medium (Chapter 3). In this chapter we address this need.

## 1 Introduction

Secure computing in untrusted environments like public clouds is an emerging requirement. A common (e.g. SGX [22], SeM) assumption is that the secure CPU chip and the user's own code are the only trusted elements. Every-thing else, including the board and off-chip memory as well as the operating system (OS) and hypervisor, is untrusted. Encryption can protect the confidentiality of the data residing in untrusted memory, and message authenticating code (MAC) can protect against forged or misplaced data; however, replay attacks (whereby old data is maliciously restored) by a privileged attacker may harm the integrity (version) of the data, and this requires additional measures. For this, various secure CPU-managed integrity trees are used; however, these only protect a program running at a single compute node, whereas many relevant application programs are parallel or distributed.

Distributed applications commonly use message passing interface (MPI) [79] or distributed shared memory (DSM) [80]. In MPI, the memory space itself is not distributed, so the single-node integrity solutions suffice. DSM, wherein threads are spawned and access a shared address space for data sharing and synchronization, is natively easier to program and thus attractive. However, this sharing requires a distributed integrity-preserving mechanism. Also, since the DSM coherence state metadata is managed by the underlying (untrusted) OS, it is vulnerable to manipulations. (E.g., preventing the invalidation of a local memory block, causing the read of an old value; or granting write permission without remote invalidations, causing incoherent blocks.) Secure CPUs that protect at application granularity (e.g., SGX, SeM) cannot protect this metadata against privileged attackers, and solutions like AMD-SEV [87] lack memory integrity protection.

In this work, we present the Integrity-Verified Local Coherence State (IVLCS) mechanism, which protects the coherence state of ***shared memory*** blocks against malicious manipulations. We then present Distributed Integrity Tree (DIT), a novel scheme that uses IVLCS, any existing inter-node secure coherent data transfer layer (a coherence preserving layer that transfers data securely, e.g. SDSM or [16]), and any single-node integrity tree to construct a corresponding distributed integrity tree. Finally, we use DIT to construct distributed versions of Merkle Tree [72], Bonsai Merkle Tree [34], and of Intel SGX's [22] Memory Encryption Engine (MEE) integrity mechanism [81]. Although we target protection at a process level, DIT may also protect at a virtual machine level.

The remainder of the chapter is as follows: Sec. 2 reviews single-node integrity trees; Sec. 3 presents our new schemes; Sec. 4 applies them to existing integrity trees; Sec. 5 evaluates them, and Sec. 6 offers concluding remarks.

## 2 Single Node Integrity Trees

In this section we overview integrity trees that are commonly used for single-node applications. All these will be extended by DIT in Sec. 4.

**Merkle Tree** [72]. A secure hash (based on a secret key) is calculated for each memory block (data or instructions); the hashes are stored in the clear in memory blocks ('hash blocks'), in the untrusted memory. For each hash block, a secure hash is calculated, and this is repeated in a tree structure until a single (root) hash value is calculated to protect the integrity of the entire tree. This root never leaves the CPU chip, so it cannot be forged by an attacker. The use of secure hashes ensures that without knowing the secret hashing key, an attacker cannot forge a data block and calculate a correct hash for it. The use of a chip-resident root hash furthermore ensures that even if an attacker restores old data blocks with their old hashes, they will fail verification when read into the CPU chip.

In many implementations, the hash blocks are cached on chip, so when a missing hash is fetched (from unprotected memory), it needs to be validated only up to the first ancestor hash that is already cached. Since the latter hash was verified when fetched (and possibly updated for subsequent sub-tree modifications), and is inside the chip so no attacker could have modified its value, it is treated as a root hash and the hash validation is completed. In case of block modifications, the corresponding hash block need only be updated upon eviction of the modified block.

**Bonsai Merkle Tree** (BMT) [34] targets systems that protect their memory using counter mode encryption, wherein each memory block has its corresponding counter value. [34] observed that instead of protecting the actual memory blocks using a secure hash tree, protecting the counters by a Merkle hash tree and using a per-block secure MAC will result in a smaller hash tree, the Bonsai Merkle Tree, which provides the same security guarantees as the original Merkle Tree. The smaller hash tree increases cache hit-rate, thereby improving performance.

Each memory block has a small MAC alongside the data, so when it is fetched into the cache and decrypted by the counter mode technique (assuming a correct counter), forged data will result in a mismatch between the fetched MAC and the one computed over the fetched, decrypted block. Forging a counter (e.g. old counter with old data and MAC) will be detected upon counter block fetch, since BMT directly protects the counter blocks (similarly to data protection by the Merkle Tree).

The BMT values are stored in the clear in the unprotected memory, and can be cached in the chip. BMT cannot, however, be used as is for protecting a distributed program, because the program's memory space spans multiple memories that are controlled by different CPUs.

**Intel Memory Encryption Engine** (MEE) [81] integrity tree is used in SGX [22]. In MEE, each data block has a *version*, and a Tag that is calculated using a MAC algorithm over the data and its version. The blocks of versions are protected similarly (using an upper level version and a Tag), and this continues in a tree structure up to a root version that is kept on chip.

Unfortunately, a distributed program's memory space spans multiple memories, controlled by different CPUs, so none of the above can be used as is to protect it.

# 3 Distributed Integrity Trees

A single-node integrity tree is responsible for its CPU's fixed set of memory blocks. In a DSM, however, valid blocks move among nodes, so a collection of local integrity trees with fixed responsibilities does not work. Using instead a **single root value** (stored in one of the secure CPUs) to maintain the integrity of the entire distributed memory is impractical, because this value must change upon any write to memory, overwhelming both the network and the CPU holding the root value. A replicated global integrity tree is even more expensive to create and maintain.

Our key observation is that, by having a secure coherent inter-node data transfer layer, each node need only 1) know with certainty which blocks are pre-sent in its local memory ("*locally resident*"), and 2) only maintain the integrity of those. Specifically, each node must maintain an ***integrity-protected coherence state*** ("locally resident + write permission"/ "not resident") of all its shared memory blocks, be they locally resident or not. However, a node need not even be aware of changes being made to a non-resident block's content or location until it is fetched again. We next exploit this key insight.

**Definition**. The *local coherence state* of a block is *integrity verified* (IVLCS) iff the coherence state of the block is protected by a local integrity preserving mechanism: when the CPU checks if a block resides locally, the results are either correct or an error is declared.

**Constructing the IVLCS** is done via the following modifications to existing integrity verified structures:

1) Assign a special value 'NR' to mark a locally non-resident block (unallocated or currently present at a different node).

2) Add a per-block write-permission bit (0-R, 1-RW).

'NR' can be kept in the data block itself, in the leaf of the hash tree responsible for the integrity of this block, in place of the counter of this block (if counter mode is used), etc.; the write permission bit can be kept along-side. NR can either be marked using a dedicated value of existing bits, or using an additional bit (incurring memory overhead). In fact, one may use a partial integrity tree protecting only the allocated and locally existing part of the memory, in which case NR can also mark an unallocated or locally unavailable subtree (i.e., a block containing solely NR’s is omitted, and its predecessor hash contains NR, recursively).

**The Distributed Integrity Tree (DIT) scheme.**

DIT uses secure CPUs that employ a secure coherent inter-node data transfer layer to protect the data while in transit, along with integrity verified local coherence state (IVLCS) and single-node local memory integrity trees to protect the data while in a secure node.

The data transfer layer, embedded in the secure CPU, is responsible for trustfully locating the current owner of a requested block (e.g. by communicating with a trusted directory), and providing encrypted (if desired) and integrity protected data transfer between the current owner and the requesting node. Any tamper attempt is detected, and an error is treated by the trusted compute node.

Each CPU has a single-node local memory integrity tree. Locally resident data blocks (along with their integrity-protecting metadata) are fetched into the CPU both for local use and for verification prior to being sent re-motely. The local memory integrity mechanism detects modifications of both data and integrity blocks (as done in single-node integrity trees). This also holds for local permission checks, which are integrity protected (by IVLCS definition). Non-resident blocks are determined properly (by IVLCS), so only the missing ones are re-quested from their remote owner. These are requested by the secure CPU, their current owner is detected, it sends them, they are received correctly, and are invalidated (when required) – all this by the data transfer layer. Having arrived, their local coherence state is trustfully up-dated in the local integrity structure, and these become locally resident blocks. **DIT can thus be used to pre-serve the integrity of the entire distributed shared memory.** It thus provides all the truly required functions of a global integrity tree at a cost and performance that are very similar to those of independent local trees.

DIT was described with cache-block sharing granularity. For memory-page granularity

DSM, an in-CPU buffer can hold the rest of the page while updating its integrity, not delaying the availability of the requested cache line. Other approaches, like bringing the entire page into the local cache and marking it 'dirty', remain for future research.

# 4 Distributed Tree Examples

In this section we apply DIT to the three aforementioned single-node trees, yielding corresponding distributed integrity trees. For each, we choose the data structure that marks a non-resident block and write permission, and discuss its behavior.

**Distributed Merkle Tree (DMT).** We let the hash corresponding to a data block (leaf hash of the Merkle Tree) contain the actual hash for an residing block, and 'NR' for unmapped blocks or ones currently residing only in other CPUs' local memory. (An absent data block could itself be marked 'NR', but its hash is smaller so hash marking is more efficient.) A per-block write permission bit is kept alongside the leaf hash, and is checked before a write operation. Since hash values are also kept in blocks, hashes of residing and non- residing blocks ('NR') may reside in the same hash block. A per-CPU local Merkle Tree is maintained normally, and each CPU maintains a local root hash of its local Merkle Tree. (See Figure 15.)

On a cache hit, a block is accessed directly; if a first write is requested, then write permission is requested from the secure coherent node-to-node data transfer layer. On a cache miss, the block's hash (and write-permission bit) is examined. If the stored hash value is not 'NR', the block is simply fetched from the local memory; if it is 'NR', a request is sent to the data transfer layer for validating and then bringing the block securely (by definition). Once the requested block arrives, it is securely stored in the cache, marked 'dirty', and its write-permission bit is updated based on the request type. Only when a 'dirty' block is locally evicted, its hash must be updated, since until then the block is simply accessed via the cache. Similarly, upper level hash blocks are updated as soon as their children are evicted. If a cached modified block is requested by others, its write permission is revoked; its local hash is either up-dated instantly or upon eviction. If a CPU invalidates a block, its hash becomes 'NR' and the local integrity structure is updated similarly.

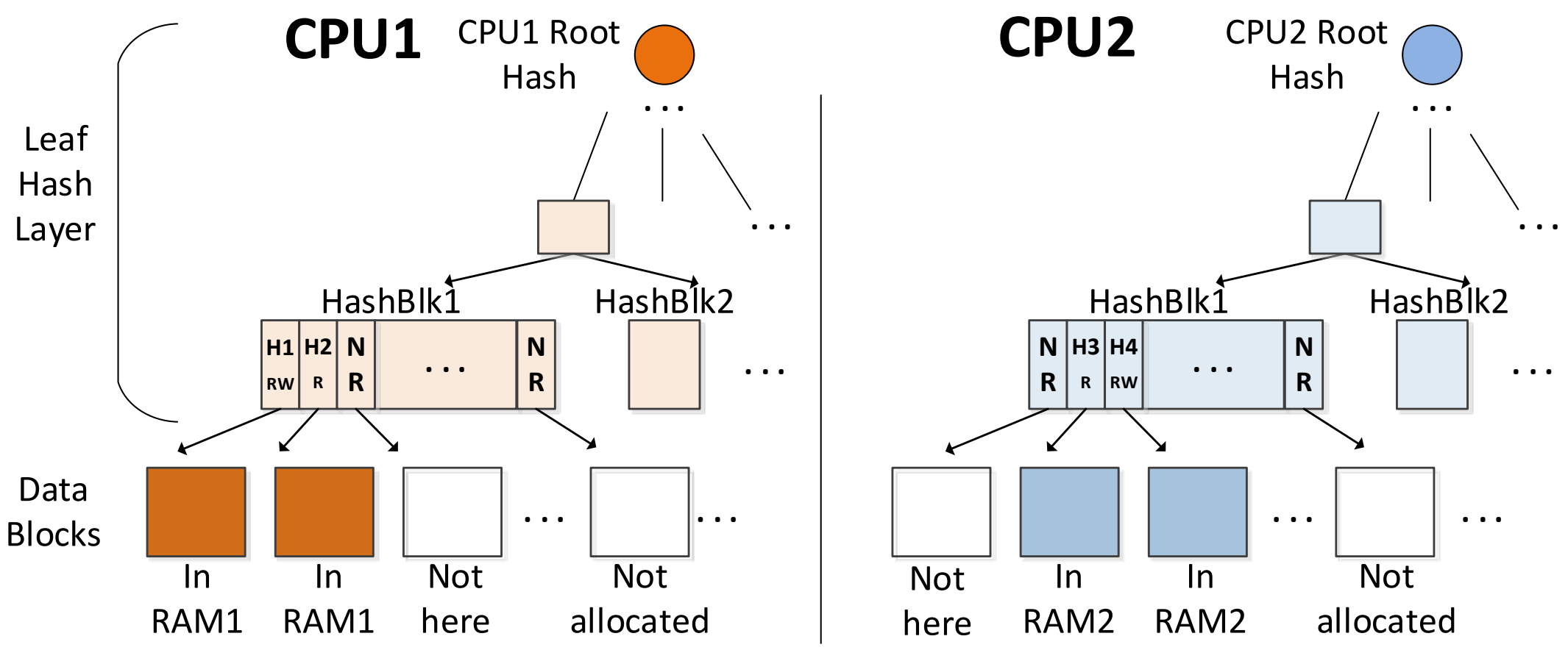

*Figure 15: DIT example: distributed Merkle Tree (DMT).*

**Distributed Bonsai Merkle Tree (DBMT).** We construct DBMT with private per-CPU encryption counters (not shared), so other CPUs cannot access them, thereby avoiding redundant management (inter-node communication for counter updates even for inactive blocks) and potential false sharing. The counter corresponding to the data block either contains the actual counter for a resident block, or 'NR' for unmapped blocks and ones residing only in other CPUs' local memory. A per-block write-permission bit is kept alongside the counter (also covered by the MAC calculation), and it is checked be-fore a write operation. A per-CPU local BMT is maintained normally, and each CPU maintains a local root hash of its local BMT. See Figure 16.

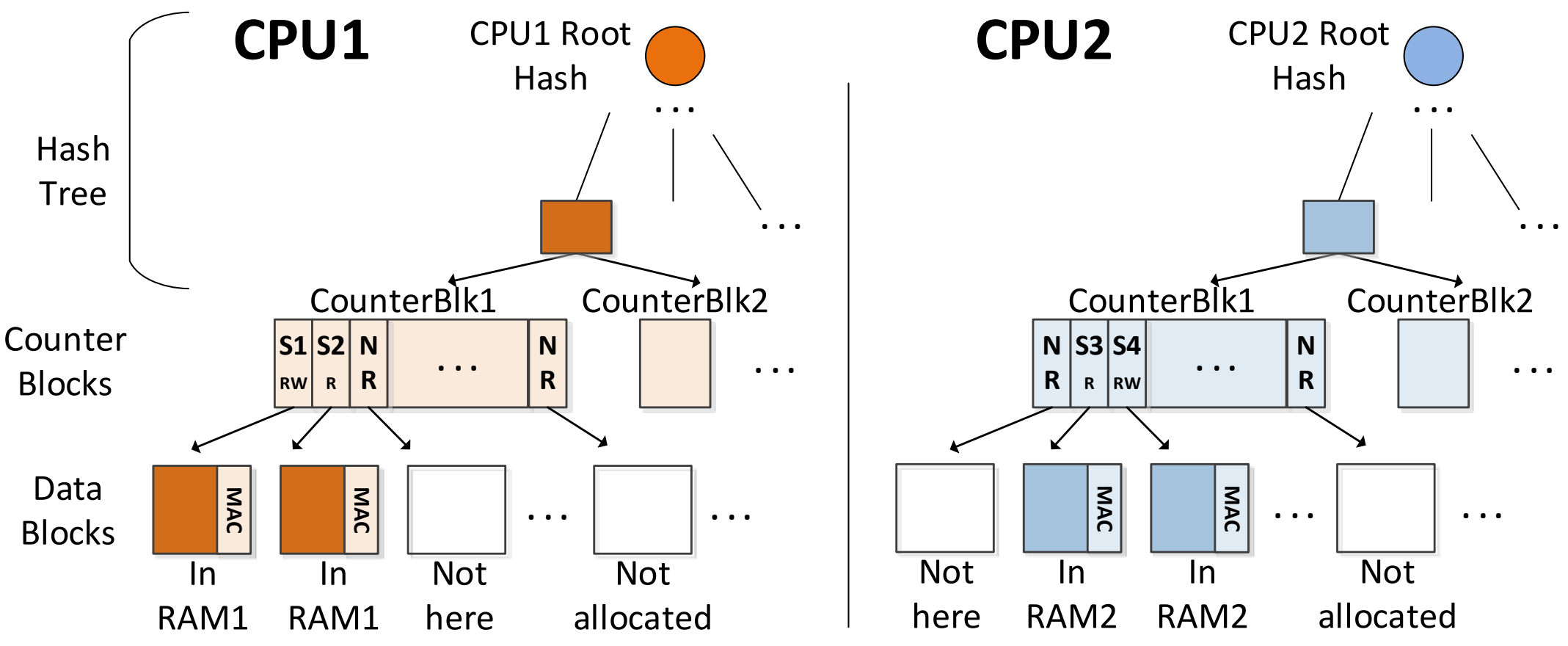

*Figure 16: DIT example: Distributed Bonsai Merkle Tree (DBMT).*

**Distributed Memory Encryption Engine (DMEE).** Although Intel SGX currently does not support distributed execution, we show that its MEE integrity tree can also benefit from DIT. The tree structure is fairly similar to BMT, so we construct DMEE with private per-CPU versions. The version corresponding to the data block either contains the actual version for a resident block, or 'NR' otherwise, alongside a write permission bit. A per-CPU local-MEE is maintained normally, and each CPU maintains a local root version of its local-MEE. See Figure 17.

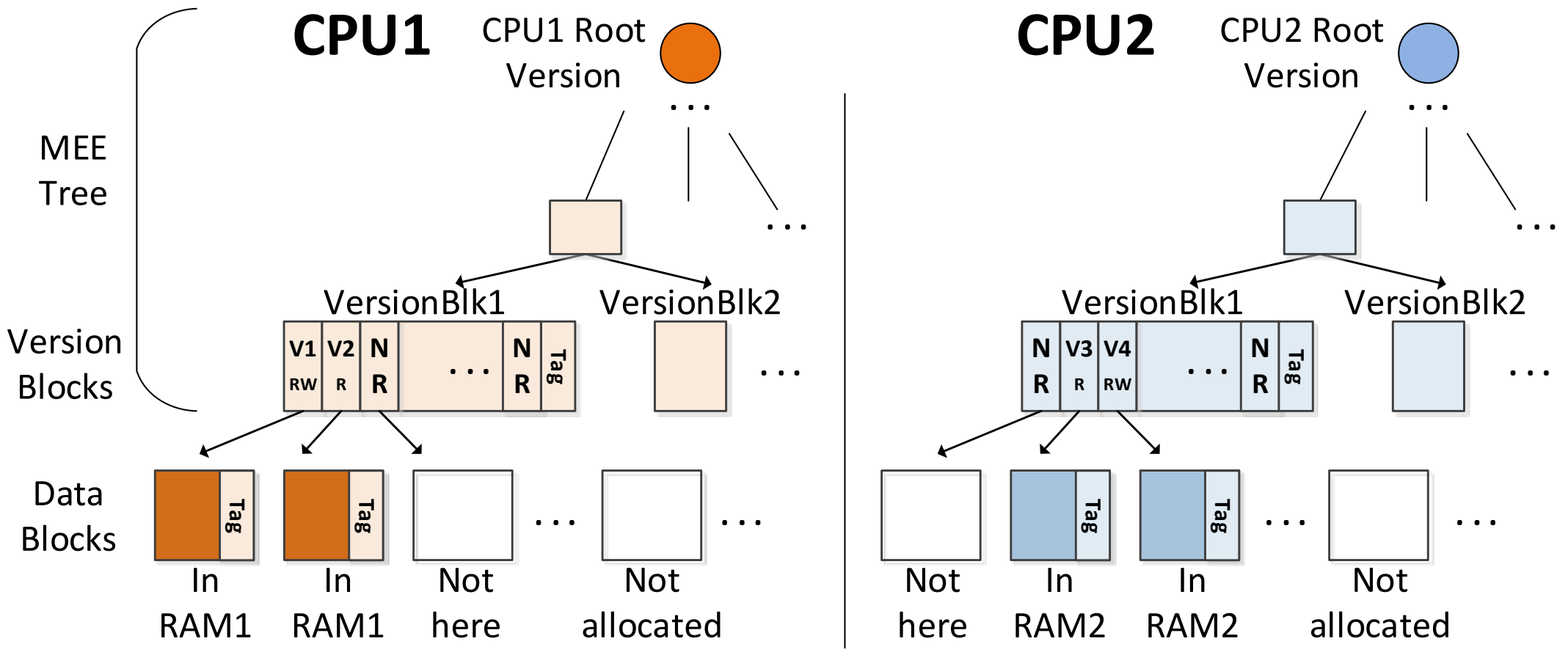

*Figure 17: DIT example: Distributed Memory Encryption Engine tree (DMEE).*

Note that in DIT, migrating a thread to a new CPU does not require upfront data movement; rather, only an empty hash/version/counter structure, a new empty local-integrity tree, and a root hash. The thread’s code and data will be subsequently fetched as shared data in the system.

# 5 Evaluation

We evaluate the additional operations performed by DIT relative to a baseline with secure coherent inter-node data transfer (SDSM) and single-node integrity trees, but without coherence integrity (and therefore no global memory integrity). We consider memory sharing at cache block granularity, though coarser granularity is possible. Therefore, the baseline's OS must verify the status of memory blocks upon access. Consider the following cases:

**Locally resident block – cache hit.** Both in the baseline and in DIT, a cached block is accessed directly for read. Before writing for the first time, write permission is first requested by the data transfer layer; however, an already modified block does not require a permission request. Therefore, performance is similar.

**Locally resident block – cache miss.** In the baseline, a missing cache block results in a local coherence lookup to check for the presence of the block. It is then fetched and verified with the local integrity system. In DIT, the value fetched for the local coherence lookup (hash / counter / version) also serves for verifying the integrity of the resident block, so no additional overhead is caused.

**Locally non-resident block.** In the baseline, a missing cache block results in a local coherence lookup to check for the existence of the block. If not resident, a data transfer request is sent, and the block arrives into the cache safely. The integrity structure is only updated when this block is evicted. With DIT, an integrity protected coherence lookup is performed, which is likely to be more costly than the local coherence lookup. Then, a remote request is performed similarly.

The average read time for the baseline is:

$$t_b = t_c + (1 - H) \cdot \left(t_{coh} + t_{fetch} + LE \cdot t_{int} + (1 - LE) \cdot t_{rem}\right)$$

Where H is the cache hit rate, $t_c$ is the cache hit time, $t_{coh}$ is the average coherence lookup time, LE is the probability of finding a cache-missed block in the local memory, $t_{int}$ is the mean integrity verification time, and $t_{rem}$ is the mean time to fetch a block from another node.

The average read time for DIT is:

$$t_m = t_c + (1 - H) \cdot (t_{int} + t_{fetch} + (1 - LE) \cdot t_{rem})$$

The difference between the two is:

$$t_m - t_b = (1 - H) \cdot (-t_{coh} + (1 - LE) \cdot t_{int})$$

Performance differs only for cache misses, and the difference depends on $t_{coh}$ versus $(1 - LE) \cdot t_{int}$.

We evaluated DBMT's overhead "synthetically" on top of a system with single-node BMTs and SDSM. We chose $t_c = 1$, $t_{fetch} = 100$ <clock cycles>. First, using $t_{coh} = 0$ (no overheads for the baseline's coherence check) and $t_{rem} = 5 \cdot t_{fetch}$, we evaluated the average memory access time (AMAT) overheads for the entire range of node miss rates (0-100%), where the high miss rates simulate intensively shared blocks; since [34] measured 1% cache miss for the BMT verification, we use 0-4% the for the DIT verification (Figure 18).

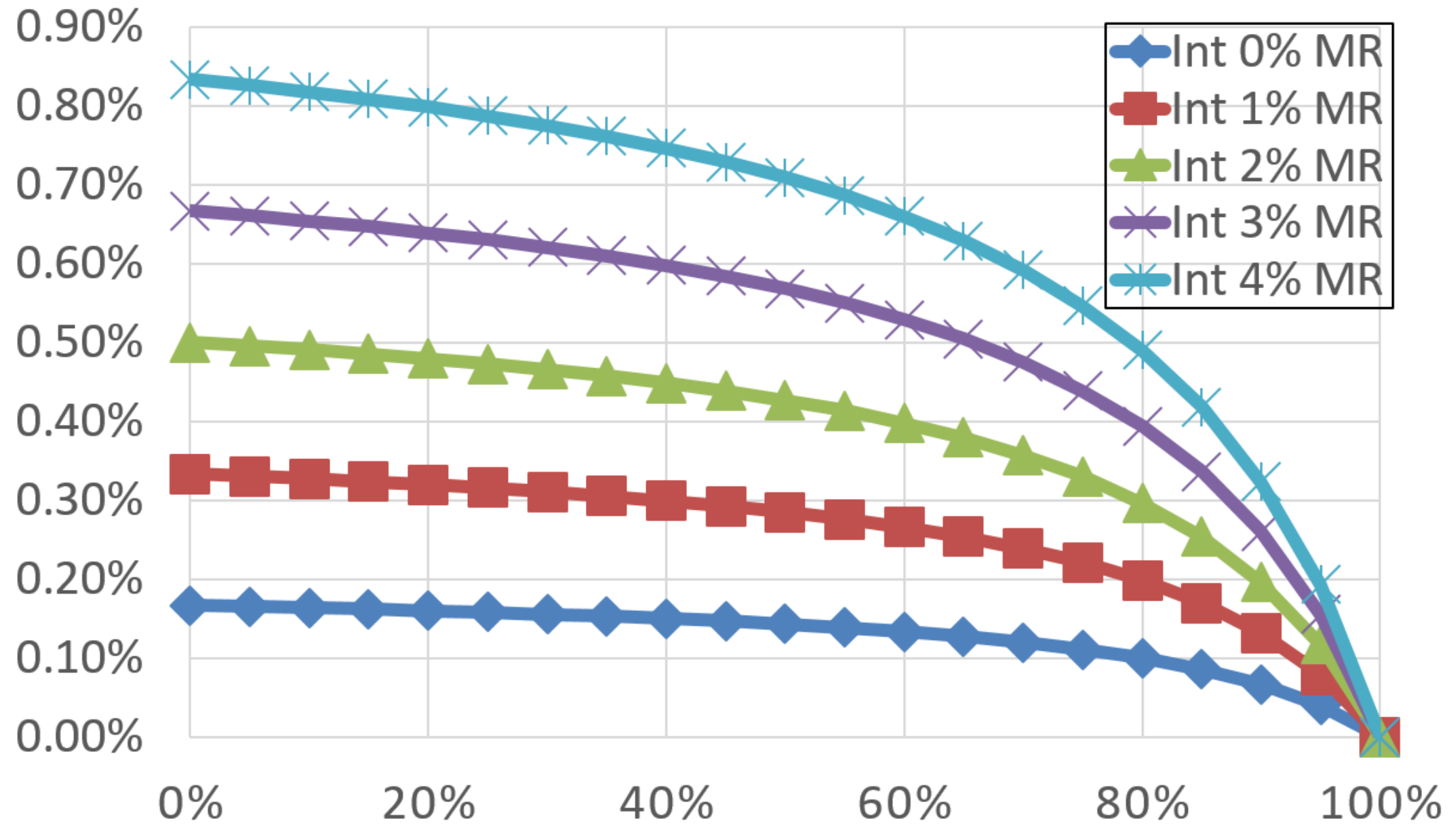

***Figure 18: AMAT Overheard at Various Integrity Miss Rates, Hop Time Between Nodes Equals to 5 Memory Access Times. Baseline Integrity is Costless.***

We then repeated for $t_{rem} = 10 \cdot t_{fetch}$ (Fig 19) and $t_{rem} = 50 \cdot t_{fetch}$ (Fig 20).

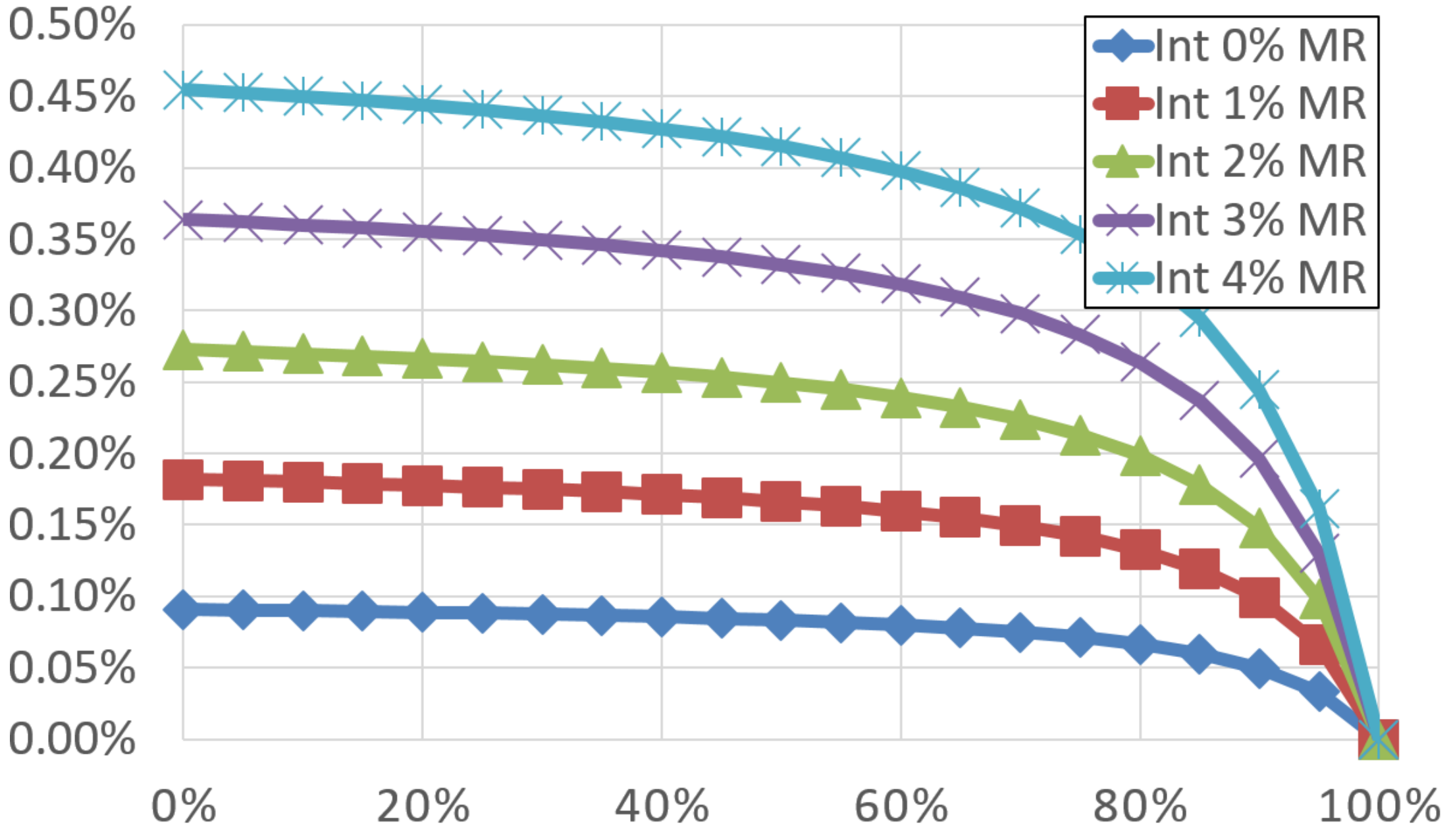

*Figure 20: AMAT Overheard at Various Integrity Miss Rates, Hop Time Between Nodes Equals to 10 Memory Access Times. Baseline Integrity is Costless.*

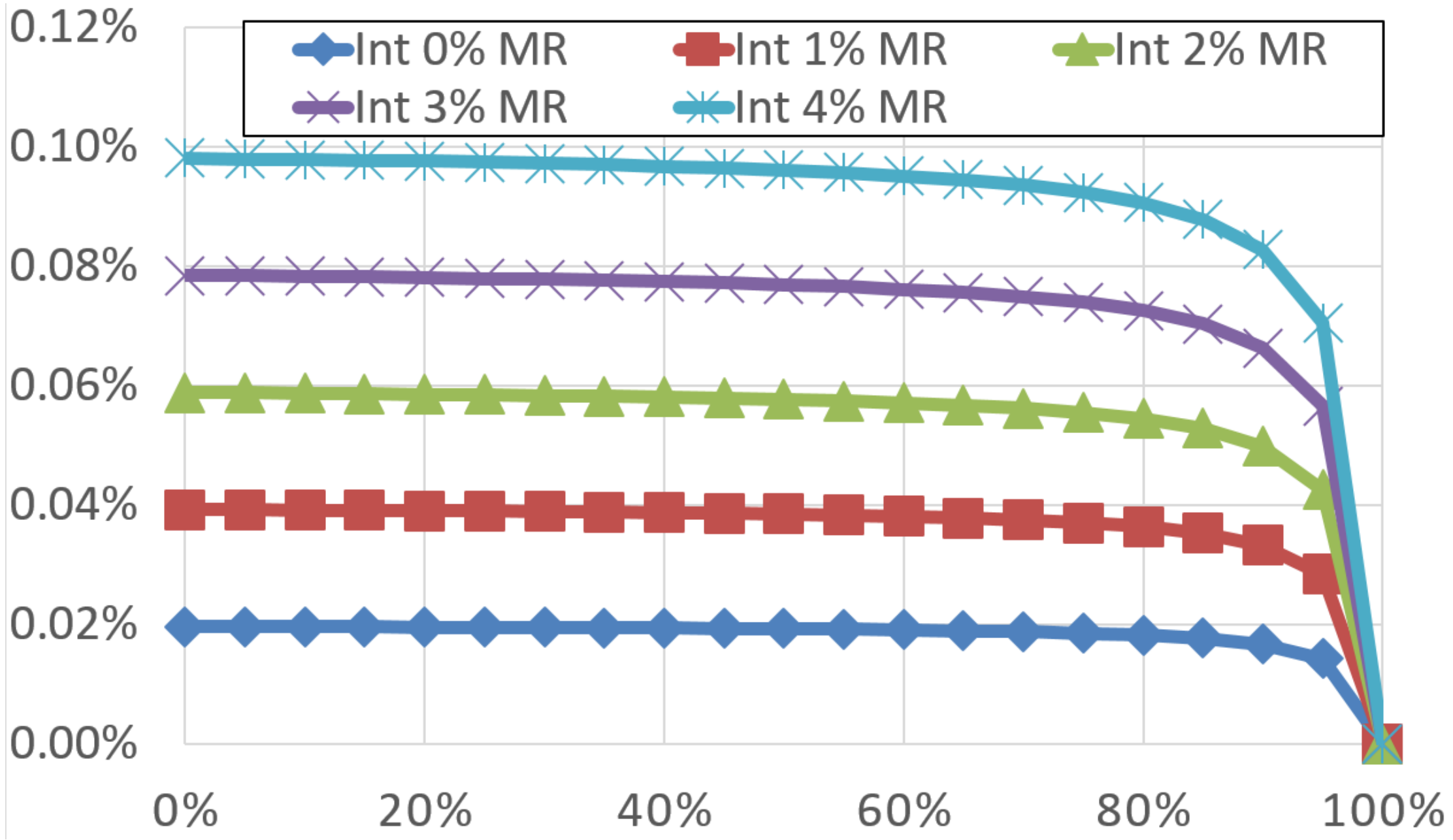

*Figure 19: AMAT Overheard at Various Integrity Miss Rates, Hop Time Between Nodes Equals to 50 Memory Access Times. Baseline Integrity is Costless.*

**Results:** not only is the AMAT overhead always less than 1%, it moreover drops to zero as the node miss rate rises.

Next, we fixed the DIT cache miss to 2% (double than [34]'s, reflecting DIT's slightly larger tree) and $t_{rem} = 5 \cdot t_{fetch}$, and evaluated $t_{coh}$ as a single memory access with miss rates of 0-4% (baseline's coherence check) and (0-100%) node miss rates (Figure 21).

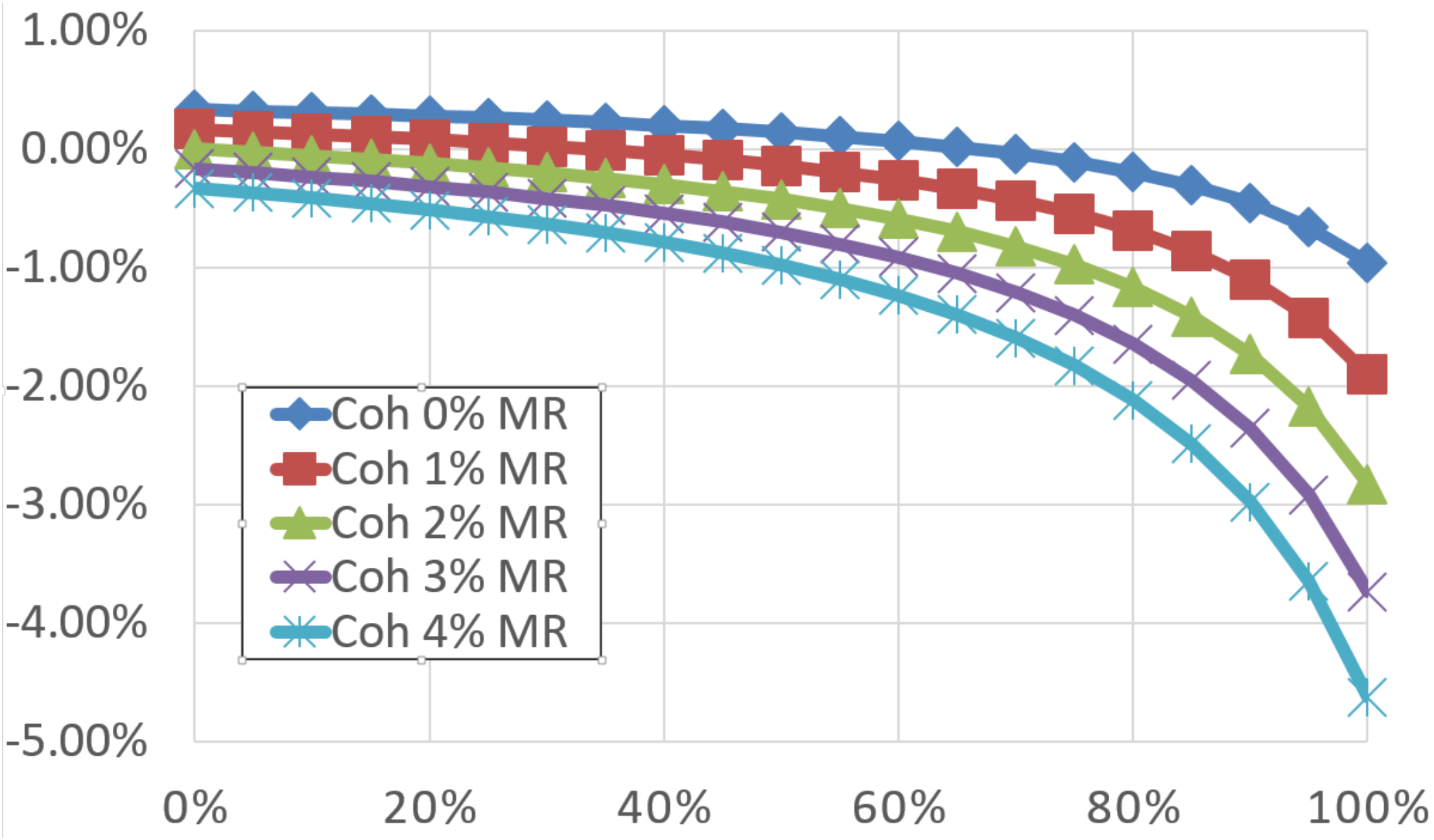

*Figure 21: AMAT Overheard Compared to Various Baseline Integrity Miss Rates, Hop Time Between Nodes Equals to 5 Memory Access Times.*

We see that, relative to a baseline with non-zero overhead for coherence check ($t_{coh} > 0$), DIT may even shorten the AMAT.

To evaluate with real workloads, we used DBMT and ran the PARSEC benchmark suit [48] with 32, 128, and 256 compute nodes. Figure 3 shows that the performance overhead is <0.5% relative to the baseline, with an average of ~0.2%. The node miss rates that were measured for PARSEC are up to 7% (*canneal*, *fluidanimate*), with memory/compute intensiveness of up to 45%.

Figure 22 shows DBMT's overhead, and per-benchmark locally non-existing rate. We see that DBMT's overhead is <0.5% compared to the baseline with the single-node BMT.

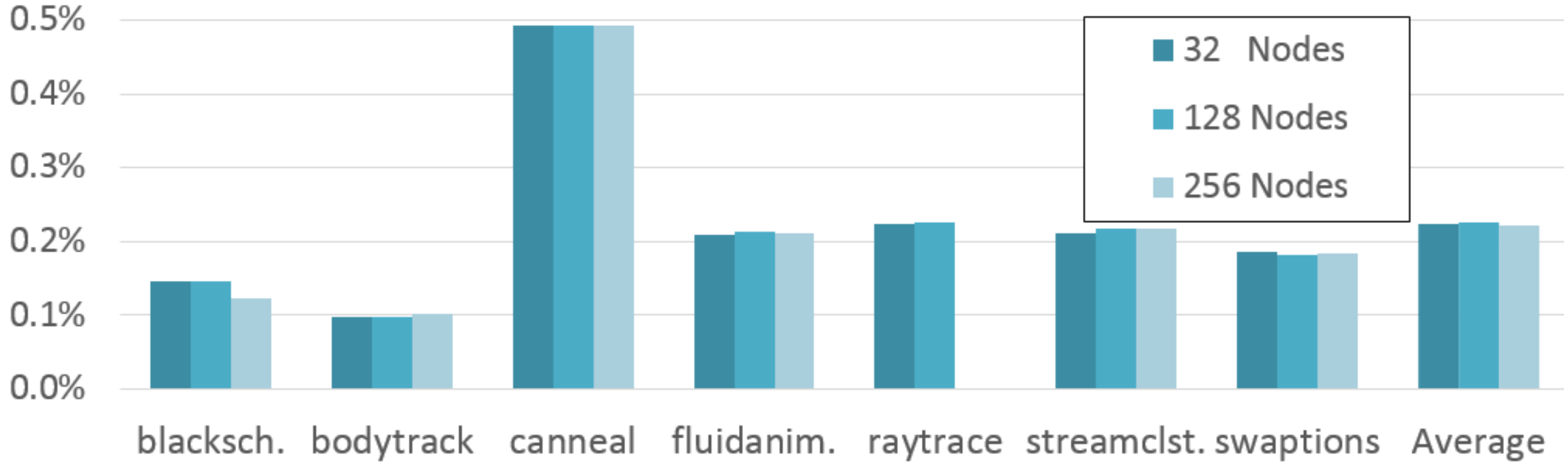

*Figure 22: DBMT performance overhead running PARSEC for different core count. Also showing percentage of misses served by other cores*

**Memory overhead.** A per-block bit in the local integrity tree's leafs is required to mark NE, and in some implementations this can be obviated (e.g., a DBMT counter / DMEE version may be set to its max value ('0xffff..') for 'NR'). As mentioned before, DIT supports partial trees by using a high level NE to compress an entire unused subtree, similarly to the integrity tree of the baseline. DIT's write-permission bit is merely moved to an integrity protected structure, since any DSM system must maintain that anyhow.

# 6 Conclusion

We presented the integrity verified local coherence state (IVLCS) mechanism, and used it with a prior-art secure inter-node coherent data transfer layer and existing single-node integrity trees to construct the Distributed Integrity Tree scheme (DIT). This enabled us to extend various single-node integrity trees (e.g., Merkle Tree, Bonsai Merkle Tree, and Intel SGX's Memory Encryption Engine) into corresponding distributed versions (DMT, DBMT, DMEE), adding the capability to preserve memory integrity of a distributed shared memory sys-tem. While providing the extra security guarantees, DIT exhibits only a slight performance reduction across a wide range of runs, and only minor additional complexity. Furthermore, the local nature of IVLCS allows DIT-supporting systems to scale well. DIT is thus a major step towards high performance, scalable secure computing.

# Chapter 8 Conclusions

In this thesis we presented the Secure Machine (SeM), a CPU architecture extension for secure execution on untrusted platforms. SeM is composed of several building blocks, some of which may also be used in other secure architectures.

The basic Secure Machine (SeM) secures existing single-threaded binaries on a single core CPU, by using a Security Management Unit (SMU) that controls the cache and registers, and guards the cache from the execution unit access. By a novel binding of code and data that are under the same ownership, SeM assures that only the owning code can access its data in a manner that no other software, privileged as it may be, can break. SeM also approaches the CPU / OS integration, discussing interrupts, context switches, system calls, signals, and etc.

To support single- and multi-threaded binaries running on single- and multi-core CPUs (CMP), machines with multiple CPUs, and many-computer systems, SeM uses SMU-to-SMU secure communication, and hardware assisted secure thread creation, migration, and termination. SeM also supports heterogeneous systems that comply with a specific set of requirements.

SeM assumes that shared libraries are embedded into the binary when preparing it for SeM, thus we discussed another approach in which shared libraries are not embedded (and therefore untrusted), and the secure program calling it will only reveals the data required for their operation. Furthermore, we provided a short discussion for the (assumingly) superior approach (in terms of security) of embedding shared libraries into the secure binary, and highlighted an inherent trade-off between these approaches.

SeM relies on important building blocks for secure data sharing and memory integrity on multi-node settings.

To enable secure data transfer between CPUs or computers that are connected via an untrusted medium, we presented Secure Distributed Shared Memory (SDSM), a high performance algorithm and architecture extension to support secure memory sharing between trusted entities that are connected through an untrusted medium. Using smart resource allocation and hiding the latency of cryptographic operations behind the inherent communication latencies, SDSM allows thousands of compute nodes to communicate securely with negligible performance reduction. Furthermore, SDSM is designed such that the amount of per node resources does not increase with the number of the participating nodes. This is a critical property, since one cannot expect a cloud owner to replace existing CPUs just to add more CPUs to its network.

To protect the memory integrity of a distributed secure program that runs on many CPUs or computers that are connected via an untrusted medium, we presented Distributed Integrity Trees (DIT), a method that extends existing single node integrity trees to preserve the integrity of a distributed memory. We have shown that DIT is easily applicable to three types of commonly used single node integrity trees, and that its added resources and performance penalty are both negligible.

We designed and implemented SeM-prepare, a tool that instruments existing binaries and makes them SeM-ready, and SeM-simulator for running SeM-ready binaries to evaluate SeM's performance. We ran the SPEC CPU2006 and PARSEC benchmark suites correctly (including C and C++ parallel workloads from various fields, and also AI, compiling and interpretation, and video and data compression). The performance reduction of SeM is less than 3%, and the added area and power are negligible.

Jointly providing security, performance, backward compatibility, and furthermore supporting parallel and distributed programs and even accelerators, SeM thus provides security in a usable manner for general purpose computing.

Topics for further study include hardware support for secure I/O, further study of the use of secure accelerators, direct support for scripting and bytecode supported languages, and using RDMA with SeM.

# Appendix A: SeM-Simulator

We implemented SeM-Simulator, to run SeM-ready binaries and evaluate their performance. While the performance results of SeM and ParSeM were reported in Chapter 2 and Chapter 3, respectively, in this chapter we focus on the simulator tool.

SeM-Simulator runs SeM-ready binaries by simulating the SMU's behavior. It was implemented using Intel Pin [39] just in time (JIT) mode that provides dynamic binary instrumentation for catching events and running handlers on those. Some of these handlers emulated the SMU behavior, and some collected statistics. Its main features are:

- Distinguishes between secure and non-secure memory regions;
- Maintains security modes (trusted / untrusted), including clear and restore the secure context on mode changes;
- Enforces the SMU's Secure Access;
- Supports two stacks, one secure and another non-secure;
- Supports all the SeM SMU instructions and enforces the trust mode required for their invocation;
- Supports special call / syscall instructions for invoking untrusted code;
- Collects statistics

Being Pin-based, SeM-Simulator does not interfere with the normal operation of the program, just like SeM doesn't. This means that the program is mostly executed normally on the machine, and only the SMU's unique mechanisms are emulated by software. To allow the use of special SMU instructions, these are implemented using multi-byte NOP instructions, which consist of many free bits that we use for encoding the new instructions. The simulator tracks these instructions and runs a software handler when these are invoked. (On a real machine, these would preferably get their own opcodes.)

**Maintaining secure memory regions.** SeM-Simulator keeps a list of secure and non-secure memory regions. All the memory sections of the binary are considered secure (determined by querying pin for the low and high address limits of the image), except for the *nosec_init* and .plt sections. Memory regions that are outside the binary's memory regions (e.g., shared functions or the code of the OS) are considered untrusted. SeM-Prepare instrumented the binary's memory allocations by an initialization instruction SMU_InitA; SeM-Simulator identifies this instruction and adds it to the secure memory region. Additionally, freeing a memory block removes it from the secure memory region.

**Security modes.** The security modes of the program are maintained, where the available modes are trusted and untrusted. Each instruction is checked for its address before it is executed; an instruction from a secure memory region is considered secure, and an instruction from a non-secure region is considered non-secure. A secure instruction that tries to run in trusted mode is simply granted, and similarly a non-secure instruction that tries to run in untrusted mode. A non-secure instruction that tries to run in trusted mode causes a mode switch to untrusted, during which the context is saved aside, the registers are cleared, and the stack pointer is changed to the non-secure stack, and the instruction is set to run. A secure instruction that tries to run in untrusted mode causes a mode switch to Trusted, during which the registers are recovered, and the instruction is set to run.

**Memory access (*Secure Access*).** Each memory instruction (load / store / push / pop) is checked for its target; instructions that try to access a memory region that does not match the instruction's source region are blocked, and the program is halted. Special memory instructions (SMU_StoreNA / SMU_LoadNA) are only allowed to run if secure, and these are allowed to access non-secure memory regions.

**Special stack operations.** SMU_PopNA and SMU_PushNA instructions are only allowed to run if they are secure. These instructions access the non-secure stack (inactive at the time of invocation), so aside from accessing non-secure memory running these also modifies the non-active stack register to be restored upon switching to untrusted mode.

**Calling untrusted code.** SeM-Simulator supports SMU_syscall and SMU_CallNoSec(i); once these instructions are invoked, the next mode switch to untrusted will not clear the desired number of argument registers, and the return address will be put in the non-secure stack (so returning is allowed). Furthermore, on the following mode switch to trusted, the return value register will not be restored, so values can be returned to the secure code.

**Non-secure code that calls secure code.** For maintaining the flow of operation of the secure program, non-secure code may only *return* to the secure code on points that it left before. E.g., secure code that invoked a non-secure function is allowed to return to the instruction following the original *call* instruction. However, secure applications will most likely start by invoking non-secure code that will use *call* to invoke the secure part of the program. On the first non-secure call that invokes a secure function, SeM-Simulator will store the return address in the secure stack, so when the secure program finishes it may return to its original non-secure caller.

**Collection of statistics.** SeM-Simulator collects many statistics on the program while being executed: number of instructions executed, mode switches, thread creations, dynamic memory allocations and the average allocated size, I/Os (*sys_read and sys_write* system calls).

# Appendix B: SeM Instruction Set

The SMU operation requires instructions that are accessible to all users but do not leak data or secret keys. As the SMU instructions can change important security settings that may or may not be accessed by subsequent instructions, they are treated as barriers in out-of-order or speculative execution. Table 3 provides their formal definitions.

| Instruction | Explanation | Only run if Auth |
|---|---|---|
| **Setup and Results** | | |
| *SMU_GenKeys ()* | Generate a pair of public-private keys (PbK, PrK). Optimization: prepare beforehand and store in a FIFO. Return Value: PbK, signed by the SMU. | N |
| *SMU_StoreKeys (PbK, enc[SymK && HashK], phash, FirstLEP)* | RSA decrypt enc[Skey && Mkey,by PbK] using PrK and store the keys, phash, and FirstLEP in an empty table entry. | N |
| *SMU_SetPID (phash)* | If no table entry with *phash* exists, report an error; else, set the current PID in the found entry. Destroy any remnants of an existing SMU entry with the same PID, and purge such blocks in the cache. | N |
| *SMU_GetResults (PID, rand)* | Returns PID's error status padded by rand and signed & encrypted w. Skey & Mkey. | N |
| **Context Switch** | | |
| *SMU_EvictContext ()* | Stores the content of the SMU Sealed Storage in the secure process memory. | N |
| *SMU_RestoreContext (PID)* | Loads the content of the SMU sealed storage from the secure process memory. | N |
| **Memory Access Instructions** | | |
| *SMU_StoreNA (address, data)* | Stores data into a memory block regardless of the block’s Auth status, and sets its Auth bit to False. | Y |
| *SMU_LoadNA (address)* | Loads data from a memory block whose Auth bit is False. | Y |
| *SMU_InitA(addr, size)* | Fills a memory block with '0's regardless of its Auth status, and sets its Auth bit to True. Used in conjunction with write-no-allocate | Y |
| SMU_PushNA | Pushes data into the non secure stack, regardless of its Auth status, and sets its memory block's Auth bit to False | Y |
| SMU_Pop_NA | Pops data from the non secure stack, regardless of its memory block's Auth status. | Y |
| **Special Call Instructions** | | |
| *SMU_syscall (argnum)* | Invokes a system call, and leaves *argnum* register-arguments in place on mode change | Y |
| *SMU_CallNoSec(i)* | Similar to *call*, with three minor differences: 1) (only) the first i register arguments will not be cleared in the next switch to *Untrusted mode*; 2) the return address is pushed into the NSS instead of the SS; and 3) the next switch to *Trusted mode* will not clear the register used as return value | Y |
| **Tread Management** | | |
| *SMU_NewThread()* | Prepares a new TSC with its following address as LEP. Returns SCID. | Y |
| *SMU_NewThreadDelete(SCID)* | Deletes a new TSC of ID=SCID | N |
| *SMU_NewThreadAttach(TID, addr)* | Attaches a new TSC with LEP addr to a thread with ID=TID | N |
| *SMU_ThreadDelete(TID)* | Deletes an existing TSC of thread with ID TID | N |
| *SMU_MigrateEntry(Phash, Target)* | Migrates a new entry with Phash to Target machine | N |
| *SMU_MigrateThread(PID, TID, addr, Target)* | Migrates thread TID (if active) or with LEP=addr (if not active yet) of process PID, to Target machine | N |

***Table 3: SeM instruction set***